\documentclass[]{aa} 
\usepackage{graphicx}
\usepackage{subfigure}
\graphicspath{ {images/} }

\usepackage{txfonts}
\usepackage{booktabs}
\usepackage{amsmath}
\usepackage{epstopdf}
\usepackage{enumitem}
\usepackage{multicol}
\usepackage{multirow}
\usepackage{chngcntr}
\usepackage{chngcntr}
\usepackage{epstopdf}
\usepackage{lscape}
\newcommand{\kms}{km s$^{-1}$\xspace}
\newcommand{\HI}{{\rm H\,{\scriptsize I}}\xspace}

\usepackage{natbib}
\usepackage[colorlinks, citecolor={blue}]{hyperref}

\begin{document}

\title{Radio continuum and OH line emission of high-z OH megamaser galaxies}

   \subtitle{}

   \author{Zhongzu Wu
          \inst{1}\fnmsep\thanks{zzwu08@gmail.com}
          \and
          Yu. V. Sotnikova\inst{2}
           \and
         Bo Zhang\inst{3}
         \and
         T.~Mufakharov\inst{2,6}
         \and
         Ming Zhu\inst{4,5}
         \and
         Peng Jiang\inst{4,5}
         \and
         Yongjun Chen\inst{3}
          \and
         Zhiqiang Shen\inst{3}
          \and
          Chun Sun\inst{4,5}
          \and
          Hao Peng\inst{1}
          \and
          Hong Wu\inst{1}
                 }

   \institute{College of Physics, Guizhou University, 550025 Guiyang, PR China
              \email{zzwu08@gmail.com}
              \and
Special Astrophysical Observatory of RAS, Nizhny Arkhyz 369167, Russia
              \and
              Shanghai Astronomical Observatory,
Chinese Academy of Sciences, 80 Nandan Road, Shanghai 200030, PR China
\and
National Astronomical Observatories, Chinese Academy of Sciences, Beijing 100101, China
\and
CAS Key Laboratory of FAST, National Astronomical Observatories, Chinese Academy of Sciences, Beijing 100101, China
\and
Kazan Federal University, 18 Kremlyovskaya St, Kazan 420008, Russia 
         }

   \date{  }


  \abstract
   {
 We present the study of arcsecond scale radio continuum and OH line emission of a sample of known OH megamaser galaxies with $z \geq$ 0.15 using archival Very Large Array (VLA) data. And also the results of our pilot Five hundred meter aperture spherical radio telescope (FAST) observations of 12 of these OHM galaxies. The arcsecond-scale resolution images show that the OH emission is distributed in one compact structure and spatially associated with radio continuum emission. Furthermore, nearly all the fitted components are likely smaller than the beam size ($\sim$ 1.4"), which indicates that the broad OH line profiles of these sources originated from one masing region or that more components are distributed in sub-arcsec scales. The radio parameters, including brightness temperature, spectral index, and q-index, show no significant differences with the low-redshift OHM galaxies, which have significantly lower OH line luminosities. Because these parameters are indicators of the central power sources (AGN, starburst, or both), our results indicate that the presence of radio AGN in the nuclei may not be essential for the formation of OH emission. Over 1/3 of OHMs in this sample (6/17) show possible variable features likely caused by interstellar scintillation due to small angular sizes. We might underestimate this value because these sources are associated with this sample's highest OH line flux densities. Those with low OH line flux densities might need higher sensitivity observations to study the variabilities. These results support the compact nature of OH maser emission and a starburst origin for the OHMs in our selected sample.
  }


  \authorrunning{Zhong zu Wu et al. }            
   \titlerunning{RADIO EMISSION FROM OHMs}  
   \maketitle
%

\section{Introduction}
OH megamasers (OHMs) are a type of luminous extragalactic maser sources that produce non-thermal emission from the hydroxyl (OH) molecules, with two main lines at 1665/1667 MHz and two satellite lines at 1612/1720 MHz \citep{2013ApJ...774...35M}. Generally, OHM emission were believed to be pumped by infrared radiation from their environment and amplification of an intense radio continuum background. Therefore, OHM galaxies were detected primarily through targeted surveys in luminous infrared galaxies (LIRGs, $\rm L_{FIR}$ $>$ $10^{11}$ $\rm L_{\odot}$), or ultraluminous infrared galaxies (ULIRGs, $\rm L_{FIR}$ $>$ $\rm 10^{12}$ $L_{\odot}$ ) \citep[see][]{1989MNRAS.237..673N,1992MNRAS.258..725S,2000AJ....119.3003D,2002AJ....124..100D,2012IAUS..287..345W} and a total number of over 119 OHMs have been found in (U)LIRGs \citep[till 2014][]{2014A&A...570A.110Z}.  

Alternatively, high sensitivity \HI surveys, such as the Arecibo ALFALFA \HI survey \citep[see][]{2018ApJ...861...49H};  the upcoming \HI surveys by Five-hundred-meter Aperture Spherical radio Telescope (FAST) \citep{2019RAA....19...22Z} and those with the Square Kilometre Array (SKA) and its precursors \citep[see][]{2021ApJ...911...38R}, can be served as blind surveys for detecting redshifted OH lines. In these untargeted emission-line surveys for neutral hydrogen, the OH lines can mimic a low-redshift \HI line which needs optical spectroscopy or other methods developed in the literature to disentangling \HI from OH \citep[see][]{2021ApJ...911...38R}. Then, the characteristics of OH line emission of known OHM galaxies at similar redshift might also be considered a complementary for these methods to identify OH emission.

One characteristic of OH maser emission is that they are compact at high-resolution observations. The emission is generally unresolved at VLA angular scales ($\sim $1.4"), but can be resolved using long baseline arrays such as the VLBA, EVN, and MERLIN \citep{2007IAUS..242..446P}. The very-long-baseline interferometry (VLBI) observations led to the suggestion that there are two different types of OH emission: the diffuse and the compact emission, which are originated with different modes or attributed to these differences in line overlap effects of compact masing clouds \citep[see][and references therein for details]{2008ApJ...677..985L}.  Meanwhile, VLBI observations of several nearby OHMs indicated that compact OHM sources are typically of the order of 100 pc in extent, while diffuse OHM emission might spread over the 100-pc scale structure \citep[see][]{2005ARA&A..43..625L,2007IAUS..242..446P}.

Second is that the OHM emission might be variable. Generally, it is believed that an OHM galaxy at z>0.1 is distant enough that many masing regions will have sufficiently small angular scales to show interstellar scintillation \citep[ISS, see][]{2001AJ....121.1278D,1998MNRAS.294..307W}. \cite{2002ApJ...569L..87D} and \cite{2007IAUS..242..417D} have presented several discoveries of time variability of OHM emission caused by ISS and provide additional evidence that narrow lines correspond to compact masing regions. 
However, the time variability of OHM emission caused by intrinsic changes in the maser environment is rarely studied. \cite{2016MNRAS.460.2180H} found that the flux density of the brightest OH maser components of IRAS 20100-4156 has reduced by more than a factor of 2 over 26 years. Similarly, \cite{2015MNRAS.447.1103M} shows that the amplitudes in the VLBI spectra of Arp 220 decreased about 40\% in 14 years. In the literature, the variation of the OHM emission in these two sources are believed to be intrinsically caused by masing regions rather than variability produced by ISS. Because the timescale of the variability in the two cases are over 10 years, new observations of the known OHM galaxies, which were detected by Arecibo telescope 20 years ago by \citep{2000AJ....119.3003D,2001AJ....121.1278D,2002AJ....124..100D}, can be helpful further in studying the time variation of the OHM line emission. 

Third is that the OHM emission are generally associated with the radio continuum emission from arcsecond-scale observations. VLA-A observations of a few nearby OHM galaxies found that the OHM maser emission coincides with one component of radio continuum very closely typically $\leq$1" \citep[e.g., Arp 220, IRAS 17208-0014, see][]{2005ARA&A..43..625L}. In the literature, optical spectrography has found that about 67 \% of the OHM galaxies can be classified as AGN or show evidence of both AGN and starburst activity \citep[e.g.,][]{1998ApJ...509..633B,2006AJ....132.2596D}. Because the optical line emission may be swamped by dust emission from the nuclei and envelope \citep{2015ApJ...799...25S}, the actual fraction of AGN in OHM galaxies might be higher. The recent results from an ongoing project using the Hubble Space Telescope (HST), Very Large Array (VLA), and Integral Field Spectroscopy (IFS) data have successively identified the presence of a previously unknown AGN in four of five OHM galaxies \citep[see][]{2020MNRAS.498.2632H,2018MNRAS.474.5319H,2018MNRAS.479.3966H,2015ApJ...799...25S}. And also, the possibility of the presence of an AGN in the fifth OHM galaxy (IRAS 17526+3253) can not be excluded from presented data \citep[see][]{2020MNRAS.498.2632H,2019MNRAS.486.3350S}. These results support a hypothesis that OHM galaxies harbor recently triggered AGN \citep[see][]{2020MNRAS.498.2632H}. 
Meanwhile, the properties of radio continuum emission, including spectral index, and brightness temperature, can also be used to classify the central energy source of OHM galaxies \citep{2005Ap.....48...99K,2005Ap.....48..237K,2006A&A...449..559B,2022MNRAS.510.2495S}. \citet{2005Ap.....48...99K} and \cite{2006A&A...449..559B} obtained the radio structure of 30 and 51 OHM galaxies in the radio continuum based on the VLA observations. They found that the OHM galaxies can be classified into different types, e.g., AGN and Starburst-dominated sources. Similarly, \cite{2022MNRAS.510.2495S} has investigated the radio multi-band spectral indexes of 74 OHMs with RATAN observations and found that 32\% of the OHM sample are flat-spectrum sources which is likely a sign of the presence of AGNs in these galaxies. 

Generally, the future blind \HI surveys could mainly detect redshifted OHM galaxies, which should be at relatively high redshift. The first sample of OHMs from the blind Arecibo \HI survey with redshift ranges at 0.16$<$z$<$0.22 \citep[see][]{2018ApJ...861...49H}. \cite{2002AJ....124..100D} and \cite{2009ASPC..407...73B} have shown that distant luminous OHM galaxies display strong and broad multiple component profiles, which differ significantly from nearby sources. Therefore, the main aims of this paper are to investigate the characteristics of the continuum and OH line emission of high-z known OHM galaxies. Then, our results will help to understand the connections with low-redshift OHM galaxies and could also help identify OHM galaxies detected from future blind \HI surveys. We presented the sample selection, observations, and data reductions in section 2. The results, discussions, and summary are in sections 3--5.

\section{Sample selection and Data Analysis}

\subsection{Sample Selection}
\label{sect:Obs}
     Based on a sample of 30 OHM galaxies with redshift ranges from z$\geq$0.15 to the highest 0.2644 from \cite{2014A&A...570A.110Z}, we have found archive VLA data for OH line and radio continuum observations, which are available for 12 and 21 sources, respectively (see Table \ref{table1} and \ref{imagepar}). The high-resolution VLA observations of the OH line emission are mainly from two VLA projects, AD462 and AD483 (PI: Darling, J.). The details about these projects for OH line and radio continuum emission are listed in Table \ref{imagepar}. Furthermore, we randomly selected 12 OHM galaxies for pilot observations with the FAST telescope to study the possible variable properties of the OH line profiles. The observational and derived properties of the selected sample are present in Table \ref{table1}.

\subsection{Reduction of the VLA archival data}

      We reduced the historical VLA data with standard procedures using the NRAO Astronomical Image Processing System (AIPS), including initial data flagging, amplitude, phase, calibrations, and the correction for bandpass. We also shifted the radial velocity to correct the effects of the Earth's rotation and its motion within the Solar System and towards the Local Standard of Rest. The EVLA data for the continuum emission were performed using the Common Astronomy Software Applications \citep[CASA][]{2007ASPC..376..127M} package pipeline (version 5.4.1). We imported all the calibrated data into the DIFMAP package \cite{1997ASPC..125...77S} to obtain the continuum and OH line channel images. The continuum images of the OH line projects were produced using line-free channels, and we made the OH line channel images after the subtraction of the clean model obtained from the radio continuum emission. 
\subsection{FAST observations and data reduction}
\begin{figure*}
   \centering
  
   \subfigure{ \includegraphics[width=0.485\textwidth,height=6.2cm]{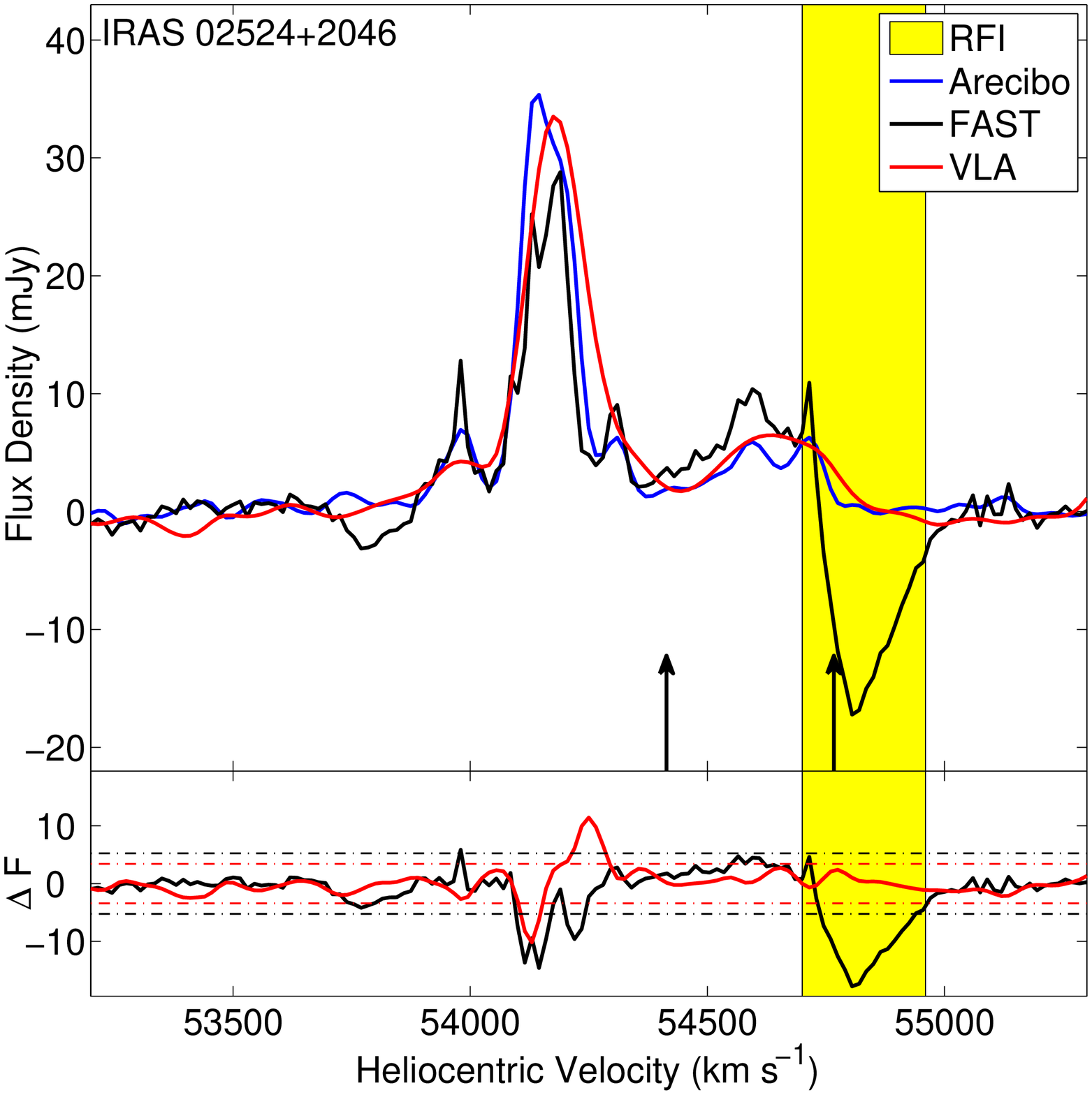}}
   \hfill
  \subfigure{ \includegraphics[width=0.485\textwidth,height=6.2cm]{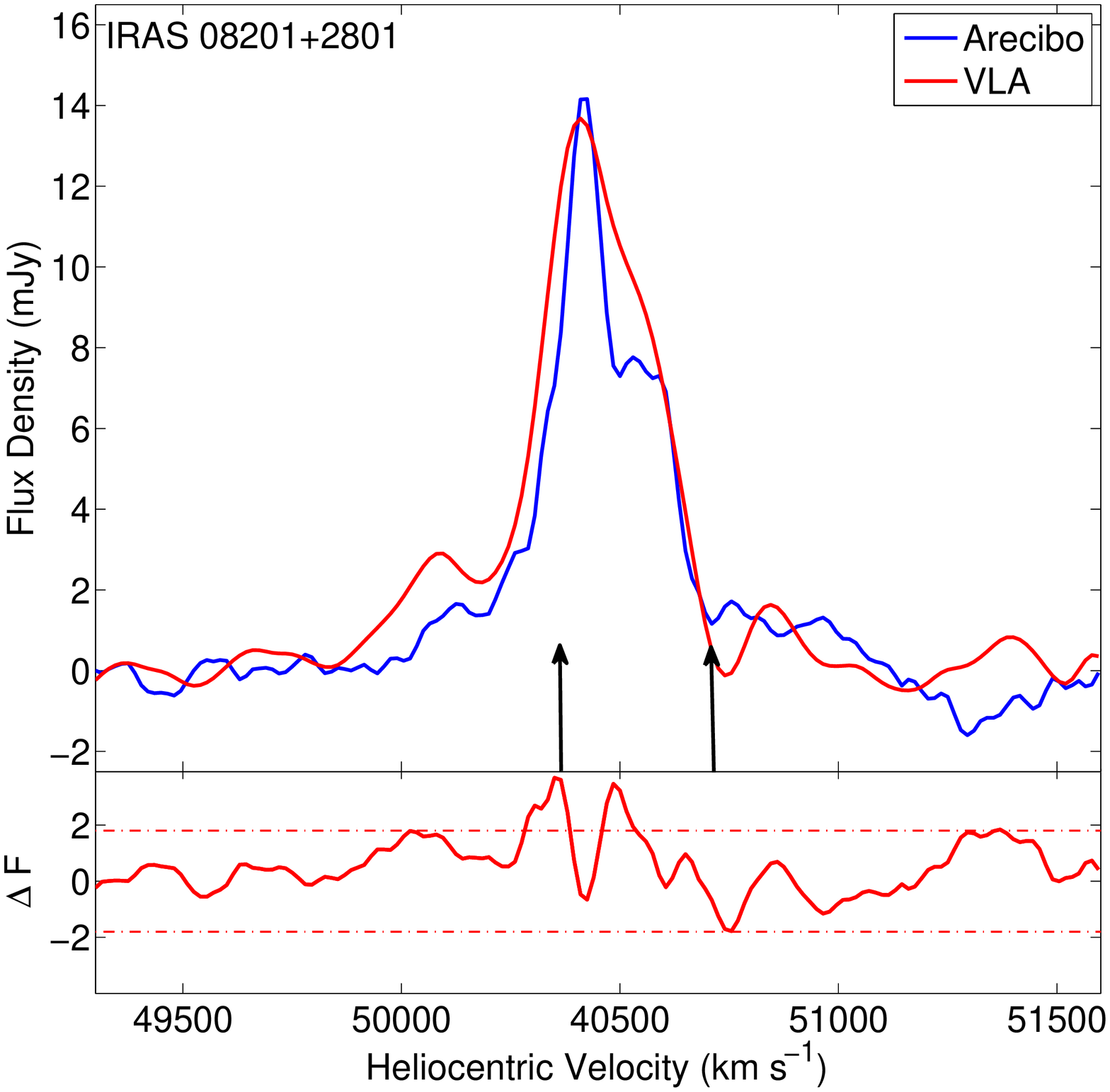}}
  \hfill
   \subfigure{  \includegraphics[width=0.485\textwidth,height=6.2cm]{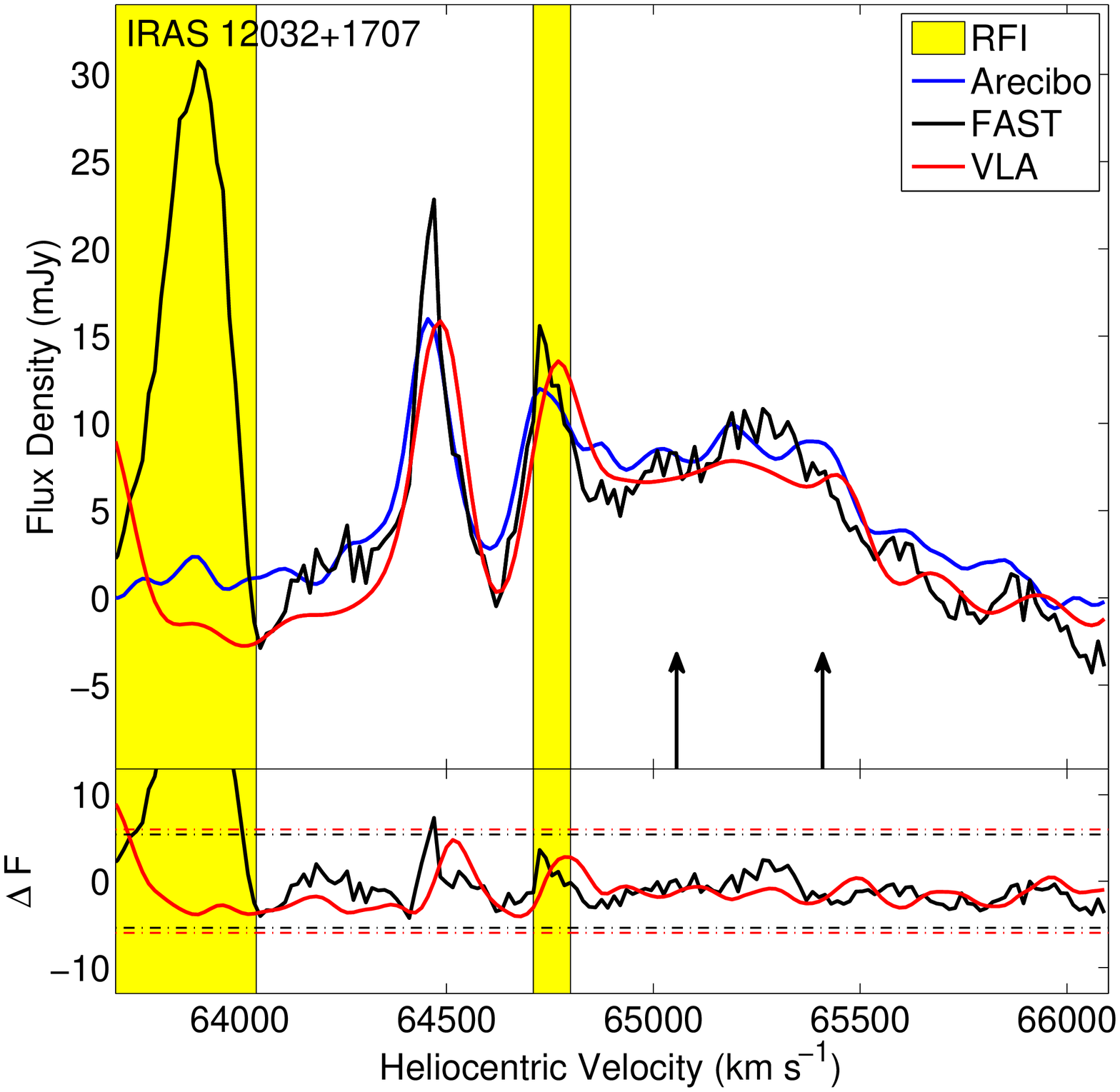}}
   \hfill
   \subfigure{   \includegraphics[width=0.485\textwidth,height=6.2cm]{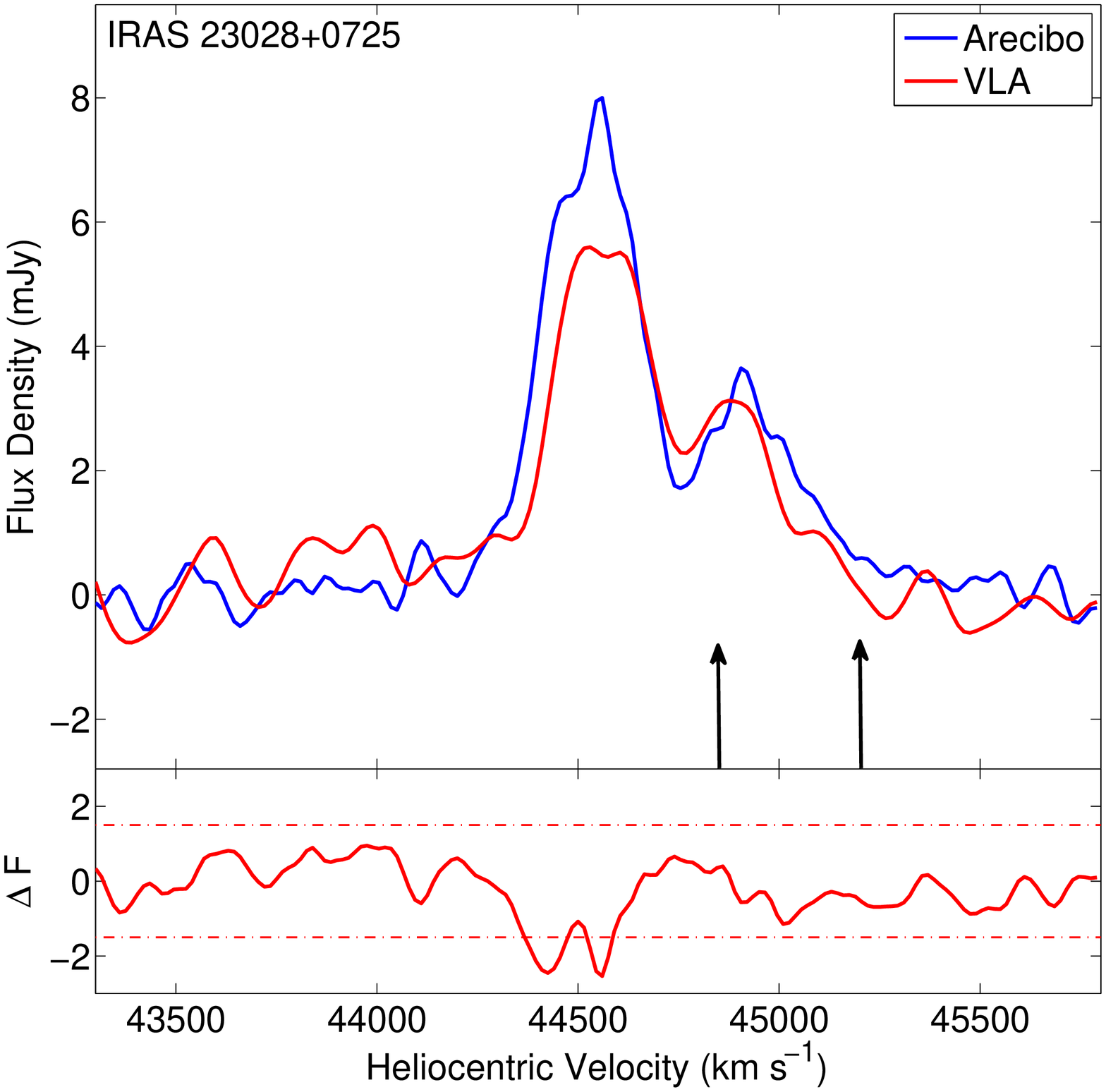}}
   \hfill
    \subfigure{  \includegraphics[width=0.485\textwidth,height=6.2cm]{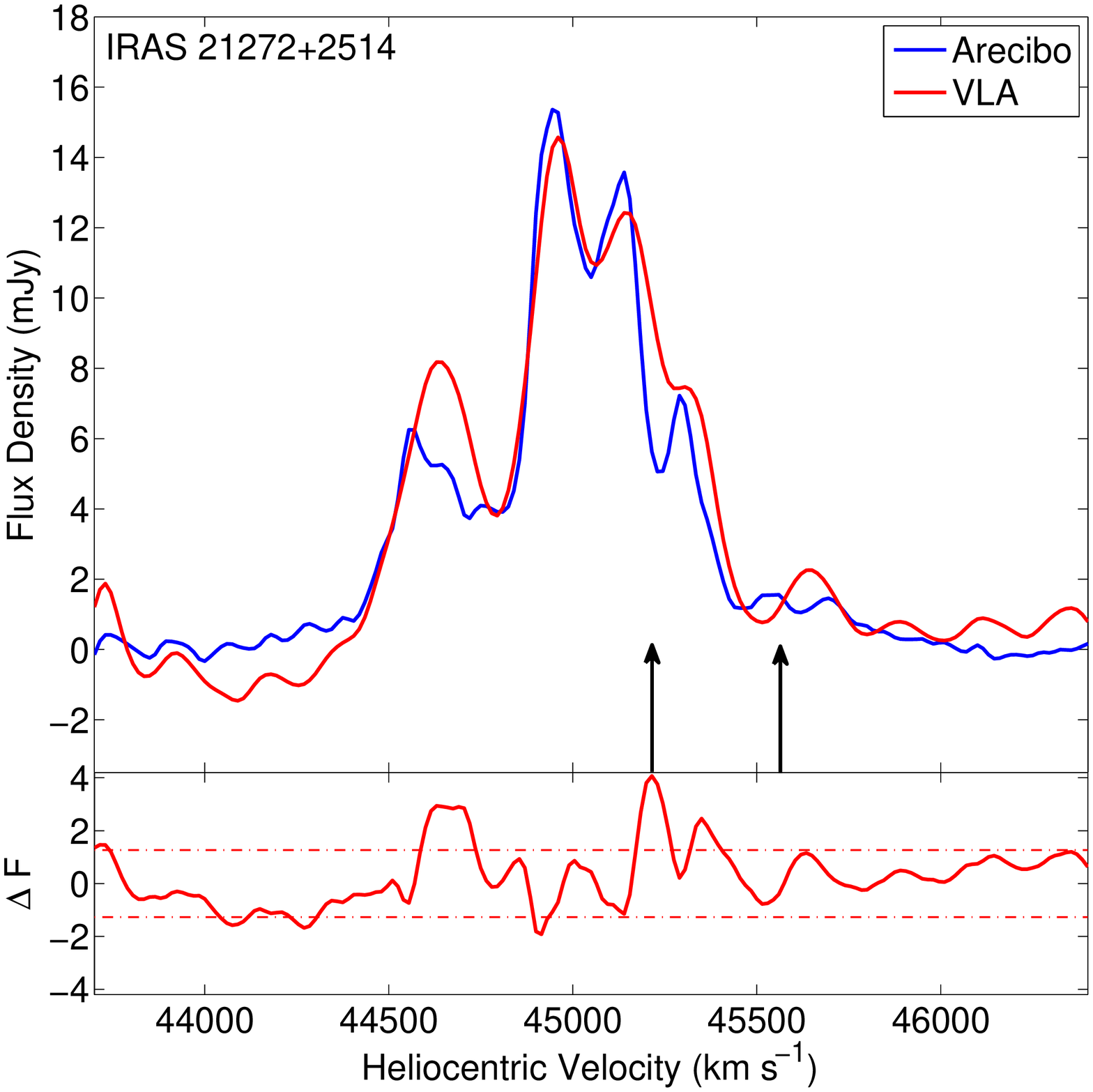}  }
    \hfill
  \subfigure{ \includegraphics[width=0.485\textwidth,height=6.2cm]{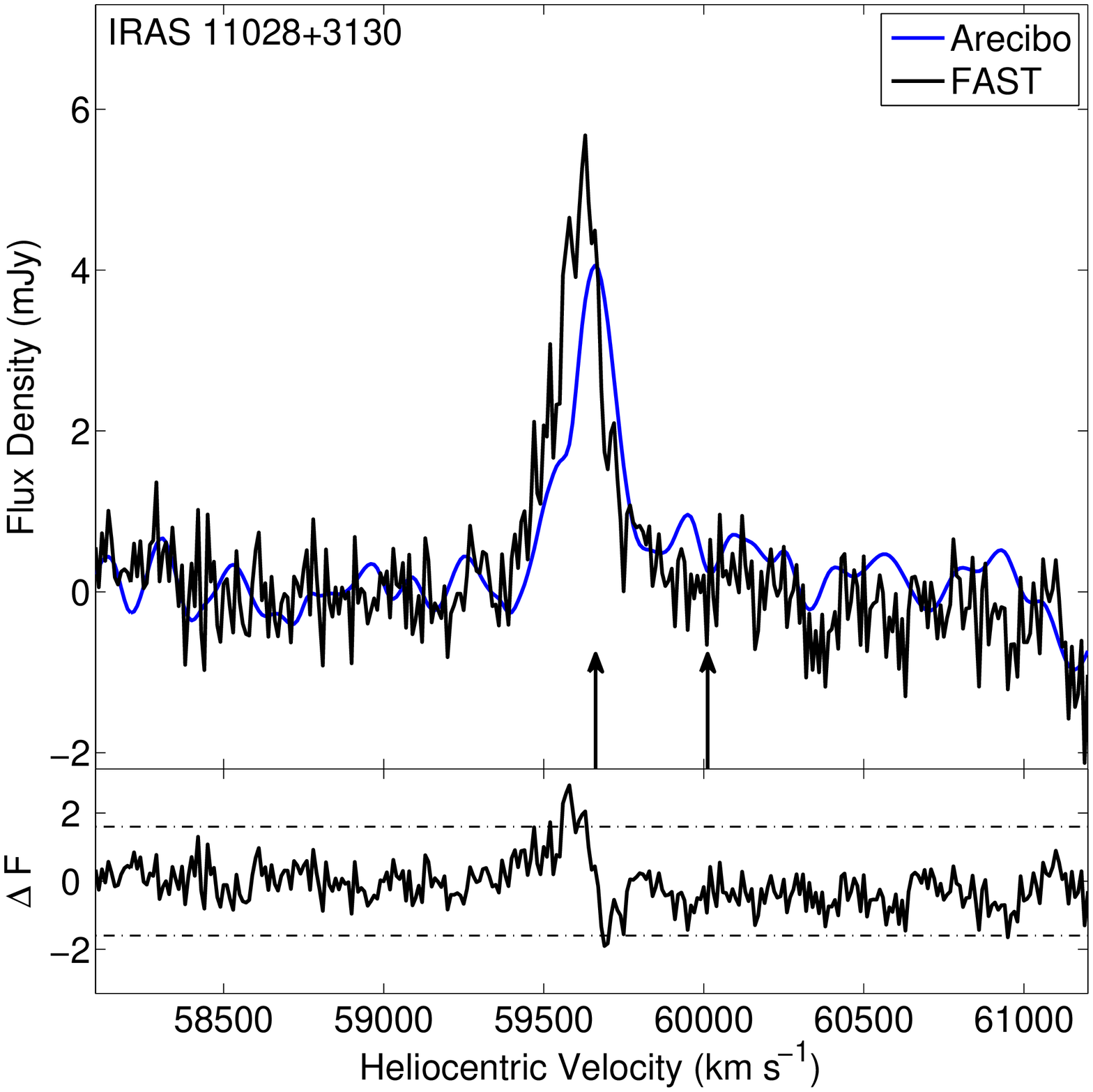}}
   \hfill

      \caption{The OH spectra use the 1667.359
MHz line as the rest frequency for the velocity scale, two arrows stand for the velocity of the 1667.359 (left) and 1665.4018 (right) MHz lines based on the optical redshift. Blue: Arecibo observations by \citep{2000AJ....119.3003D,2001AJ....121.1278D,2002AJ....124..100D,2008ApJ...680..981R}. Red: VLA-A observations as listed in Table \ref{imagepar}. Black: FAST observations as listed in Table \ref{table2}. For each figure, the top panel presents the observed OH spectra, and the bottom panel presents the $\Delta$F=$\rm OH_{VLA/FAST}$-$\rm OH_{Arecibo}$ to show the variations between the two or three epoch results. The colors red and black stand for the spectra from VLA and FAST, respectively. The dashed lines represent the 3 $\sigma$ level estimated from the standard deviation of $\Delta$F using the line-free channels. 
      }
         \label{ohline1}
   \end{figure*}

    We have done pilot FAST observations of 12 high-z OHMs and listed the observational parameters in Table \ref{table2}. We use the central beam (M01) of the 19-beam receiver with ON/OFF mode. We record the data every 1.00663296 s and inject a 10 K continuum signal using a noise diode every 32.21225472 s. Our data reduction procedures are as follows: 
    
    First, we convert the spectral signals into units of K using the nearest noise diode from the data. And we have adopted the average value of about 11.5 K for the injected signal, which is the mean value in "median\_20190115.Tcal-results.HI\_w.high.sav" measured by the FAST group. Second, we further converted the signals into units of Jy with the equation:   Gain=$\sim$16.0 K/Jy, which is from the fundamental performance parameters of the FAST telescope when zenith angle $\theta_{ZA}$<26.4 $^{\rm o}$ (available in the webpage \footnote{https://fast.bao.ac.cn/cms/article/97/}). Third, we generated the spectra of target-ON and target-OFF by combining the first two polarizations and then did ON–OFF subtraction (see Fig. \ref{fastonoff}). Because the baseline of the ON-OFF spectra still contains ripple caused by standing waves, we did the baseline fitting with equation \begin{equation} \label{eq1}
\rm a_{0}+\sum_{i} (a_{i}*cos(i*x*w)+b_{i}sin(i*x*w))
\end{equation}, 
where 'i' ranges from 1 to 6 for different sources, and finally did the baseline subtraction and got the spectrum for each target as present in Fig. \ref{ohline1} and \ref{ohline2}. The frequency interference (RFI) affected channels were labeled in yellow colors, which were identified by visually comparing the ON and OFF spectra, where both the on and off spectra show bumps higher than the typical waves (see Fig. \ref{ohline1}, \ref{fastonoff} and \ref{ohline2}).

  \begin{table*}                                                                                                                                                                             
       \caption{The properties of the sample. }                                                                                                                                   
     \label{table1}                                                                                                                                                                        
  \centering                                                                                                                                                                                 
  \begin{tabular}{l c c c r r l l l l l l l l l l l}     
 \hline\hline                                                                                                                                                                               
 IRAS Name           & z       & type    & f$_{60\mu m}$ & f$_{100\mu m}$ & $\rm L_{OH}$ & T$\rm _{B}$  & $\alpha\rm _{total}$ & $\rm \alpha\rm _{peak}$ & q(1.4)& q(4.8)      &  $\rm \beta\rm _{c} $ & $\Delta_{lc}$ \\
                                 &           &           &       (Jy)       & (Jy)  &     $\rm L_{\odot}$               &     (K)                  &       &              & &    &     &    arcsec      \\ 
\hline                                                                                                                                                                                                                         
01562+2528 & 0.1657  & $\rm H^{1}$     &    0.81         &  1.62             &3.2  & 4.6   & -0.62                &    -0.67            &2.32 &  2.66  &     0.07          &  0.13 $\pm$ 0.05               \\  
02524+2046 & 0.1814  & $\rm H^{1}$    &    0.96         &  <4.79         & 3.7&3.0   & -1.01                &    -1.08            &2.88 &  3.43  &     1.05          &  0.14 $\pm$ 0.09               \\  
08201+2801 & 0.1678  & $\rm H^{1}$     &   1.17         &  1.43             & 3.4& 3.7   & -0.58                &    -0.83            & 2.67 & 2.98  &     0.52         &  0.07$\pm$ 0.10                 \\  
08279+0956 & 0.2086  &  $\rm L^{1}$    &   0.59         &  <1.26        &3.2 &3.8   & -0.36                &    -0.94            & 2.62& 2.81  &     0.55          &  0.23$\pm$ 0.16          \\ 
09531+1430 & 0.2153  &  $\rm S2/H^{1}$ &   0.78         &  1.04         & 3.4&3.9   & -0.50                &    -0.45            &2.59  &2.87  &     0.09          &  0.15$\pm$ 0.09         \\ 
10339+1548 & 0.1972  & $\rm S2^{1}$    &    0.98         &  1.35           & 2.6 &4.8   & -0.71                &    -0.75            & 2.33 &2.71  &     0.11          &  -                          \\  
12032+1707 & 0.2178  & $\rm L^{1}$     &   1.36         &  1.54         &4.1 &5.5   & -1.18                &    -0.89            & 1.78 &2.37  &    -0.03          &  0.06$\pm$       0.03          \\  
13218+0552 & 0.2028  &$\rm S1^{2}$ &   1.17         &  0.71          &3.4 &3.4   & -0.74                &    -0.81            &2.61  &2.88  &     0.54          &  0.58 $\pm$ 0.23               \\   
14586+1432 & 0.1477  & -     &    0.57         &  1.07                & 3.4&5.2 & -0.24                &    -0.22            & 1.89 &1.99  &    -0.69         &  0.19 $\pm$ 0.07         \\  
21272+2514 & 0.1508  & $\rm S2^{1}$    &    1.08         &  <1.63         &3.6 &3.7   & -0.33                &    -0.25            & 2.67 &2.84  &    -0.06          &  0.09 $\pm$ 0.04        \\ 
23028+0725 & 0.1494  & $\rm L^{1}$     &    0.91         &  <1.37         &3.3 &4.4  & -1.06                &    -0.87            &1.88  &2.43  &     0.23          &  0.06 $\pm$ 0.05             \\  
23129+2548 & 0.1789  & $\rm L^{1}$     &    1.81         &  1.64          & 3.2&4.1   & -0.28                &    -0.39            & 2.70 &2.85  &    -0.02          &  0.13 $\pm$ 0.05                \\ 

01355-1814 & 0.1920  &      $\rm cp^{2}$         &       1.40            &    1.74             &   2.8$^{a}$   &    2.4           & -0.31 
                     &      0.27    &  - &3.21     &    -0.12               &- \\
03521+0028A &  0.1519  &$\rm L^{1}$                 &    2.64            &     3.83            &  2.4    &     4.9          &     0.16                &    0.21                  &  2.90&2.97     &       -0.72            & -\\
07572+0533 & 0.1900  &  $\rm L^{1}$               &      0.96            &   1.30           & 2.7     &        4.5       &     0.90                &    1.33                  &2.06 &2.46      &    -1.82               & -\\
08449+2332 &  0.1515 &   $\rm H^{1}$              &    0.87             &  1.20            & 3.2     &    3.7           &    -0.86                 &  -0.86                    &2.54 &2.71      &    0.47               & -\\
11180+1623A & 0.166 &      $\rm cp^{2}$            &        1.18          &      1.60            &  2.3    &  2.8             &  -1.29                   &    -0.45                  & 2.73 &3.13     &     0.43              & -\\
12018+1941$^{*}$ & 0.1686 &       $\rm L^{3}$            &       1.76           &    1.77           &  2.6    & 3.5             &      -0.63               &     -0.62                 & 2.55 &2.89     &     0.35              &- \\
14070+0525$^{*}$ & 0.2644  &     $\rm S2^{2}$         &    1.45            &   1.82             &  4.1    &    2.9          &   -0.56                  &     -0.56                 & 2.54 &2.84     &  0.41                &- \\

21077+3358 & 0.1767 &     $\rm L^{1}$          &       0.89         &        <1.55       & 3.2     &          3.6     &        -1.02             &       -1.12               & 2.23 &  2.73  &   0.74     & -\\
22088-1831$^{*}$ & 0.1702 &        $\rm H^{2}$          &       1.73         &     1.73           &   3.3   &     4.1          &   0.04                 &       0.27               &2.34  &2.32     &     -0.78              & -\\

 \hline

   \end{tabular} 
\vskip 0.1 true cm \noindent Notes. The format of the Columns is as follows:
     (1) source name, the "*" stands for the radio parameters ($T_{B}$, peak and total $\alpha$ and $\beta_{c}$) are collected from \cite{2006A&A...449..559B}. (2) optical redshift. (3): the optical spectroscopic classification in the literature: 1 \cite{2006AJ....132.2596D}, 2 \cite{2010ApJ...709..884Y}, 3 \cite{1998ApJ...509..633B}. Symbols:  H,  {\rm H\,{\scriptsize II}} region (starburst); L, LINER; S2, Seyfert type 2; S1, Seyfert type 1. cp = starburst–AGN composite galaxies. (4)-(5) are the IRAS 60 and 100 $\mu$m flux densities in Jy. (6) luminosity of the OH 1667 MHz line from \cite{2014A&A...570A.110Z}. (7) brightness temperature from Table \ref{imagepar} with lower uncertainties. (8) and (9) are the spectral indexes derived from the integrated and peak flux densities of Table \ref{imagepar}. (10) and (11), the FIR-radio luminosity ratio derived from equation \citep[][]{2006A&A...449..559B} (q={\rm log(3.36$\times$$10^{2}$*(2.58 $\rm f_{60\mu m}$+$\rm f_{100\mu m}$)/$\rm F_{radio}$)}) using radio flux density at 1.4 GHz and 4.8 GHz, respectively. Most radio flux densities are from Table \ref{imagepar}.  For sources that do not have a 1.4 GHz flux density, we used the flux densities from the NVSS survey. While sources with not 4.8 GHz, we derived the flux density from fitting the multi-band radio flux densities from Table \ref{imagepar}.
(12) The nuclear Activity Factor \citep[][]{2006A&A...449..559B} derived from equation: $\rm \beta_{c}$ = 0.308[q-0.75$\times$$\rm T_{b}$+3$\times$$\rm \alpha_{peak}$-1]. Positive values indicate a starburst signature for the dominant nuclear power source, and negative values indicate an AGN signature. (13) The relative positions between OH line and continuum peak positions from Table \ref{imagepar}. The error on relative positions were estimated using equation from \cite{2011A&A...525A..91T}:  $\rm \sigma_{rel}$=$\rm \sqrt{(\rm \theta_{line}/(\rm 2*SNR_{line}))^2+(\rm \theta_{cont}/(\rm 2*SNR_{cont}))^2}$, where $\theta$ denotes the beam FWHM of the map, and line and cont refer to line and continuum emission. 
   \end{table*}   
\begin{table*}
       \caption{Parameters of FAST observations. \label{table2} }
    
  \centering
  \begin{tabular}{c c c c c c c c c c c c}     
  \hline\hline

   IRAS Name      & z    & Epoch & Freq.  &  $t_{ON}$  &  Repeat Num& $\delta$V & $\sigma_{1667}$&$\sigma^{10}_{1612}$  &$\sigma^{10}_{1720}$ &$\rm F^{INT}_{1612}$  & $\rm F^{INT}_{1720}$\\ 
                   &    & (MHz)  & (second)  &   & \kms & (mJy) & (mJy)& (mJy) & Jy \kms & Jy \kms \\
    \hline                                                                                                                                   
    02524+2046&0.1814 &   2021.06.27    & 1411.35  & 180    &  4    & 1.93   &  0.85 & -    & 0.49 & -      & <0.07  \\ 
07572+0533&0.1906 &   2022.05.23    & 1401.14  & 180    &  12   & 1.96   &  0.54 & 0.23 & 0.26 & <0.07  & <0.08 \\  
08279+0956&0.2086 &   2021.06.23    & 1379.58  & 200    &  3    & 2.04   &  1.12 & 0.55 & -    & <0.20  & -      \\ 
09531+1430&0.2153 &   2021.06.22    & 1374.01  & 180    &  7    & 2.05   &  0.65 & 0.37 & 0.43 & <0.34  & <0.40  \\ 
10339+1548&0.1972 &   2021.06.22    & 1392.72  & 170    &  5    & 1.99   &  0.98 & -    & 0.51 & -      & <0.07 \\  
11028+3130&0.1988 &   2022.05.25    & 1391.09  & 170    &  5    & 2.00   &  0.66 & 0.28 & 0.30 & <0.07  & <0.08  \\ 
11524+1058&0.1787 &   2022.05.25    & 1414.57  & 180    &  5    & 1.92   &  0.70 & 0.33 & 0.30 & <0.14  & <0.12 \\  
12032+1707& 0.2178&   2021.06.28    & 1369.16  & 192    &  5    & 2.05   &  1.14 & -    & 0.50 & -      & <0.57  \\ 
13218+0552&0.2028 &   2021.06.26    & 1383.59  & 180    &  10   & 2.01   &  0.90 & -    & -    & -      &  -     \\ 
14070+0525&0.2644 &   2021.06.26    & 1318.69  & 180    &  3    & 2.23   &  0.82 & -    & -    & -      &  -     \\ 
22116+0437&0.1938 &   2021.06.27    & 1396.68  & 190    &  11   & 1.97   &  0.72 & 0.52 & 0.47 & <0.38  & <0.34  \\ 
23129+2548&0.1789 &   2021.06.23    & 1414.21  & 192    &  5    & 1.94   &  0.59 & -    & 0.36 & -      & <0.23  \\

      \hline
       \end{tabular}
       \vskip 0.1 true cm \noindent Notes. Col (1) source name. Col (2) optical redshift. Col (3) the observational epoch. Col (4) the central frequency of the OH 1667 MHz line.
       Col (5) and (6) the scan length for each target-ON observation and the number of cycles for On/Off observations. Col (7) and (8) the spectral resolution in \kms and the 1 $\sigma$ noise level around the OH 1667 MHz line. Col (9) and (10) the  $\sigma$ noise near the binned OH 1612 and 1720 MHz lines with velocity resolution about 10 \kms. Col (11) and (12) the upper limit of integrated 1612 and 1720 MHz lines estimated with equation 1 in \cite{2013ApJ...774...35M}.
   \end{table*}

\begin{table*}
       \caption{Statistical parameters of high and low redshift OHM galaxies. The two samples were collected from Table~\ref{table1} of this work and  \cite{2006A&A...449..559B}, respectively. \label{table3} }
    
  \centering
  \begin{tabular}{c c c c c c c c c c}     
  \hline\hline

   property           & $L_{OH}$ & $L_{1400MHz}$  &  $L_{IR}$ &q(1.4)&q(4.8)&T$\rm _{B}$ & $\alpha\rm _{total}$& $\rm \alpha\rm _{peak}$ &  $\rm \beta\rm _{c} $  \\ 
   property           &     $L_{\odot}$         & W Hz$^{-1}$    & $L_{\odot}$  &  &    \\
    \hline                                                                                                                 K-S p-value& 1.10E-05  &  7.25E-06 & 1.35E-06  &   0.35  &  0.29  &  0.87 &  0.71  & 0.28    & 0.96  \\              
    $\rm mean^{H}$     &  3.20     & 23.7      & 12.1      &  2.44   &  2.77   & 3.9  & -0.53  & -0.46    & 0.06  \\              
       $\rm mean^{L}$    &  2.23     & 23.1      & 11.6      &  2.49   &  2.80   & 4.0  & -0.58  & -0.55    &  0.12  \\    
    $\rm median^{H}$   &  3.20     & 23.6      & 12.1      &  2.55   &  2.84   & 3.8  & -0.58  & -0.62    & 0.11   \\            
         $\rm median^{L}$   &  2.32     & 22.9      & 11.6      &  2.60   &  2.90   & 4.0  &  -0.55 & -0.55    &  0.16 \\

      \hline
       \end{tabular}
   \end{table*}


\section{Results}
 
\subsection{The arcsecond scale OH line emission}
We have detected OH line emission from 11 of the 12 sources using the VLA archival data. We generated the OH spectra (see Fig. \ref{ohline1}) by integrating the flux densities from a 3 $\sigma$ region defined from the combined line emission images as shown in Fig. \ref{lowfreq}.
The OH emission images all show core-dominated structures. We fitted these images with one gaussian component, and the fitted component sizes are smaller than one beam (see Table \ref{imagepar}). As shown in Fig. \ref{ohline1}, the OH line profiles indicated that we had recovered all the OH line flux found from single-dish Arecibo observations. These results suggest that the dominant OH emissions are compact (see Fig. \ref{lowfreq}) and distributed in a region with sizes less than the arcsecond scale. 

\subsection{The OH line profiles from FAST observations}
We present the OH spectra from pilot FAST observations for 12 sources (see Fig. \ref{ohline1} and \ref{ohline2}).   The OH profiles from FAST observations agree with the results from previous Arecibo observations in the literature \citep{2000AJ....119.3003D,2001AJ....121.1278D,2002AJ....124..100D} or archive VLA data investigated in this work. Six sources are likely showing OHM emission variability for their narrow line components (see Fig. \ref{ohline1}), including three known variable sources \citep[02524+2046, 12032+1707 and 21272+2514, see][]{2002ApJ...569L..87D,2007IAUS..242..417D}. 

The OH satellite lines in OHM galaxies are rarely detected and we also tried to search for the OH satellite lines (OH 1612 and 1720 MHz lines) based on our pilot observations. The binned spectra of each source with a resolution of about 10 \kms are shown in Fig. \ref{OHsatellite}. We have presented an upper limit of the two OH satellite lines for ten sources (see Table \ref{table2}), which are lower than the values measured by \cite{2013ApJ...774...35M}. Our results agree with the view from \cite{2013ApJ...774...35M} that there is no evidence for a significant population of strong satellite line emitters among OHMs.

\subsection{The radio structure of the continuum emission}
\begin{figure*}
   \centering
     \includegraphics[width=0.485\textwidth,height=6.2cm]{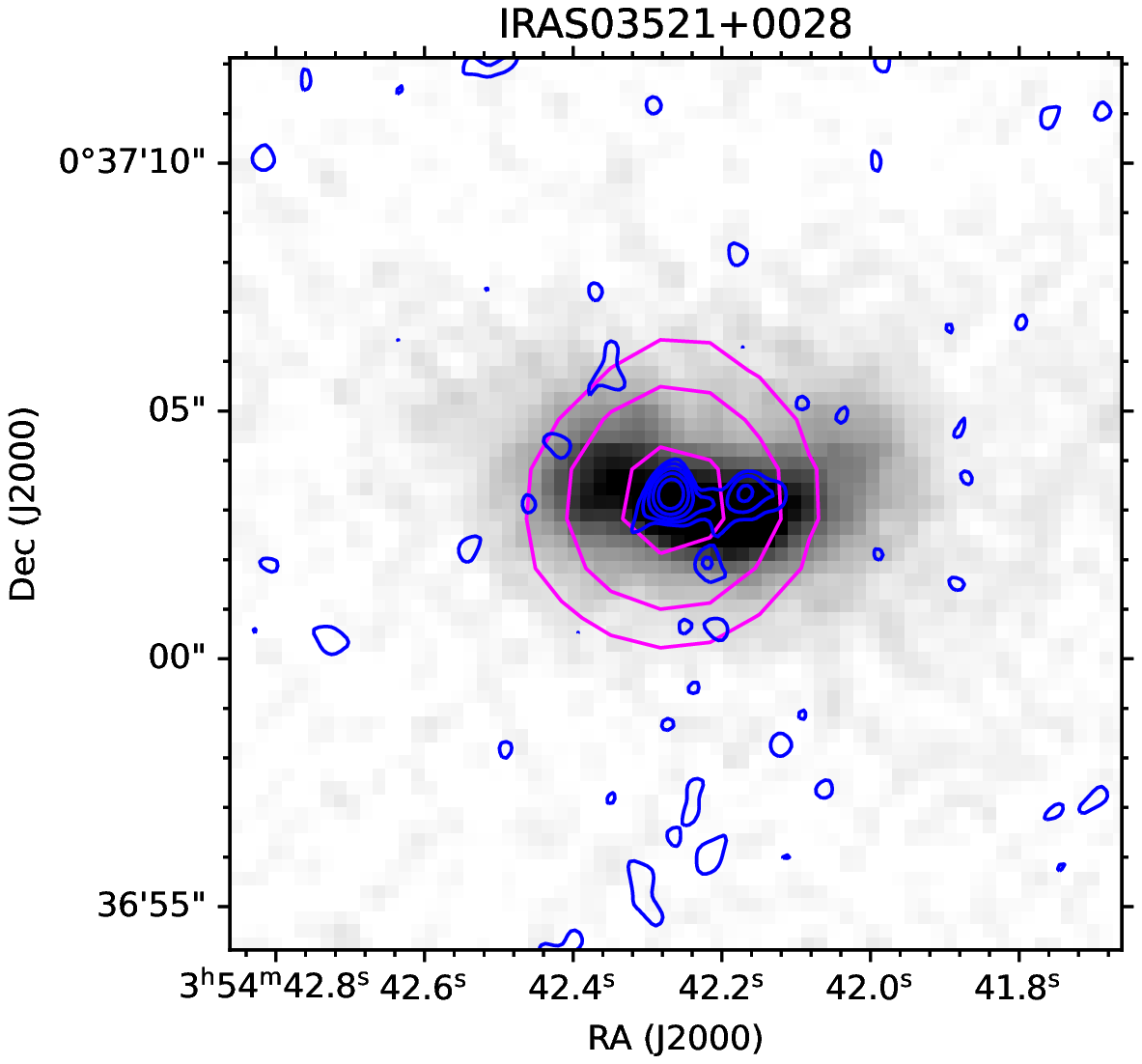}
    \includegraphics[width=0.485\textwidth,height=6.2cm]{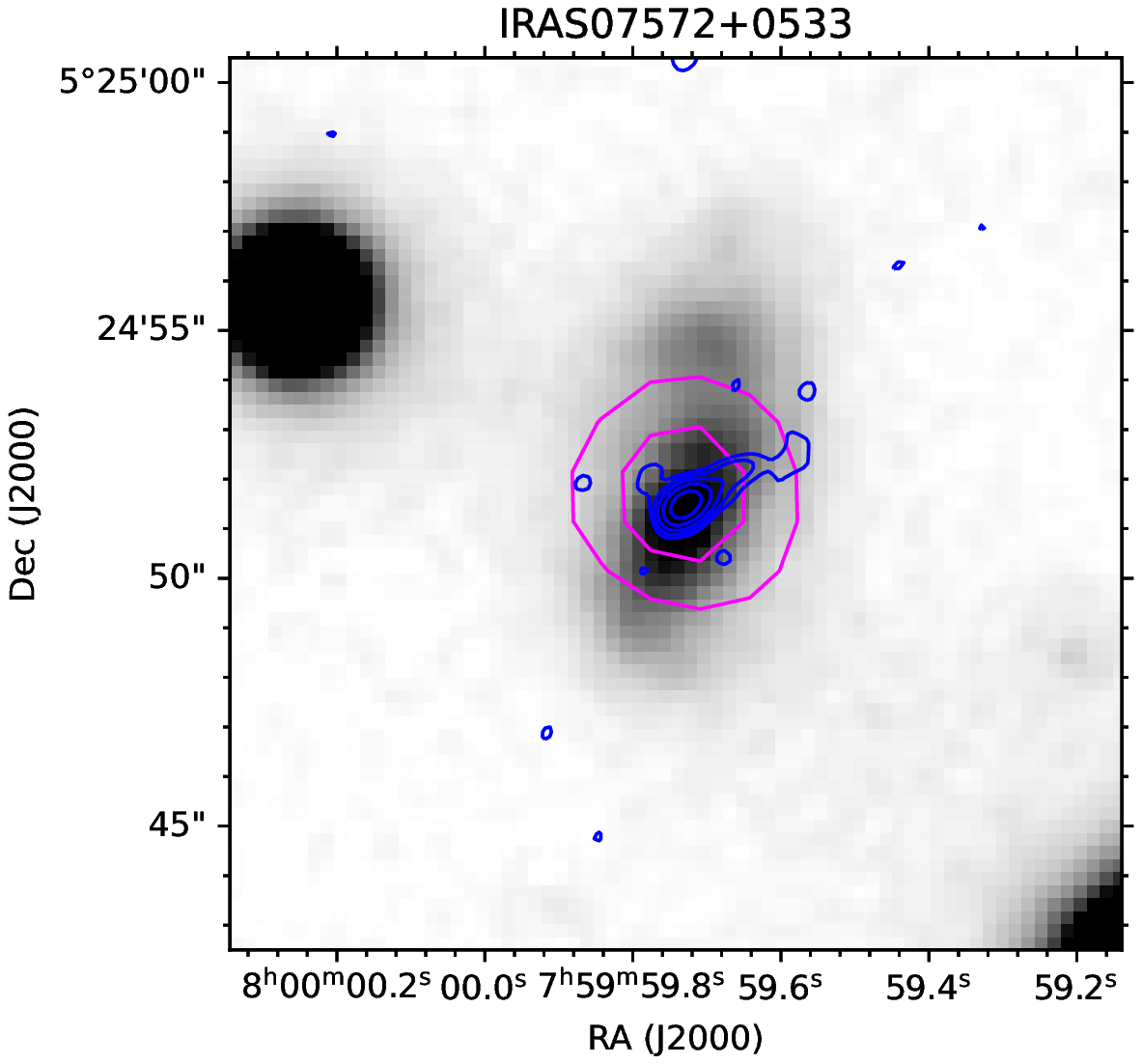}
     \includegraphics[width=0.485\textwidth,height=6.2cm]{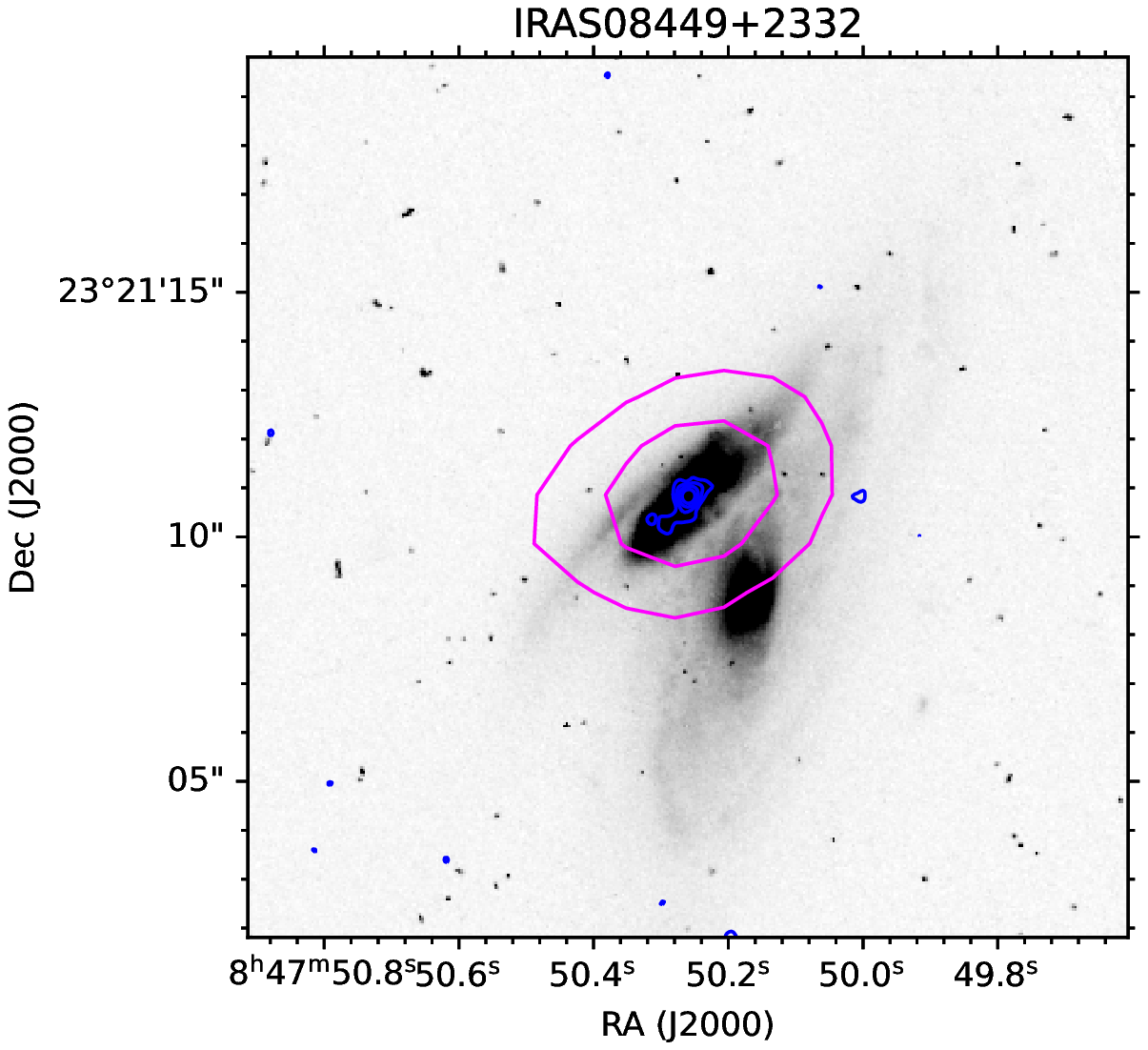}  
    \includegraphics[width=0.485\textwidth,height=6.2cm]{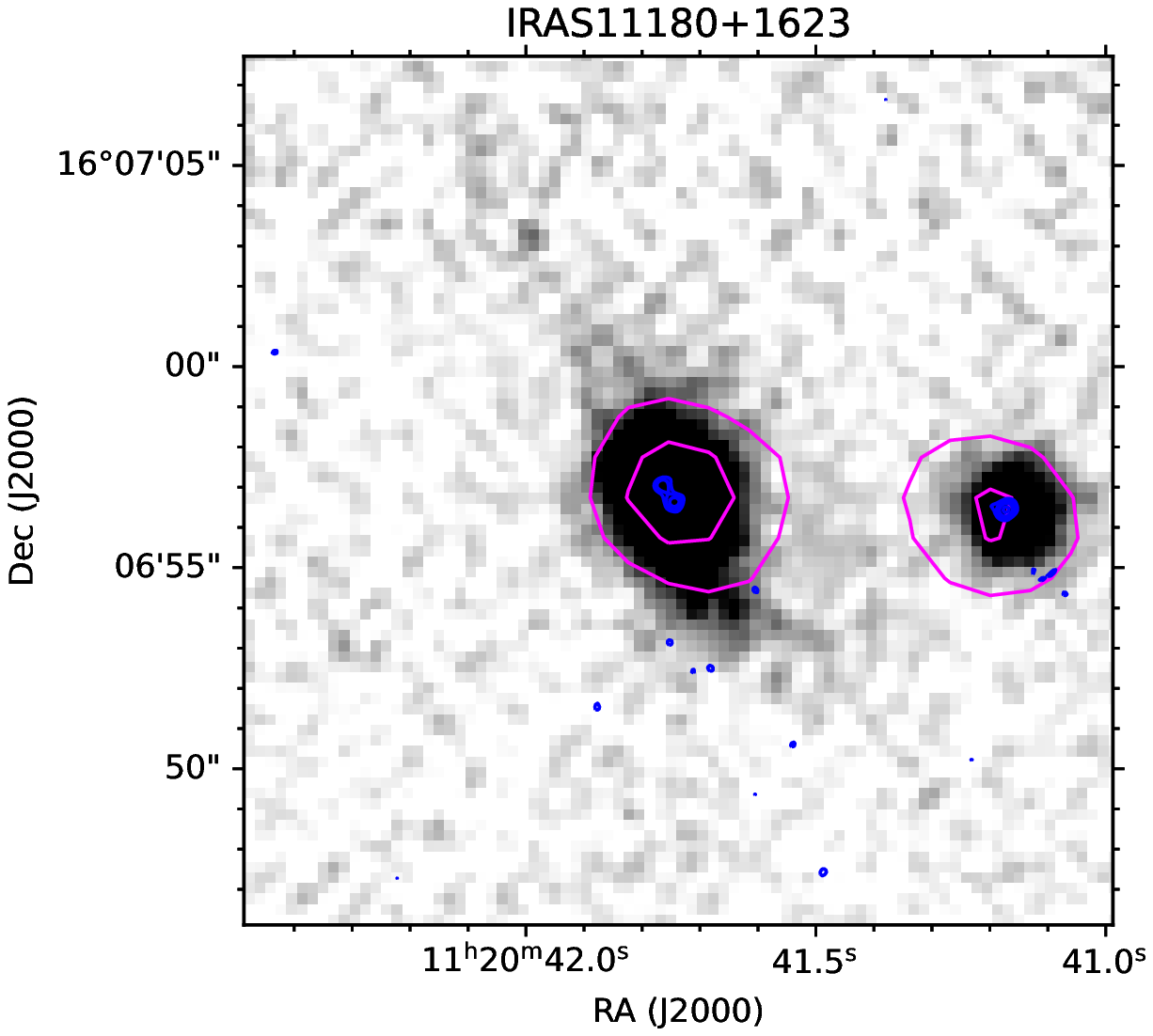}   
       \includegraphics[width=0.485\textwidth,height=6.2cm]{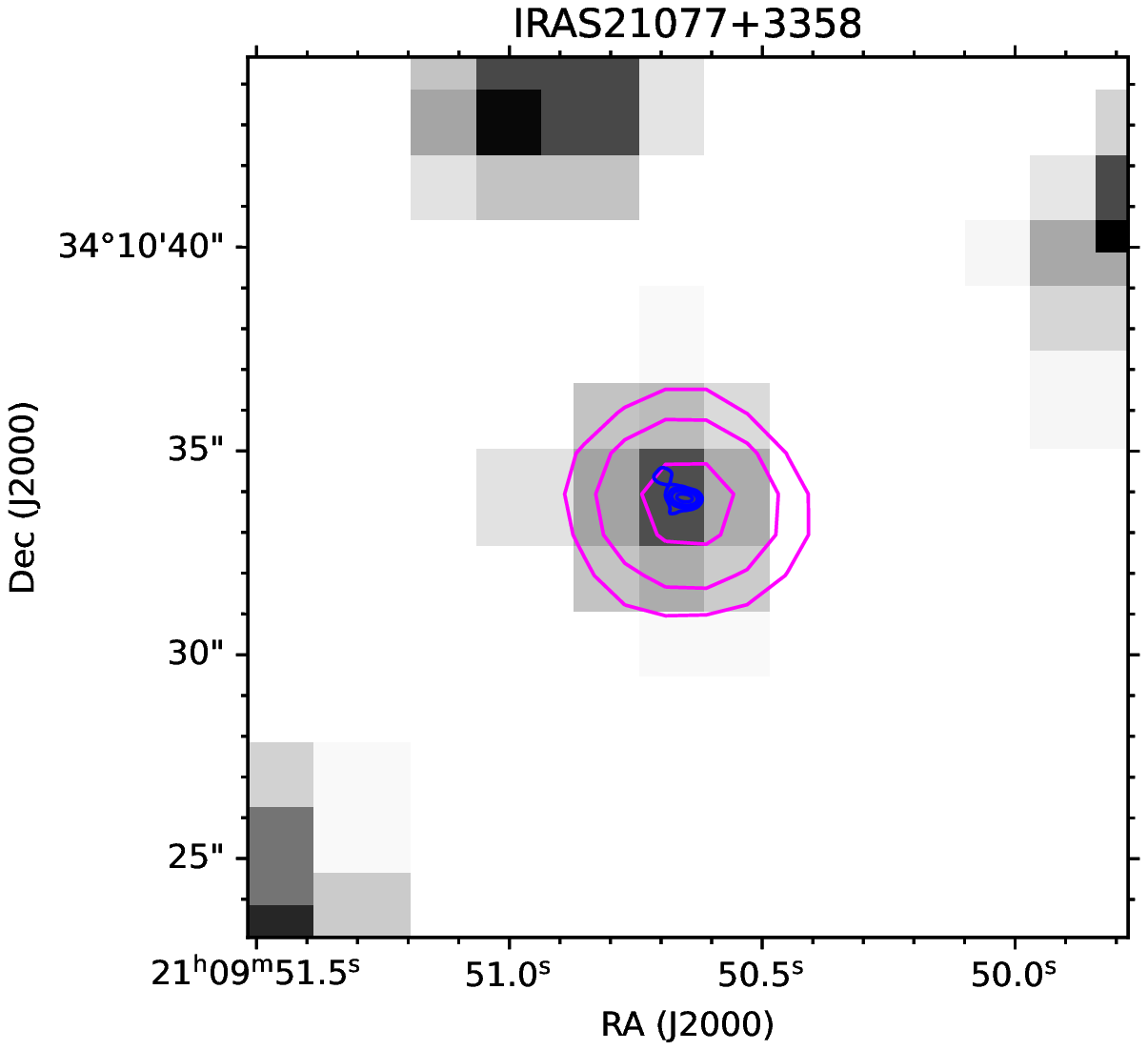}

     \caption{VLA contours overlaid on SDSS or HST R-band gray image of OHMs. Magenta: The contour map of the radio continuum emission from the VLASS survey. Blue: The C or X band contour map of the radio continuum emission: IRAS 03521: C band at epoch 2002Jan2. IRAS 07572: C band at  2002Jan2. IRAS 08449: 9GHz at 2016Dec02  IRAS 11180: X band at 2004OCT1. IRAS 21077: 9 GHz at 2014May23. The image parameters, including map peak, beam FWHM, and  3 $\sigma$ noise level, were present in Table \ref{imagepar}.
      }
         \label{highfreq1}
   \end{figure*}

The arcsecond-scale radio structure of OHM galaxies is available for 21 sources (see Table \ref{imagepar}). The radio images for sources with arcsecond-scale VLA line observations and extended structures are presented in Fig. \ref{highfreq1}, \ref{lowfreq} and \ref{highfreq2}). We found that the radio continuum emission of these sources shows no clear extended structure from the VLASS survey and L-band VLA-A observations (see Fig. \ref{lowfreq}). Five sources show possible extended structure at sub-arcsec scale from C or X band VLA data (see Fig. \ref{highfreq1}), which might be caused by possible jet or starburst regions. However, all the radio continuum emissions can be fitted with one gaussian component with a size smaller than the beam FWHM (see Table \ref{imagepar}).  

\section{Discussion}

\subsection{The high-resolution OH line emission}

Because the majority of these galaxies are likely showing multi-nuclei from optical images (see Fig. \ref{lowfreq}), there is a possibility that the broad OH line profiles of these galaxies are caused by the combination of maser emission from multi-masing-nuclei (see notes of each source in Sect. 4.7). We have presented the arcsec-scale structure of OH line emission for 11 of the 12 sources with available VLA-A observations. Our results have recovered all the OH maser emissions detected by a single-dish telescope in the literature or our new FAST observations. It indicated that the majority of the fitted component sizes of the OH emission are much smaller than the beam size of VLA-A ($\sim$ 1.4"), and the actual physical sizes should be smaller than the kpc-scale.

It is also possible that these OH masing regions contain two nuclei within 1 kpc (e.g., the separation of the two nuclei of the well-known OHM galaxy Arp 220 is about 0.4 kpc),  which needs higher resolution observations of these sources. However, VLBI observations of four galaxies in this sample are available in the literature, showing that the compact OH maser emission is distributed in hundreds pc scale (IRAS 02524, 08201, 12032, 14070, see Section 4.7 for the notes of each source and references therein). It means that other factors should cause the broad OH profiles, including orbiting disk, inflows/outflows, or particular merging stage \citep[see][]{2005ApJ...618..705P}.

\subsection{The variability of OH line emission}
Generally, variability in OHMs provides additional evidence that narrow lines correspond to compact masing regions as it is believed to be caused by interstellar scintillation \citep{2002ApJ...569L..87D}.  \cite{2016MNRAS.460.2180H} and \cite{2015MNRAS.447.1103M} have found the existence of intrinsic variability in amplitudes of OH features related to the masing region rather than that produced by interstellar scintillation. And one characteristic of the inherent variability caused by masing clouds is a significant decrease in OH line flux density during more than ten years \citep{2016MNRAS.460.2180H,2015MNRAS.447.1103M}. Nearly all the OHM galaxies in this study were first detected by Arecibo OH surveys 20 years ago. However, our new FAST observations find that the OH flux densities are consistent with previous observations, likely showing no similar cases caused by intrinsic changes. Because the intrinsic variability is unclear, decreasing OH flux densities might only respond to the compact masers found in VLBI observations \citep{2016MNRAS.460.2180H}, which needs high-resolution interferometric observations to study these variabilities further. 

The VLA and FAST observations have shown that the OH profiles of six sources are variable. 
The results confirm the arguments from \cite{2007IAUS..242..417D} that IRAS 12032+1707 shows variability in at least two line components \citep{2007IAUS..242..417D}. Two other known variable sources in OH line profiles \citep{2007IAUS..242..417D}, IRAS 02524+2046 and IRAS 21272+2514, also show variabilities by comparing the multi-epoch observations. We also found that three new sources, including IRAS 08201+2801, IRAS 23028+0725, and IRAS 11028+3130, were also variable in their OH profiles. Three variable sources are available with VLBI observations for OH line emission, and all show compact structures of OH emission  (IRAS 02524, IRAS 08201, and IRAS 12032, see notes of each source in Sect. 4.7). So, the OH emission variability is consistent with the small size of masing clouds. These six sources detected with OH line variability show this sample's highest line flux densities. Other sources with weak OH emission might need higher sensitivity observations to catch their variabilities.
\subsection{The radio continuum emission and association of OH line}
We have presented the radio structure of continuum emission for 21 sources and OH line emission for 11 sources. We found that our selected sample's arcsecond-scale radio continuum and OH line emission show no significant extended structures. The fitted image component size is smaller than the beam size (see Table \ref{imagepar}). We estimated the relative positions between OH line and continuum peak positions ($\delta_{lc}$) (see Table \ref{table1}). The $\delta_{lc}$ ranges from 0.06 to 0.58 arcsec and are all smaller than the 3 $\sigma$ of the uncertainties. It means that $\delta_{lc}$ are not significant, which is similar to the results for low-redshift OHM galaxies \citep[see][and references therein]{2005ARA&A..43..625L}.

\subsection{The classification for the dominant power source for the radio activity of OHM galaxies}
Generally, the FIR-radio (q)-ratio, brightness temperature ($T_{b}$), and spectral index are three independent radio parameters for classifying the radio activity in the nuclei. Based on these three parameters, \cite{2006A&A...449..559B} defined an activity factor $\beta_{c}$ to provide a consistent picture of the radio nature of the nuclei. The OHMs and candidates in \cite{2006A&A...449..559B} were detected before the Arecibo OH surveys \citep{2000AJ....119.3003D,2001AJ....121.1278D,2002AJ....124..100D}, which are nearly all known OHMs at low redshift (z<0.1). We selected the low-redshift OHMs from \cite{2006A&A...449..559B} based on their OH luminosity available in \cite{2014A&A...570A.110Z}, including 34 low-redshift OHMs (30 sources with z<0.1 and four sources with z<0.136 ) as the control sample.

We have calculated the radio parameters for our selected high-redshift OHMs following the methods in \cite{2006A&A...449..559B} and reported the results in Table \ref{table1}. We performed the two-sided Kolmogorov–Smirnov (K–S) statistical test to compare the properties of the low and high redshift sample; the results are presented in Table \ref{table3}. We can see that the $L_{OH}$, $L_{1.4GHz}$, and $L_{IR}$ of the two samples are significantly different, with p-values much more minor than 0.01. However, the p-values for q(1.4), q(4.8), $\beta_{c}$ and $T_{b}$ are all much higher than 0.01, which indicates that the two samples do not show significant evidence of being drawn from different distributions for these radio parameters.

The radio continuum emission of OHMs shows extended emission might be related to a central disk, possible jet structures, or multiple nuclei structures. Because most OHMs show compact radio emission for our selected high-z samples and low-z samples from \cite{2006A&A...449..559B}, we only compare the sources with large extended structures.  \cite{2006A&A...449..559B} have measured the physical scale of four sources with significant extensions, which are about 3.0, 4.3, 6.4, and 16.8 kpc. About five sources in this work show clear structures ( see Fig. \ref{highfreq1}). And the scales (the distance between peak centers or peak center to the farthest edge measured from 3$\sigma$ radio contour images) are about 3.5, 8.6, 2.3 1.8 (21.8), 2.4  kpc (see Fig. \ref{highfreq1} and Notes of each source in Sect. 4.7). 
Then, there also seem to be no significant differences in the scale of extended structures between the low and high redshift OHMs in our selected sample.

\subsection{The radio properties and the OH emission detection}
\cite{2014A&A...570A.110Z} have found that the AGN fraction (based on empirical criteria from Wide-Field Infrared Survey Explorer (WISE) color)  is similar for both OHM and non-OHM samples with a value of 40\%. Similarly, \cite{2006AJ....132.2596D} also found no distinguishing spectral properties to differentiate OHMs from nonmasing mergers using optical spectrophotometry studies.
Based on the optical classification listed in Table \ref{table1}, about 71\% of sources (15) can be classified as AGN or AGN candidates. We also found that there are 62\% of the low-z sample (21 of 34 OHMs) are AGN or AGN candidates. Although the AGN rate is slightly higher in the high-redshift sources, the AGN fraction of the low and high-redshift sample seems to have no significant differences if considering the uncertainties ($\sim$ 16-19\% derived based on the 95\% confidence level of a binomial proportion).   \cite{2022MNRAS.510.2495S} showed no significant correlation between $L_{OH}$ with spectral indexes of the radio continuum emission. We further found that there are no significant differences in the classification of radio properties between the low and high-redshift OHM galaxies. We noted that the $L_{OH}$ of our selected high-redshift sample is, on average, much higher than the low-redshift sample. These results show that the nuclei's classifications of the power source do not significantly affect the luminosity of detected OH line emission. 

Furthermore, high-resolution VLBI observations are available for only several sources in this sample: IRAS 08201+2801, IRAS 10339+1548, IRAS 12032+1707, IRAS 02524+2046, IRAS 14070+0525, and all these sources show no radio continuum emission above a 3$\sigma$ level  (see section \ref{notes} and references therein). \cite{2003MNRAS.343..585F} show that the radio flux correlates with the starburst luminosity and the gas masses rather than the AGN in the nuclear regions governs the total luminosities of ULIRGs. It means that the origin of the radio continuum emission might be from starburst rather than the central AGN, which is consistent with the view from \cite{2003MNRAS.342..373R} that the existence of an AGN in OHM galaxies may well be a random phenomenon not associated with their masing properties. High-resolution optical imaging and spectroscopy observations of OHM galaxies have also shown the presence of plenty of star-forming regions in the nuclei of these galaxies \citep[see][]{2020MNRAS.498.2632H,2011MNRAS.416.1267M}. These results support a starburst origin for the OHMs, and the OH masing regions are largely believed to be star-forming \citep[see][]{2021A&ARv..29....2P,2022A&A...661A.125W}. Because of the existence of star-forming regions, the signature of AGN in OHM galaxies may be lost when using standard diagnostic diagrams \citep[see][]{2011MNRAS.416.1267M}. Meanwhile, an ongoing project using high-resolution, optical–IR multi-wavelength imaging and spectroscopy analysis has studied five from a sample of 15 OHM galaxies known to be dominated by star formation. However, the results showed the presence of a previously unknown AGN in four sources, which support the hypothesis that OHM galaxies harbor faint recent triggered AGN \citep[see][and references theirin]{2020MNRAS.498.2632H}. Although compact radio jets might be with much shorter lifetimes than the triggered AGN \citep[see][]{2006A&A...454..125E,1999ApJ...521L.107E}, further high-sensitivity radio interferometry and VLBI observations of a large number of OHM galaxies can also be useful in finding evidence of radio AGN and their connections with the megamaser emission \citep{2020A&A...638A..78P}.

Because the OHM emission region of our sample is compact at the kpc-scale, it is also possible that there are multi-masing regions or nuclei at the sub-kpc scale similar to Arp 220 \citep[the separation of two cores is about 0.4 kpc ][]{2003MNRAS.342..373R}. However, we do not find evident two isolated components from sub-arcsecond scale radio images of 12 galaxies at C and X band (see Fig. \ref{highfreq1} and Table. \ref{imagepar}). Because of the association of OH emission and radio continuum, likely, the OH emission may also be compact at sub-arcsecond scale observations. However, OH line observations at these resolutions are needed to further confirm the OH emission structure.

\subsection{Implications from the results of our pilot FAST OH observations }
We have randomly selected 12 high-redshift sources for Pilot FAST observations from 30 known high-z OHM galaxies. We have performed a K-S test that shows no significant differences between our chosen sample and the rest 18 high-redshift OHM galaxies, including OH line peak, FWHM, and $L_{OH}$. The 12 sources for our pilot observations might give us clues about the properties of a complete sample of high-redshift OHM galaxies available in the literature: 
1) The detection of intrinsic change of masing regions-related variability might need observations of a much larger sample or blind surveys because the detection rate should be lower than 1/10. 
2) The detection rate of variability caused by ISS is about 1/3, and the actual detection rate might be higher than this value if we set a flux density lower limit. It supports the view that the OHM emission clouds are small; the actual time scales of varying will need shorter spacing observations to detect. 
3) The OH 1665 MHz line existed for all the sources. However,  over 2/3 of these sources show broad multi-peak OH 1667 MHz profiles blended with the OH 1665 MHz line. The detection of two OH satellite lines needs significantly lower sensitivity observations of these OHM galaxies.

\subsection{Notes on individual sources}

\label{notes}

\subsubsection{01562+2528}  
The SDSS image of this source shows two nuclei located in the Northwest (NW) and the southeast (SE) region, respectively (see Fig. \ref{lowfreq}). \cite{2006AJ....132.2596D} measured the optical redshift of the two nuclei and found that the velocity of the NW nucleus is consistent with the OH redshift. However, our results show that the radio continuum and OH line emission are spatially associated with the SE nucleus. The optical velocity of the SE nucleus is about 49385 \kms \citep{2006AJ....132.2596D}, which is about 280 \kms smaller than the NW nucleus, and about 429 \kms smaller than the heliocentric velocity centroid of the 1667 MHz OH line.  
The OH spectrum from VLA is roughly consistent with that from Arecibo observation \citep{2002AJ....124..100D}; they both show that the 1665 MHz line is adjacent to the OH 1667 MHz line. A minor difference is that the VLA spectra also indicate a possible peak of the 1665 MHz profile (see Fig. \ref{ohline2}).
\subsubsection{02524+2046}
The OH line profiles of this source show strong and narrow 1667 and 1665 MHz lines \citep{2002AJ....124..100D}, and one of the known variable OHM galaxies \citep{2007IAUS..242..417D}. The VLA and FAST observations also indicated that this source is a variable source for the narrow line components (see Fig. \ref{ohline1}).
VLBI observations of the OH line and continuum emission by \cite{2020A&A...638A..78P} showed that the OH emission of IRAS 02524+2046 is distributed in a region about 210 $\times$ 90 pc, also detected no significant continuum emission at 3 $\sigma$ level of 30  $\mu$Jy/beam the brightness temperature is also in order of $10^{6}$ K.

\subsubsection{03521+0028}
This is a close pair separated by only 3.5 kpc from optical and Near-Infrared Images \citep{2002ApJS..143..315V}. The C-band VLA observations also show two separated components (see Fig. \ref{highfreq1}). 

\subsubsection{07572+0533}
\cite{2001AJ....121.1278D} found a blueward feature about ($\sim 3\sigma$) 400 \kms against the main 1667 MHz line. Our line profiles from FAST also detected the blueward feature. We also detected a second narrow component from the OH 1667 MHz line. The 1665 MHz line component is also consistent with that from Arecibo observations. We detected an extension of the radio continuum emission with scales of about 8.6 kpc 

\subsubsection{08201+2081}
The HST image of this source shows two nuclei \citep{2001AJ....121.1278D}. Our results show that the radio continuum and OH emission are likely both peaks at the southern nuclei (see Fig. \ref{lowfreq}).
The more accurate position of the compact OH emission (RA= 08 23 12.614, Dec=27 51 39.670) from EVN observations \citep{2004PhDT.......145R} is consistent with the position we measured from the VLA-A project (see Table \ref{table2}). No compact radio continuum emission is detected above a 3$\sigma$ level of about 0.45 mJy/beam from EVN observations \citep[see][]{2004PhDT.......145R}.

\subsubsection{08279+0956}
The optical image of this source shows the presence of three tidal tails \citep[see Fig. \ref{lowfreq} and][]{2005MNRAS.364...99V}. The VLA spectrum is consistent with the Arecibo spectra from \cite{2001A&A...377..413P},  which contain both 1667 MHz line and 1665 MHz line. The OH spectrum from FAST observation only detected part of the 1667 MHz line, while over half of the line profile is covered by possible RFI (see Fig. \ref{ohline2}). The FAST spectra also show a feature at a velocity around 61800 \kms, but RFI might cause it because the on and off spectra both show a bump slightly higher than other features related to standing wave ( see Fig. \ref{fastonoff}).
\subsubsection{08449+2332}
The radio continuum emission is compact from arcsecond-scale VLASS and L-Band VLA observations (see Table \ref{imagepar}). The X-band VLA images show an extended structure of about 2.3 kpc, no radio continuum emission is detected from the accompanying galaxy (see Fig. \ref{highfreq1}). 

\subsubsection{09531+1430}
The SDSS image shows that this galaxy is an interacting system that contains eastern and western components \citep[see Fig. \ref{lowfreq} and][]{2005MNRAS.364...99V}. The OH spectrum of this object shows two peaks in 1667 MHz emission, which might originate from two nuclei \cite{2001AJ....121.1278D}. We have imaged the radio continuum and two line structures separately, and we found that the peak positions have no offset between these images. The radio continuum and OH emission regions are likely distributed between the two nuclei (see Fig.~\ref{lowfreq}).
\cite{2001AJ....121.1278D} show a significant depression in the weights spectrum at the location of a potential 1665 MHz line. The VLA spectrum shows the existence of the 1665 line at the exact location, and the FAST spectrum shows no detection of the OH 1665 line because it contains possible RFI at this velocity range (see Fig.~\ref{ohline2}).

\subsubsection{10339+1548}
 The Arecibo observations by \cite{2001AJ....121.1278D} suspect that the 1665 MHz line might be a broad feature and cannot be distinguished from a standing wave in the bandpass. The FAST observations likely show a 1665 MHz line feature with an upper limit of the line flux density of about 1.5 mJy. There is also an absorption feature at about V=58750 \kms. Because the ON and OFF spectra of this source are unstable, the baseline of the ON-OFF spectrum is complex (see Fig. \ref{fastonoff}), and we can not exclude the possibility that the baseline fluctuations might also cause the OH 1665 line and the absorption feature. We failed to detect the OH emission from the VLA project as listed in Table \ref{imagepar}, which might be caused by a bright, confusing source near this source.
 
 \subsubsection{11028+3130}
This single-nucleus source shows a broad tidal
feature to the west \cite{2002ApJS..143..315V}.
The OH spectrum from FAST observation is consistent with the result from \cite{2001AJ....121.1278D}, which includes features of the variable 1667 MHz line and the possible existence of the 1665 MHz line (see Fig. \ref{ohline1}).  

\subsubsection{11180+1623}
This source is a wide (21.8 kpc) pair of
galaxies \citep{2002ApJS..143..315V}. The optical velocity of the eastern component measured by \cite{1998ApJS..119...41K} is consistent with the heliocentric velocity centroid of the 1667 MHz OH line, and the OHM emission is more likely associated with this component (11180+1623A). We found that both nuclei show radio continuum emission,  11180+1623A show much higher radio flux than the western component and an extended structure over 1.8 kpc (see Fig. \ref{highfreq1}).  

\subsubsection{11524+1058}
\cite{2001AJ....121.1278D} identified a spectral feature as the 1665 MHz line at the predicted location, the FAST spectrum (see Fig. \ref{ohline2}) likely confirmed the existence of the 1665 MHz feature at about 2 $\sigma$ level. The two narrow peaks of OH 1667 line are also consistent with the results from Arecibo observations by \cite[][]{2001AJ....121.1278D} (see Fig. \ref{ohline2}). 
\subsubsection{12032+1707}

IRAS F12032+1707 is among the most luminous OHMs detected, \cite{2002ApJS..143..315V} have revealed the presence of two interacting galaxies separated by 12.0 kpc, which can also be seen from the SDSS image as showed in Fig. \ref{lowfreq}.
\cite{2001AJ....121.1278D} indicated that OH line complexes emitted from this source might originate in two nuclei. 
\cite{2005ApJ...618..705P} have observed the IRAS 12032+1707 with VLBA and found that almost all OH emission previously detected by single-dish observations has been recovered on a compact scale < 100 pc, which has ruled out the predication about the combined OH maser emission from two nuclei. Although this OHM host has a large 1.4 GHz continuum flux density (28.7 mJy), the VLBA observations have detected no significant continuum emission at 3 $\sigma$ level of 0.3 mJy/beam and imply no detection of compact radio AGN in this galaxy.  \cite{2007IAUS..242..417D} indicated that the two narrow components of the OH line profiles are variable. The FAST and VLA observations have shown that the brightest narrow component is variable (see Fig. \ref{ohline1}). However, the second brightest one is affected by RFI from FAST observation, and the VLA observations are likely to confirm the variabilities at about 2 $\sigma$ level (see Fig. \ref{ohline1}).

\subsubsection{13218+0552}
The OH spectrum of 13218+0552 shows two prominent broad emission peaks with a separation of 490 \kms in the rest frame, which may be associated with multiple nuclei. Meanwhile, it also shows tidal features and possibly a double nucleus with separation less than 1 kpc \citep{2002AJ....124..100D}. 
The VLA observation of the OHM and radio continuum emission show no extended emission (see Fig. \ref{lowfreq}), but they offer an offset of about 0.58 arcsec  (see Table \ref{table1}).

\subsubsection{14070+0525}
This is an OH gigamaser and the most distant OHM known at z = 0.2655; it shows multiple broad peaks of the OH line profiles, which might suggest multiple masing nuclei in this object \citep{1992ApJ...396L..99B,2002AJ....124..100D}. VLBA observations show that a significant fraction of the OH emission is resolved out and also detected no compact radio emission associated with an AGN; meanwhile, the offset between the two detected peaks is not significant \citep{2005ApJ...618..705P}. The OH profile from the FAST observation shows no significant variations compared with previous single-dish observations (see Fig. \ref{ohline2}).

\subsubsection{14586+1432}
\cite{2002AJ....124..100D} showed that the spectrum of this OHM is extremely broad and complicated, which might be caused by two masing nuclei in this object. We have investigated the channel images and found no apparent offsets among the peaks of channel images from low to high velocities. The integrated images of the OH line emission (see Fig. \ref{lowfreq}) show that the OH emission in this source is compact and has no extended structures. Similarly, the multi-band radio images of the continuum emission show only a compact core and no extended emission.

\subsubsection{21077+3358}
The X-band radio continuum image of this galaxy shows an extended structure of about 2.4 kpc.  
\subsubsection{21272+2514}
This OHM is the first OH megamaser observed to vary in time at a redshift of z=0.1508, and the spectrum exhibits OH multiple peaks. The VLA-A data have recovered the OH flux densities from single-dish observations and show significant peak variability. The combined channel images show the OH emission is compact at VLA-A observations; a possible weak extended structure toward the western direction might need further high-sensitivity observations to confirm it.

\subsubsection{23028+0725}
The OH spectrum from VLA data show that the OH 1667 and 1665 MHz line profiles are consistent with that from Arecibo observation by \cite{2001AJ....121.1278D}. Meanwhile, the narrow peak of 1667 MHz line might show variations (see Fig. \ref{ohline1}). We also found that the OH 1667 MHz line and 1665 MHz line emission are compact, which are distributed spatially coincide with the radio continuum emission (see \ref{lowfreq}).

\section{Summary}
\label{sect:discussion}
We have investigated the radio continuum and OH line emission of a sample of known OHM galaxies with z $>$ 0.15. The OH emission with VLA-A observations shows that the OHM emission regions are compact, and all the OHM emissions are recovered from less than 1". Meanwhile, the OHM emission is associated with the radio continuum emission with no significant offset. Although several sources show broad OH line profiles, there is only one masing region of these galaxies on arcsecond-scale images. It means that the OH emission regions might be single, or the physical size of the masing areas is much smaller than kpc.  
Based on the radio continuum emission, we have derived the parameters, including brightness temperature, spectral index, q index, and nuclei activity factor $\beta$; these parameters show no significant differences with the low redshift OHM galaxies. Because the high-z sample shows significantly higher $L_{OH}$ and these parameters stand for the classification of the power source of the nuclei, it indicated that the presence of radio AGN might not affect the $L_{OH}$ of OHM galaxies. The new FAST observations of 12 known galaxies found that six sources are variable and have no intrinsic variability caused by the environments, which require a significant decrease in OH line flux densities in long-time scales. Meanwhile, the FAST observations present a lower limit of the OH satellite lines of these galaxies.

\begin{acknowledgements}
We thank the anonymous referee for his/her useful comments and suggestions on the manuscript.
The study was funded by RFBR and NSFC, project number 21-52-53035 ``The Radio Properties and Structure of OH Megamaser Galaxies''.
  This work is supported by the grants of NSFC (Grant No.U1931203,12111530009,11763002).
  The National Radio Astronomy Observatory is operated by Associated Universities,
  Inc., under cooperative agreement with the National Science Foundation.
This research has made use of the CIRADA cutout service at URL cutouts.cirada.ca, operated by the Canadian Initiative for Radio Astronomy Data Analysis (CIRADA). CIRADA is funded by a grant from the Canada Foundation for Innovation 2017 Innovation Fund (Project 35999), as well as by the Provinces of Ontario, British Columbia, Alberta, Manitoba and Quebec, in collaboration with the National Research Council of Canada, the US National Radio Astronomy Observatory and Australia's Commonwealth Scientific and Industrial Research Organisation.

\end{acknowledgements}
\bibliography{vlaoh1223.bbl}
\appendix
      \section{Online material}
\begin{figure*}
   \centering
  
   \subfigure[IRAS 02524+2046]{ \includegraphics[width=14.6cm,height=5.5cm]{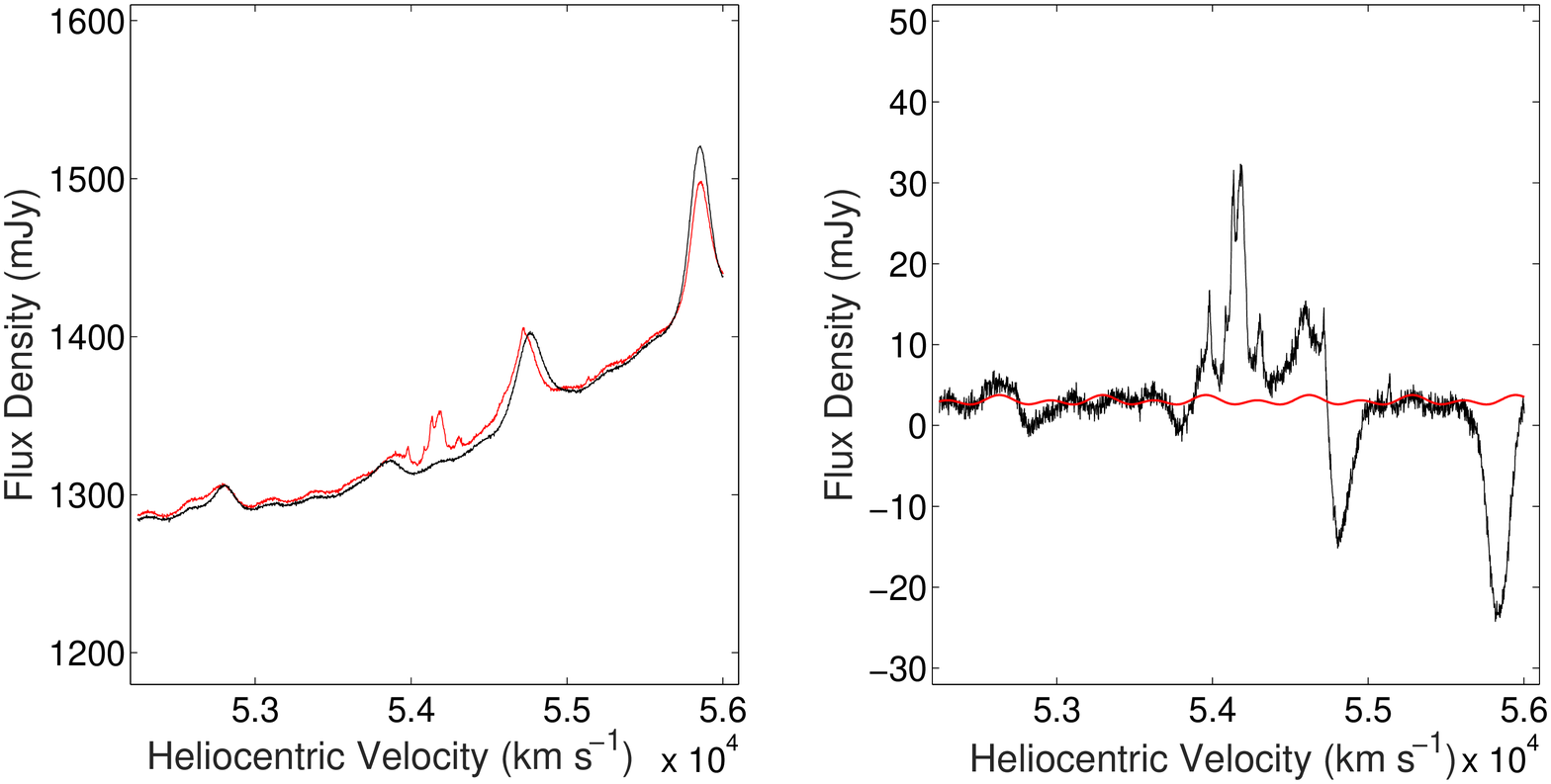}}
   \hfill
     \subfigure[IRAS 07572+0533]{ \includegraphics[width=14.6cm,height=5.5cm]{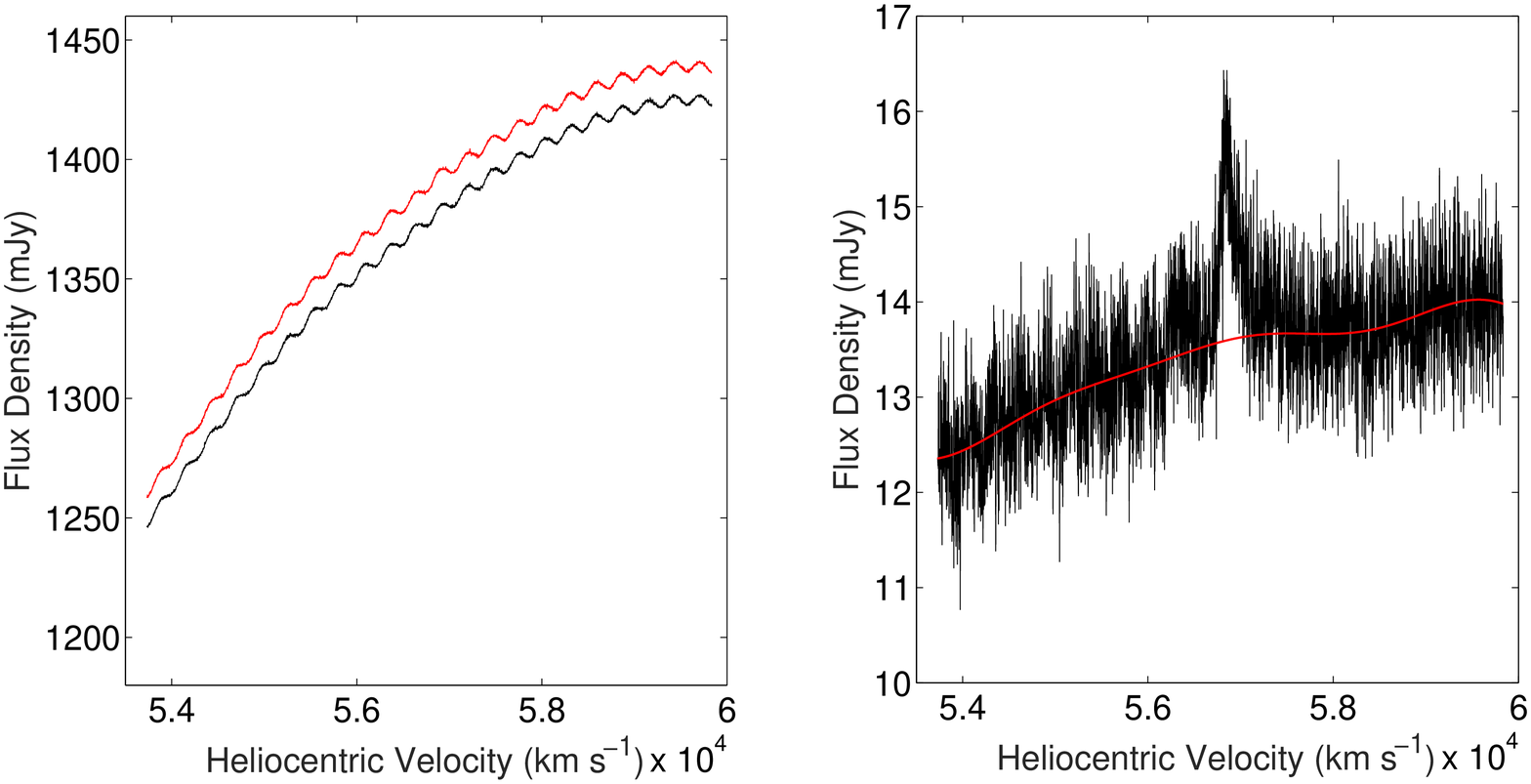}}
     \hfill
    \subfigure[IRAS 08279+0956]{\includegraphics[width=14.6cm,height=5.5cm]{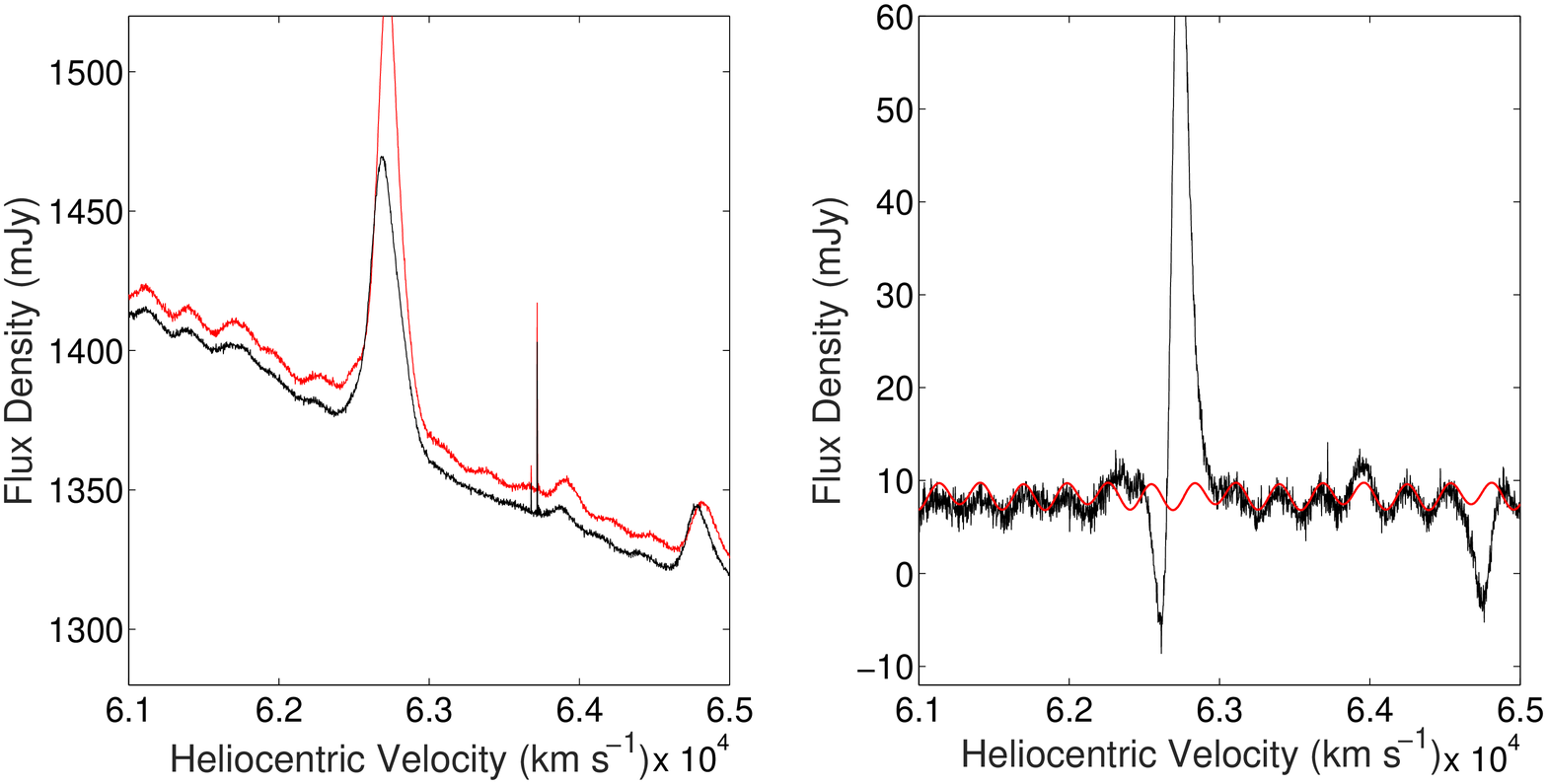}}
    \hfill
    \subfigure[IRAS 09531+1430]{\includegraphics[width=14.6cm,height=5.5cm]{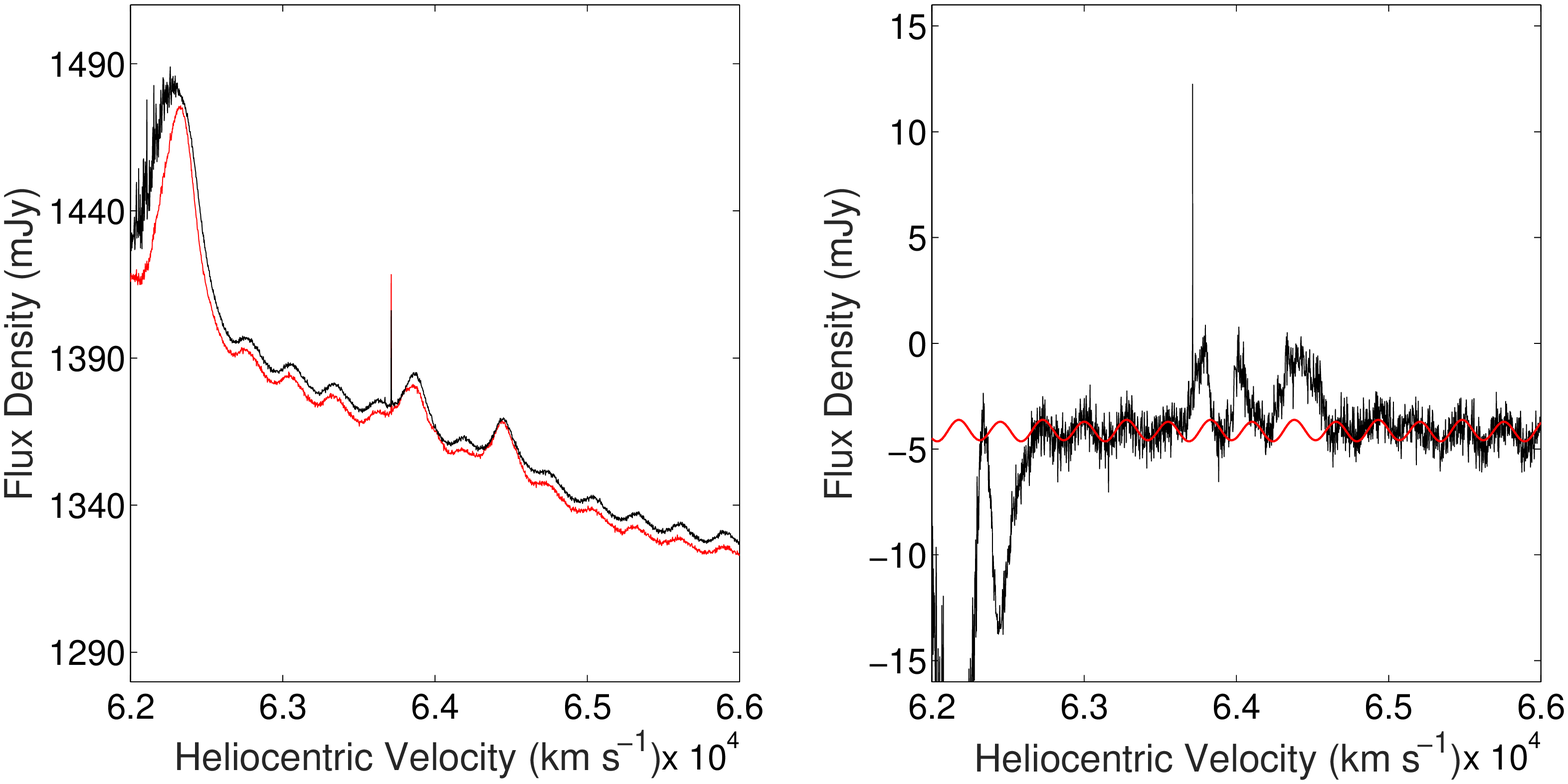}}
    \hfill
       \caption{The OH 1667 MHz line spectrum from FAST observations. Left panel: The black and red spectrum represent the target source ON and OFF spectrum, respectively. Right panel: The black and red spectrum stands for the target ON-OFF spectrum and fitted baseline, respectively.
      }
         \label{fastonoff}
   \end{figure*}  
\addtocounter{figure}{-1}
\begin{figure*}
   \centering   

  \subfigure[IRAS 10339+1548]{ \includegraphics[width=14.6cm,height=5.5cm]{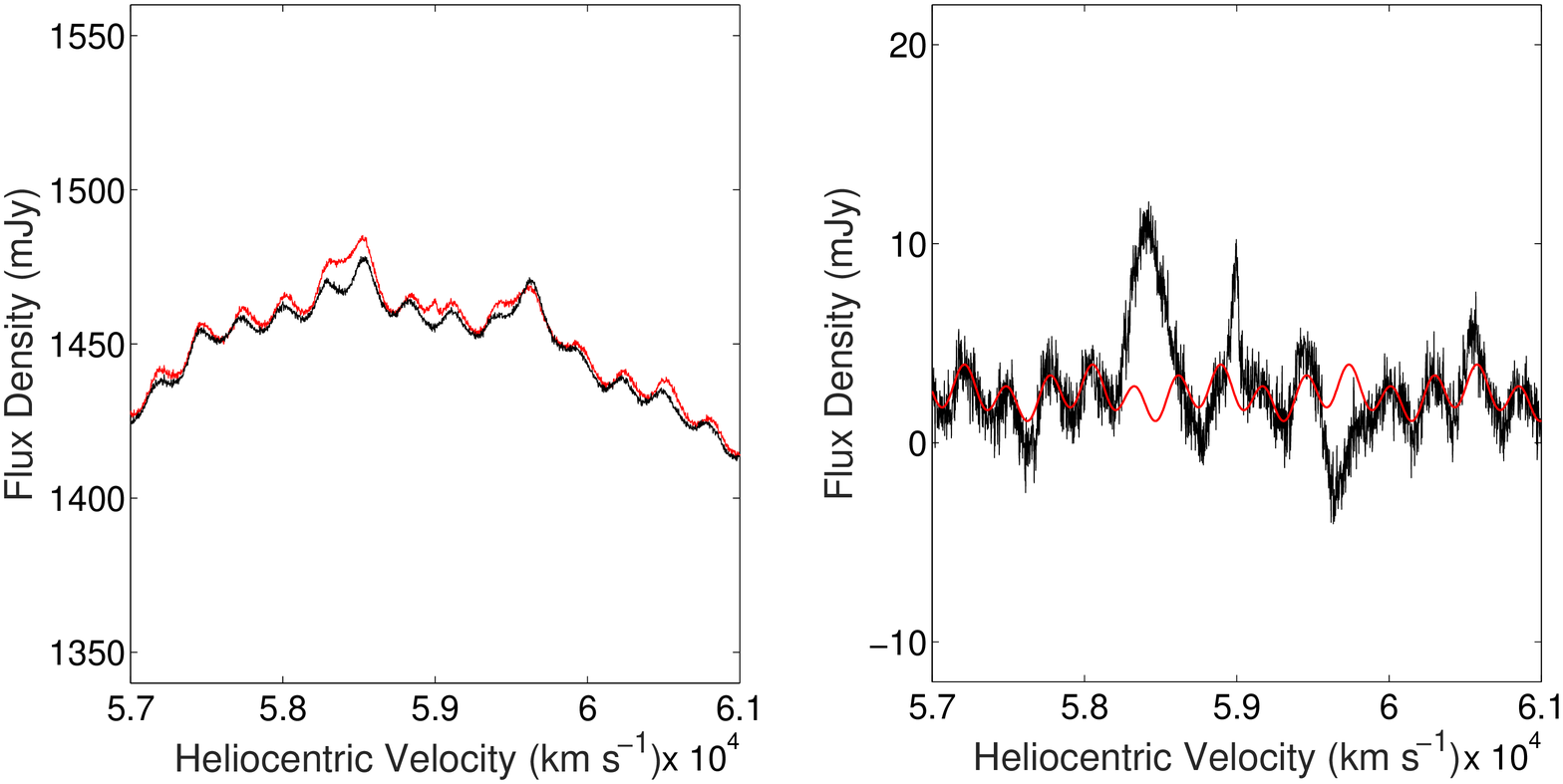}}
  \hfill
  \subfigure[IRAS 11028+3130]{ \includegraphics[width=14.6cm,height=5.5cm]{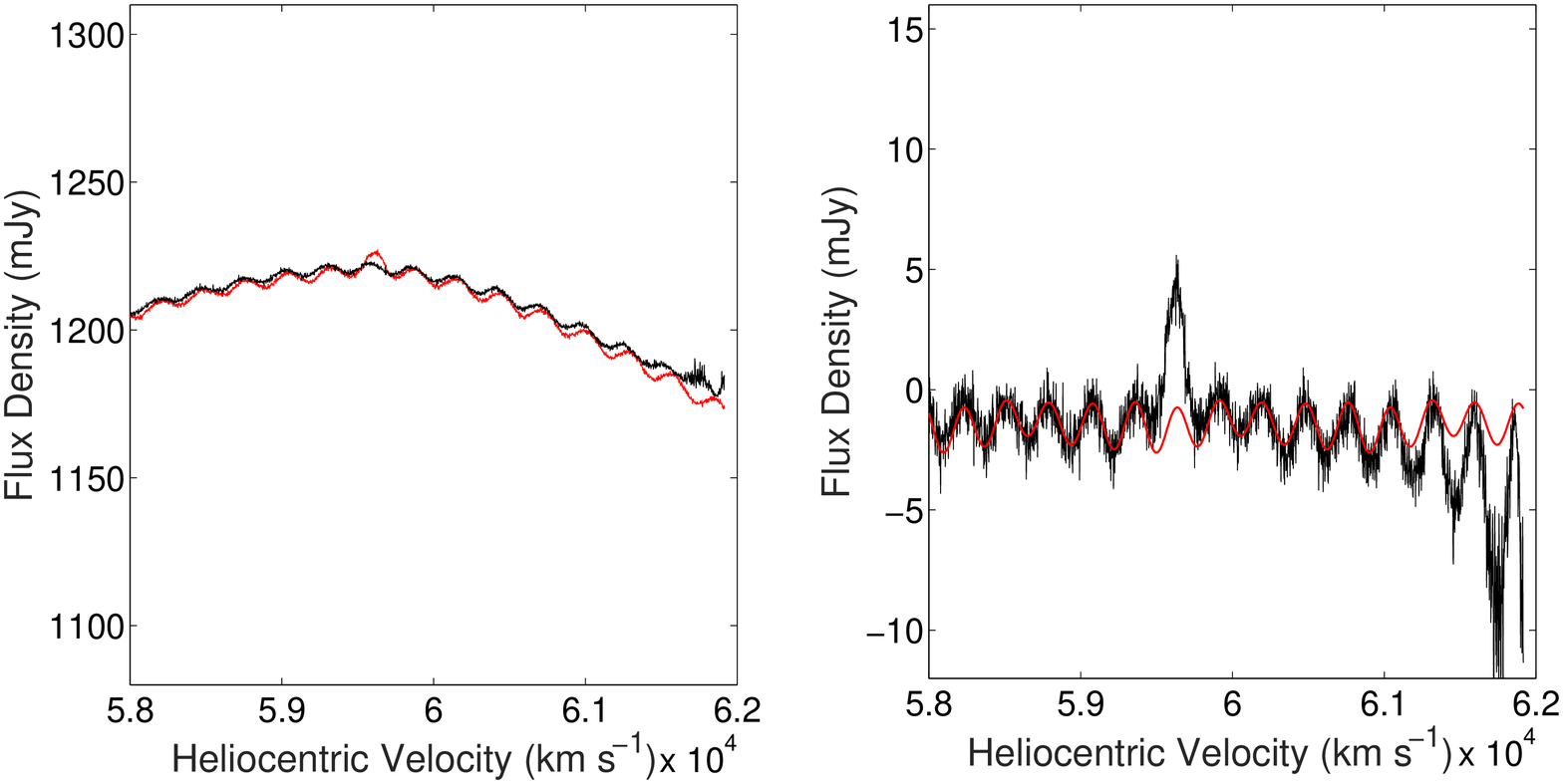}}
  \hfill
   \subfigure[IRAS 11524+1058]{\includegraphics[width=14.6cm,height=5.5cm]{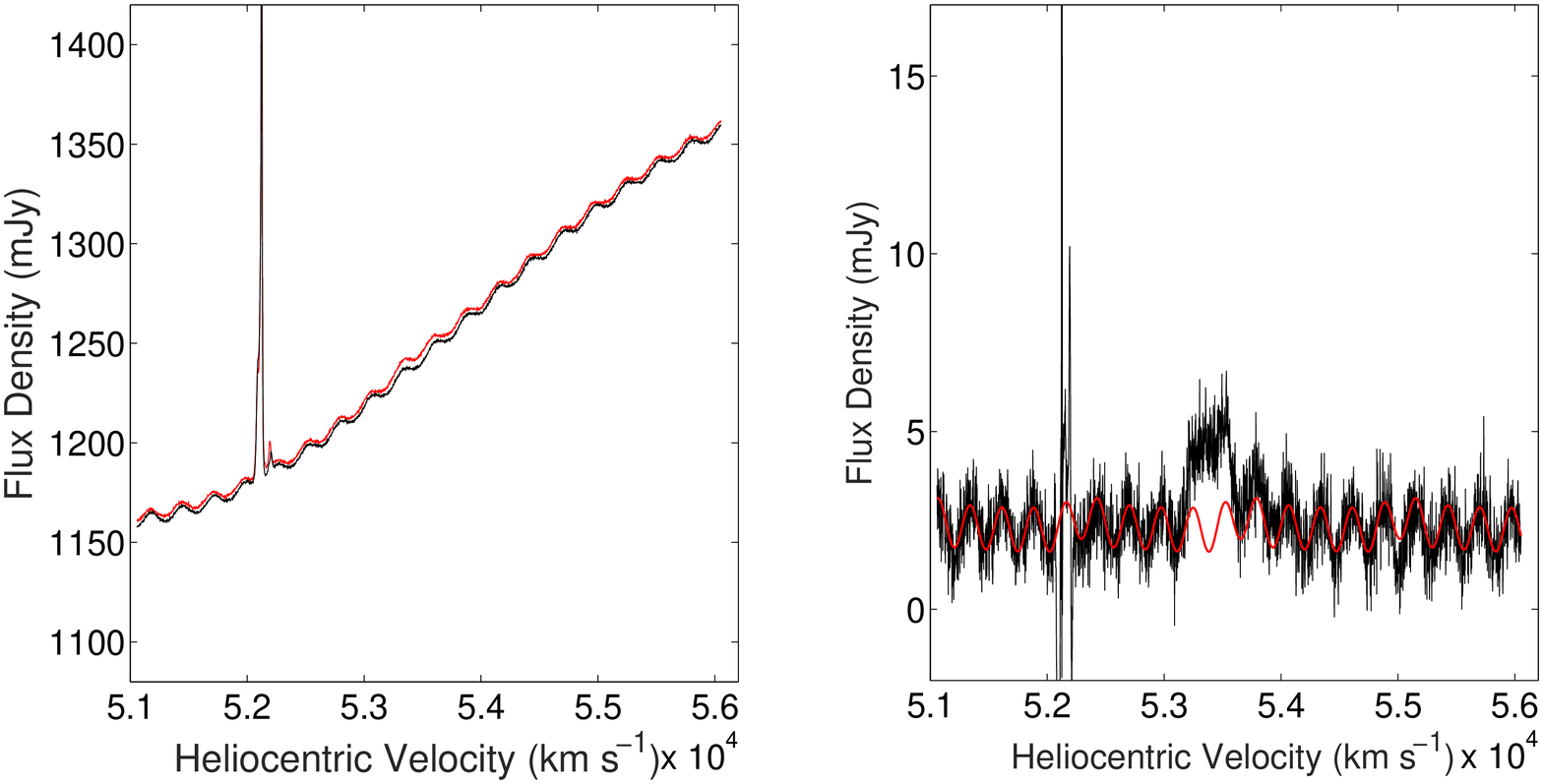}}
   \hfill
   \subfigure[IRAS 12032+1707]{\includegraphics[width=14.6cm,height=5.5cm]{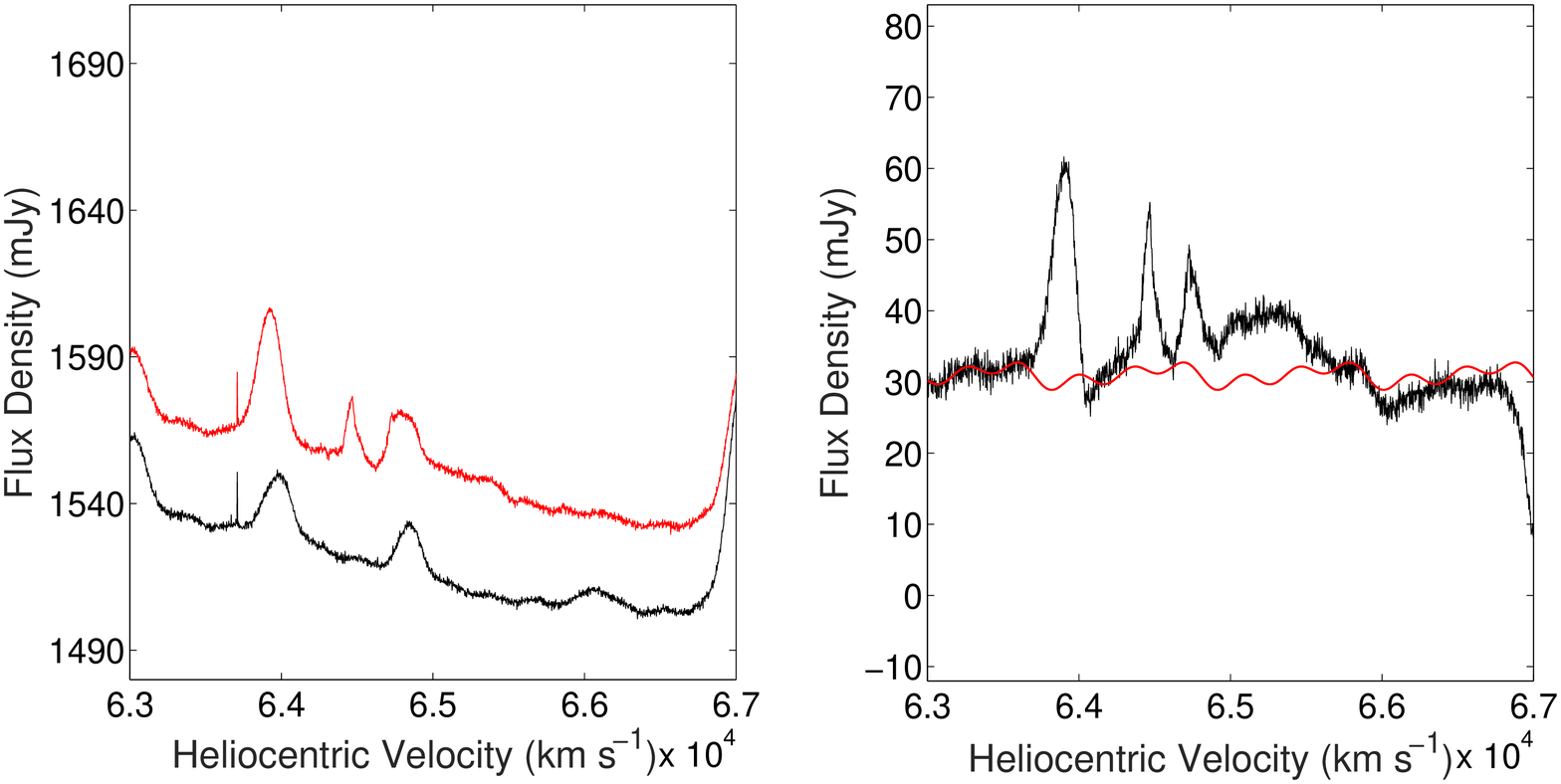}}
   \hfill
       \caption{Continued}
         \label{fastonoff}
  \end{figure*} 
\addtocounter{figure}{-1}
\begin{figure*}
   \centering   
   \subfigure[IRAS 13218+0552]{\includegraphics[width=14.6cm,height=5.5cm]{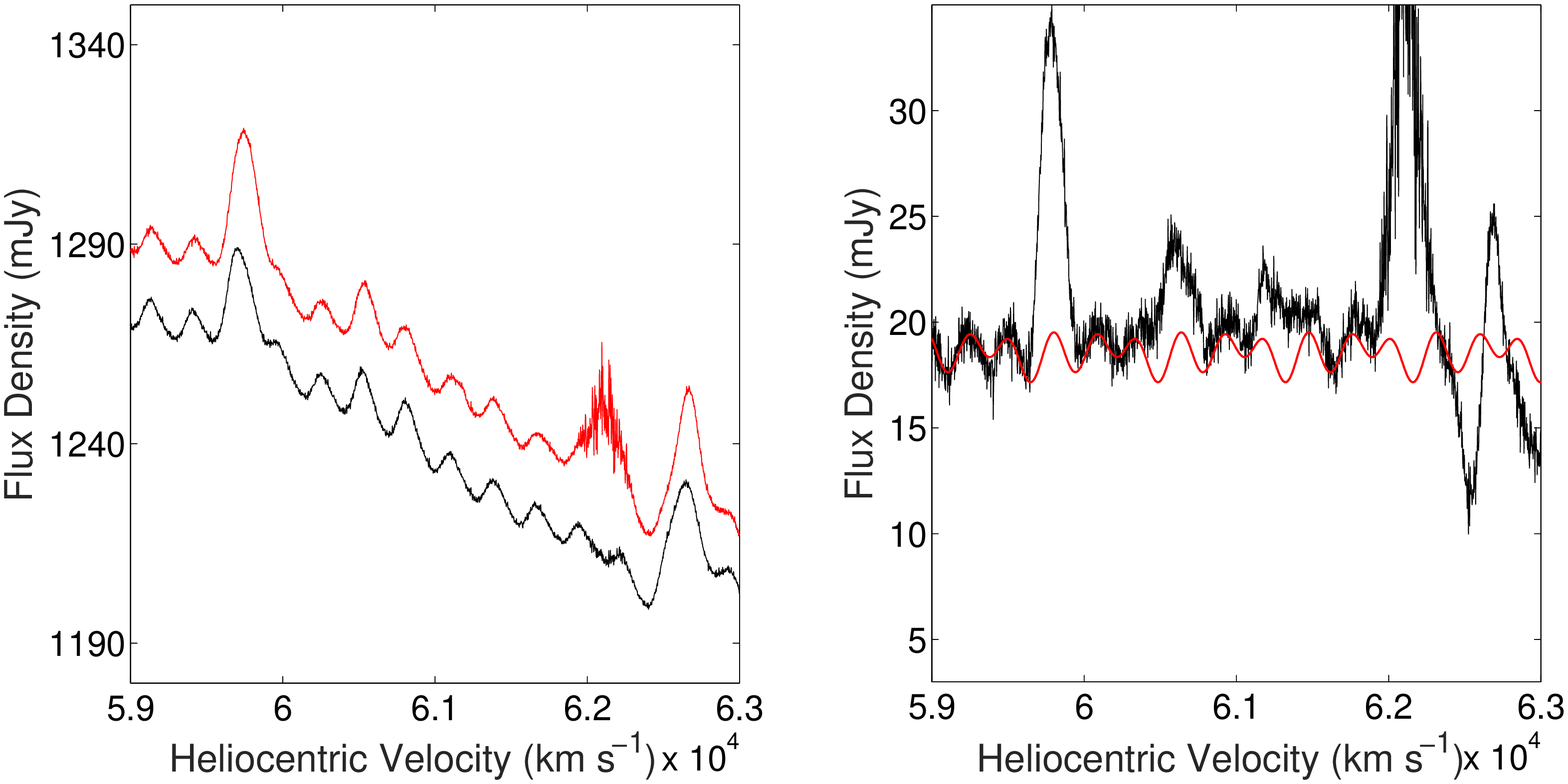}}
   \hfill
   \subfigure[IRAS 14070+0525]{\includegraphics[width=14.6cm,height=5.5cm]{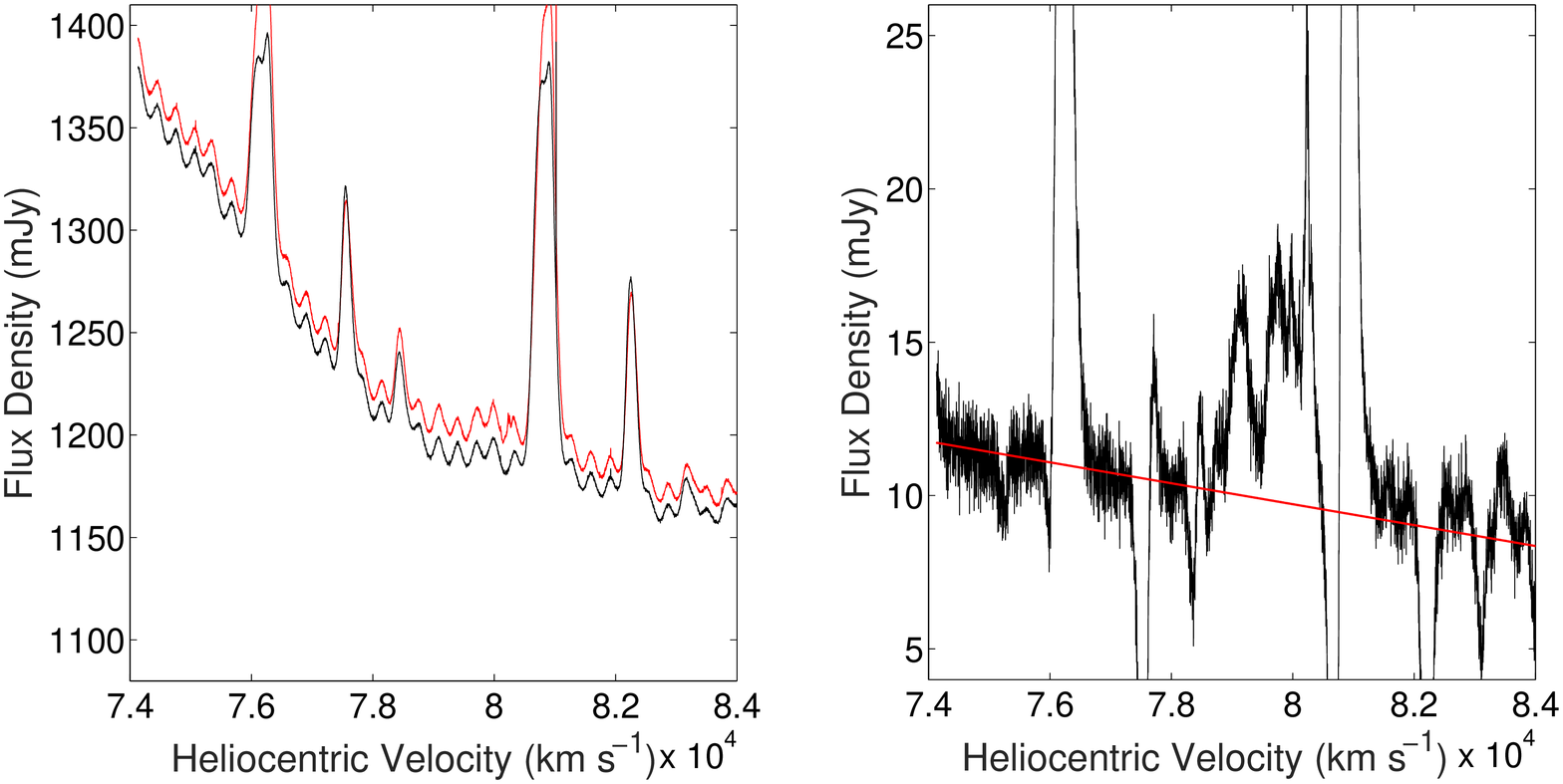}}
   \hfill
   \subfigure[IRAS 22116+0437]{\includegraphics[width=14.6cm,height=5.5cm]{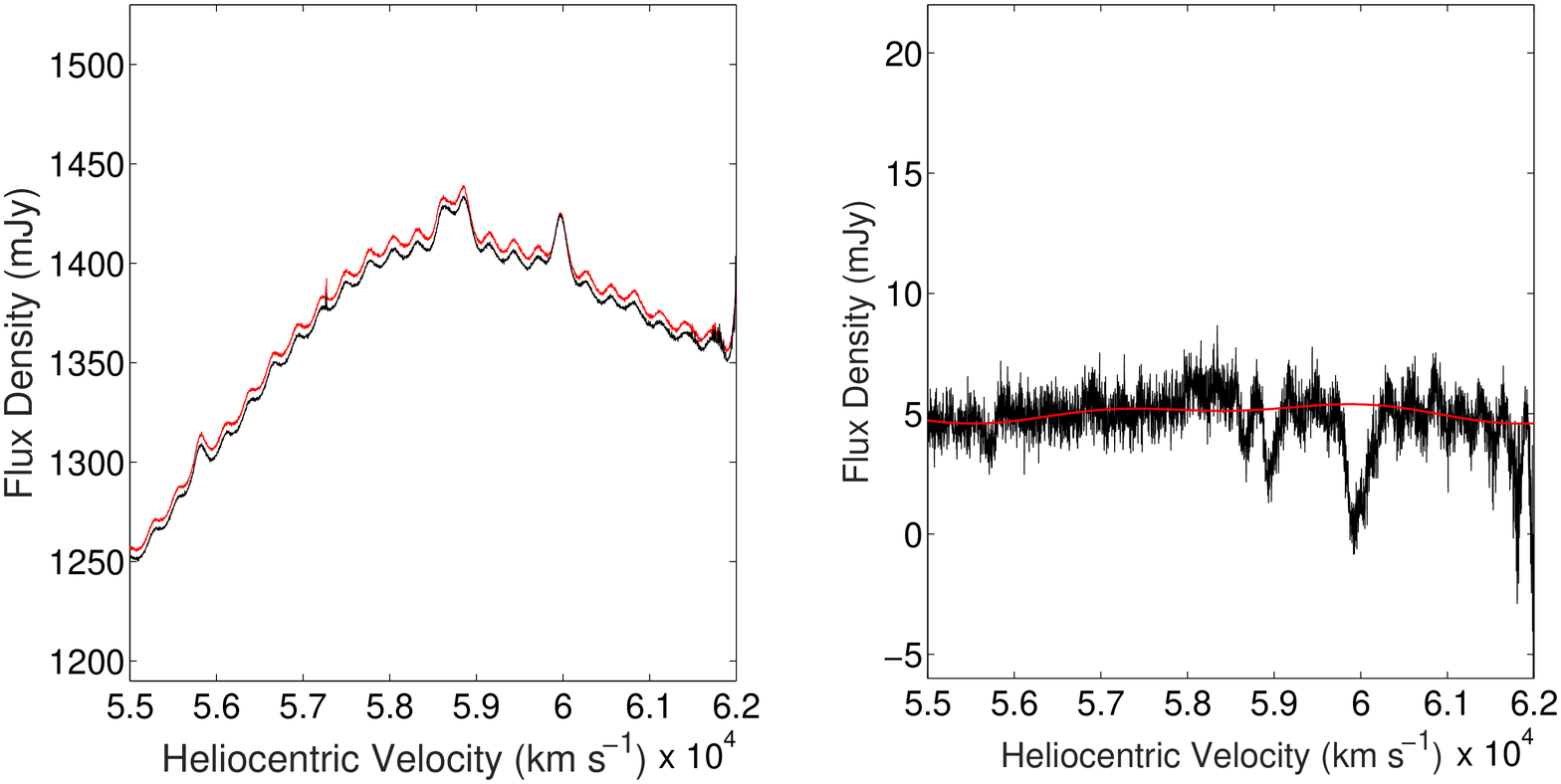}}
   \hfill
  \subfigure[IRAS 23129+2548]{\includegraphics[width=14.6cm,height=5.5cm]{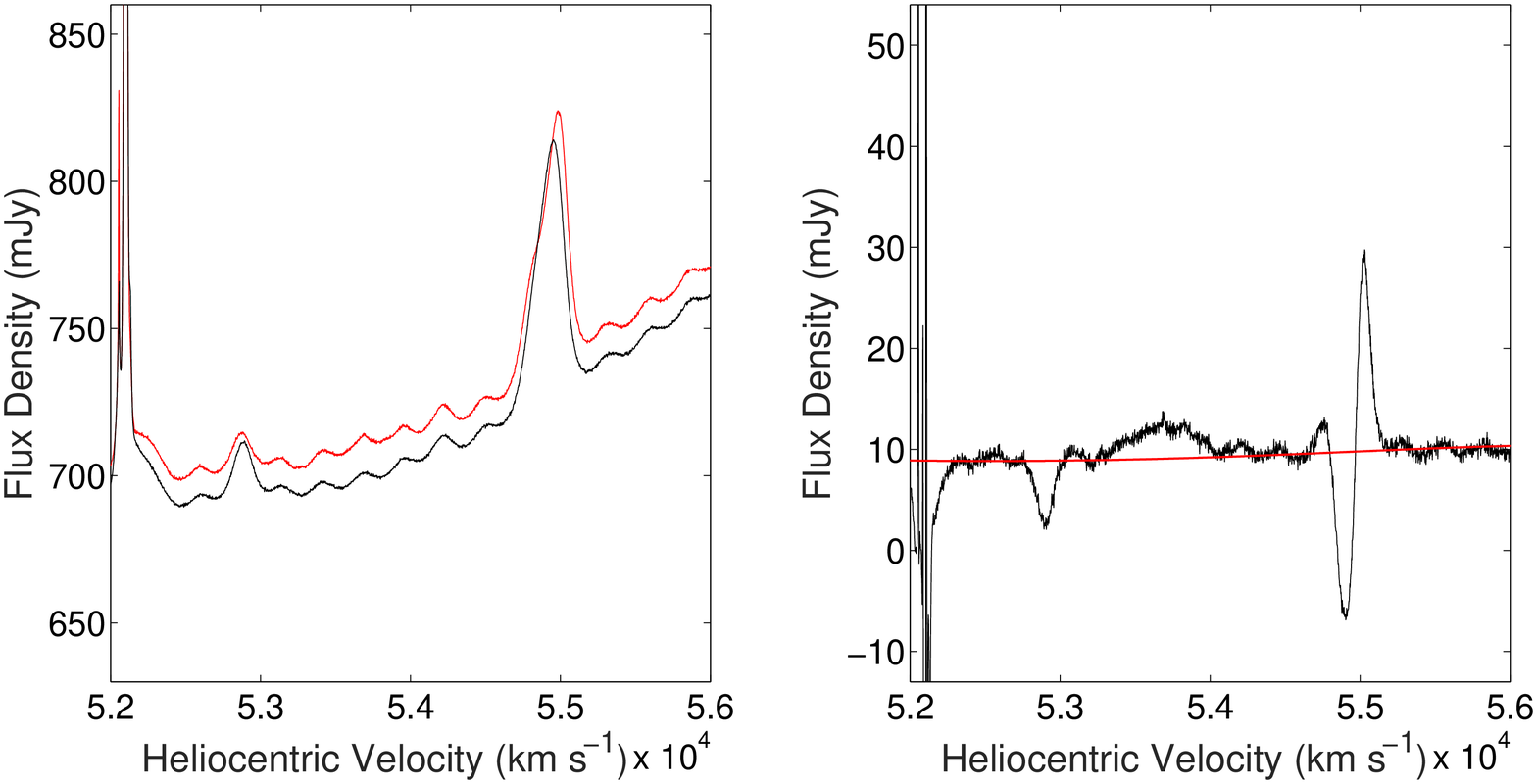}}
   \hfill
       \caption{Continued
      }
         \label{fastonoff}
  \end{figure*} 
\begin{figure*}
   \centering

    \subfigure{ \includegraphics[width=0.485\textwidth,height=6.2cm]{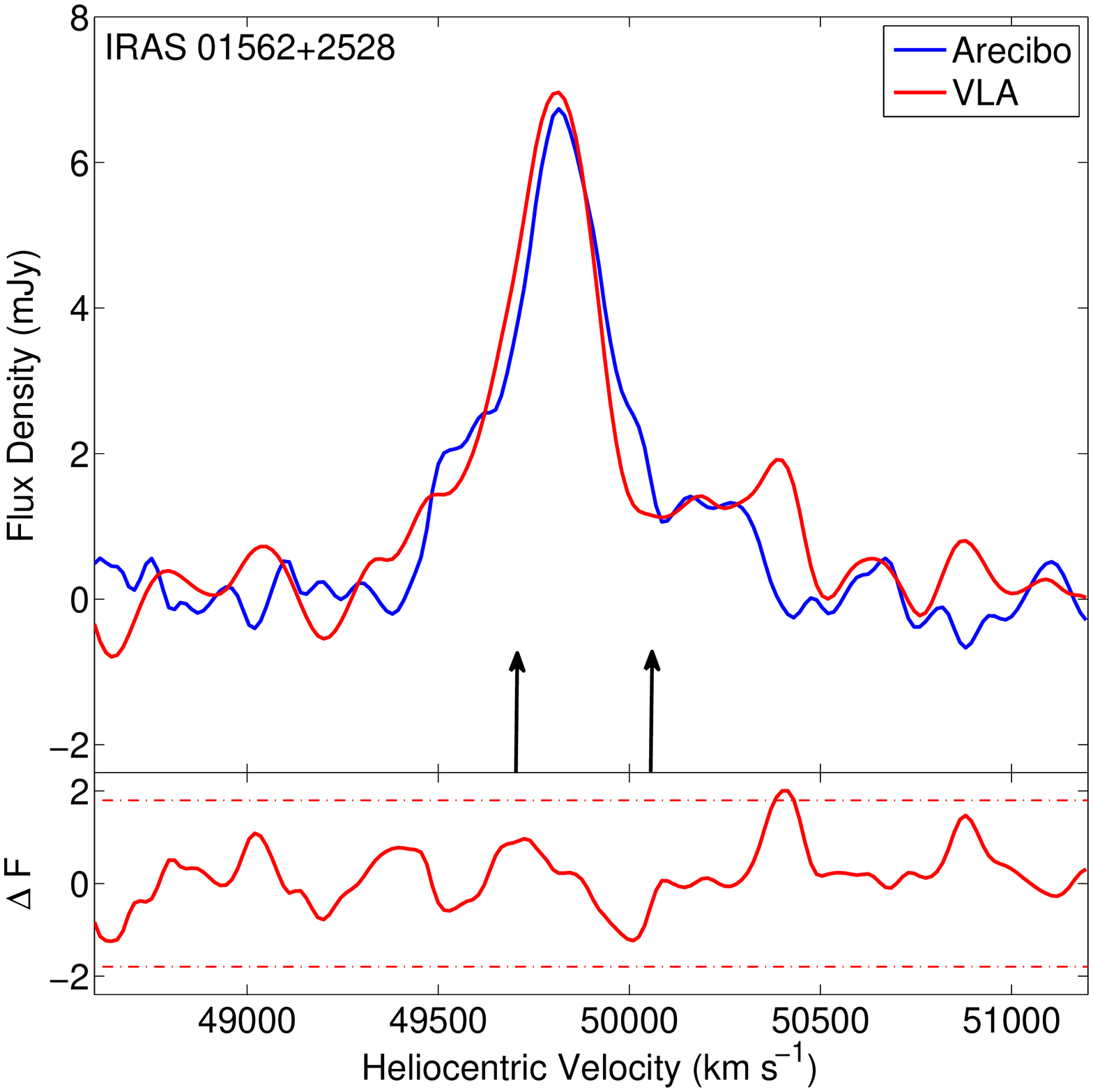}}
    \hfill
   \subfigure{ \includegraphics[width=0.485\textwidth,height=6.2cm]{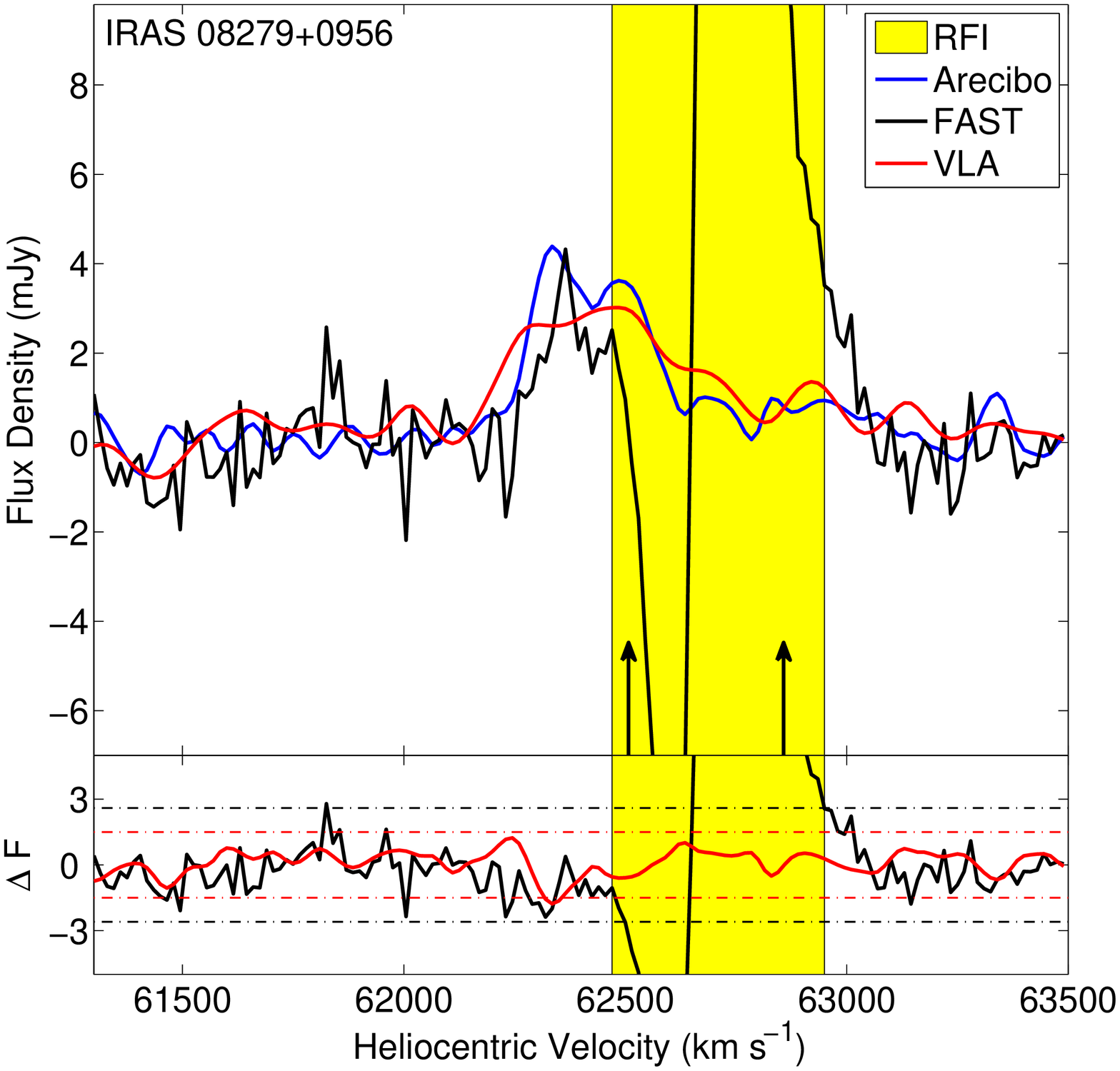}}
   \hfill
  \subfigure{ \includegraphics[width=0.485\textwidth,height=6.2cm]{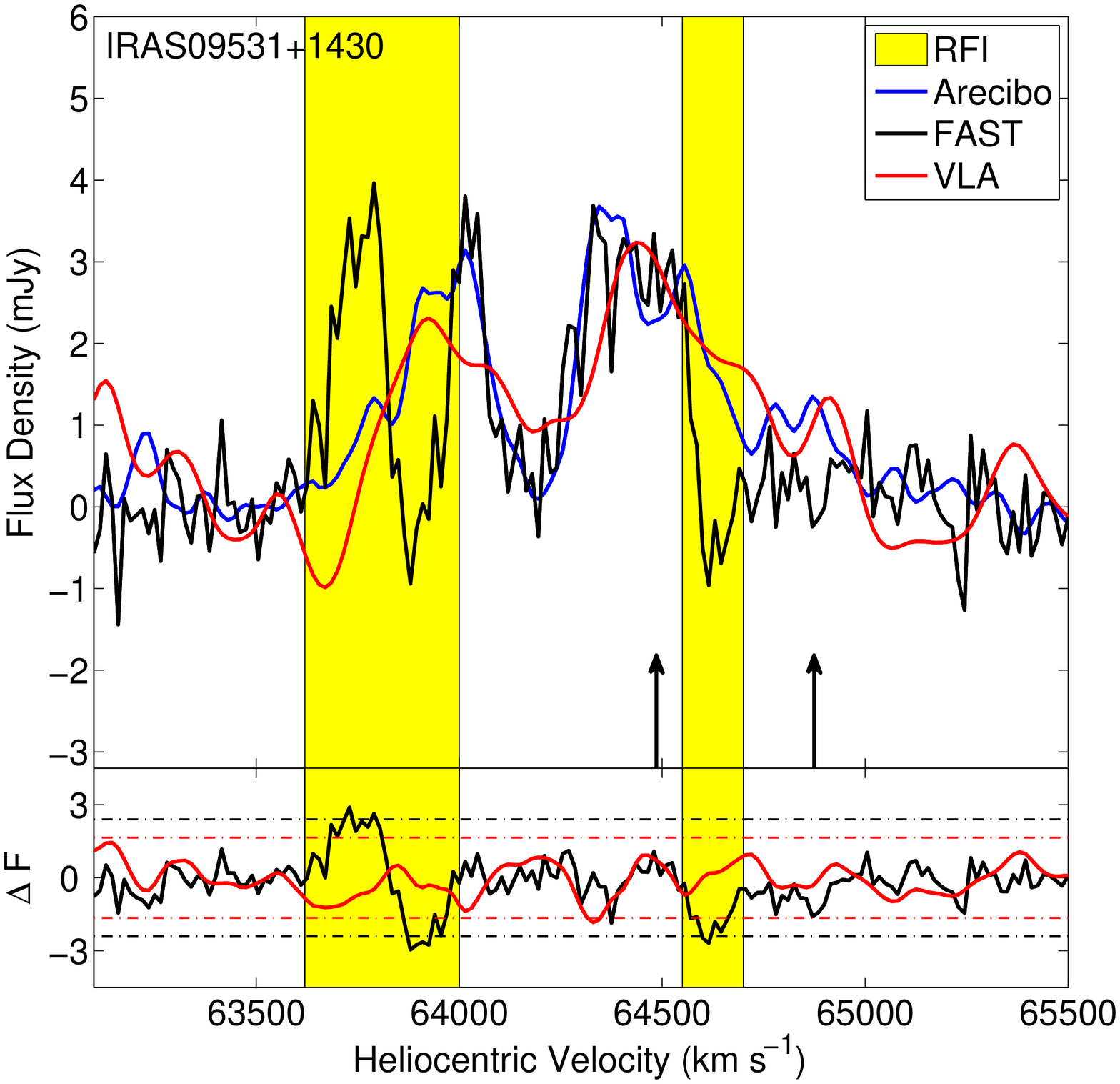}}
  \hfill
  \subfigure{ \includegraphics[width=0.485\textwidth,height=6.2cm]{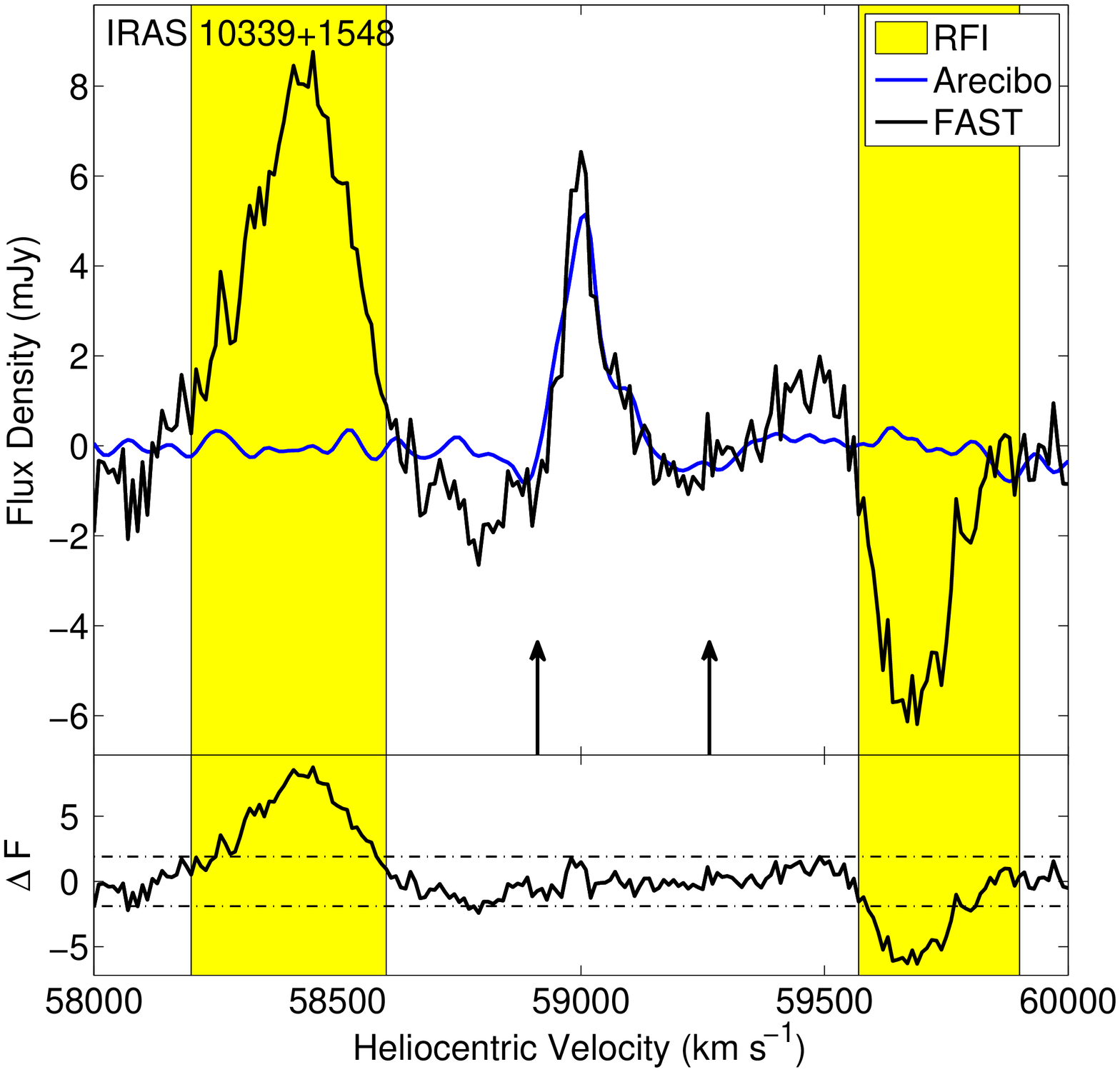}}
  \hfill
   \subfigure{\includegraphics[width=0.485\textwidth,height=6.2cm]{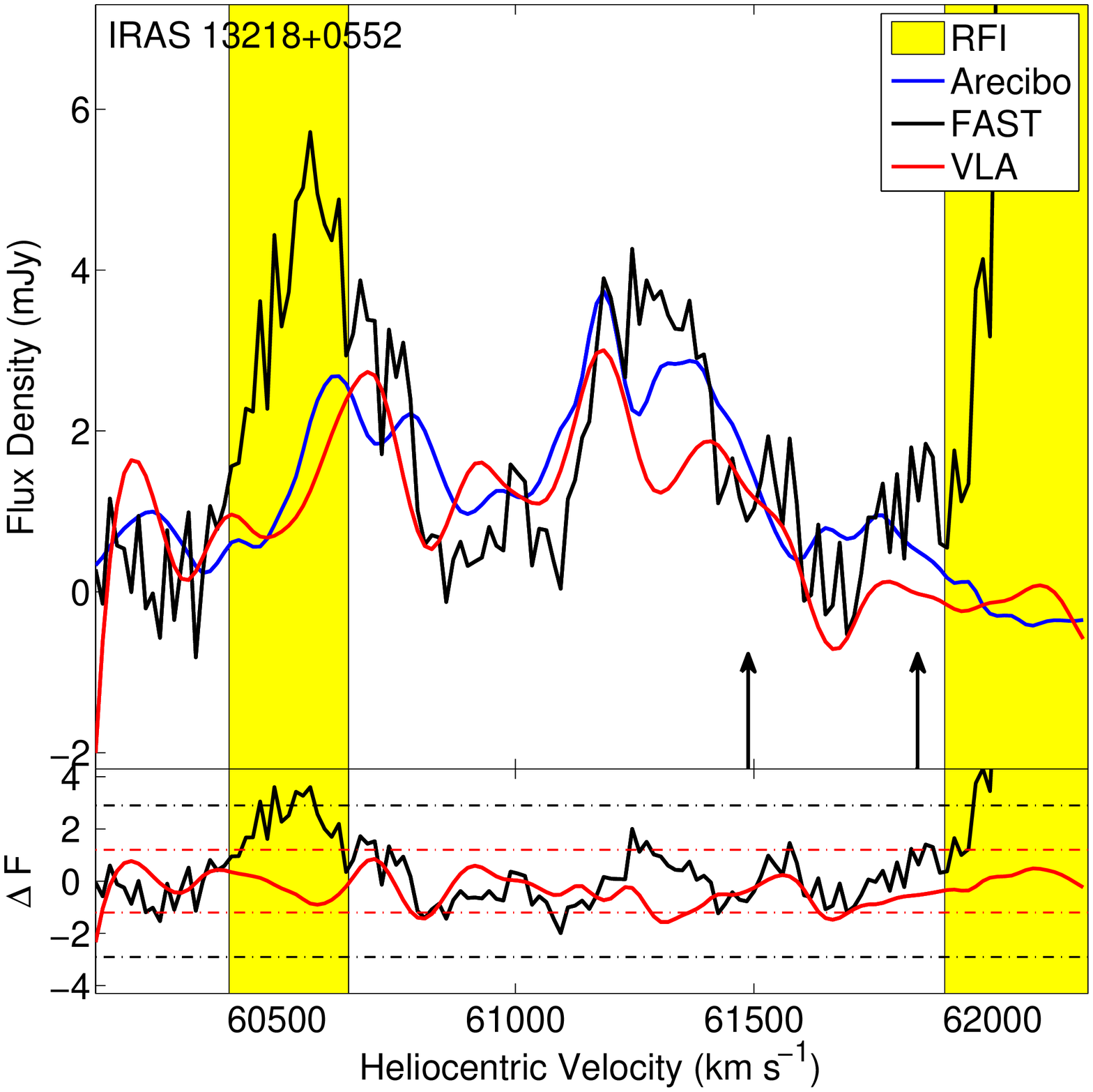}}
   \hfill
  \subfigure{ \includegraphics[width=0.485\textwidth,height=6.2cm]{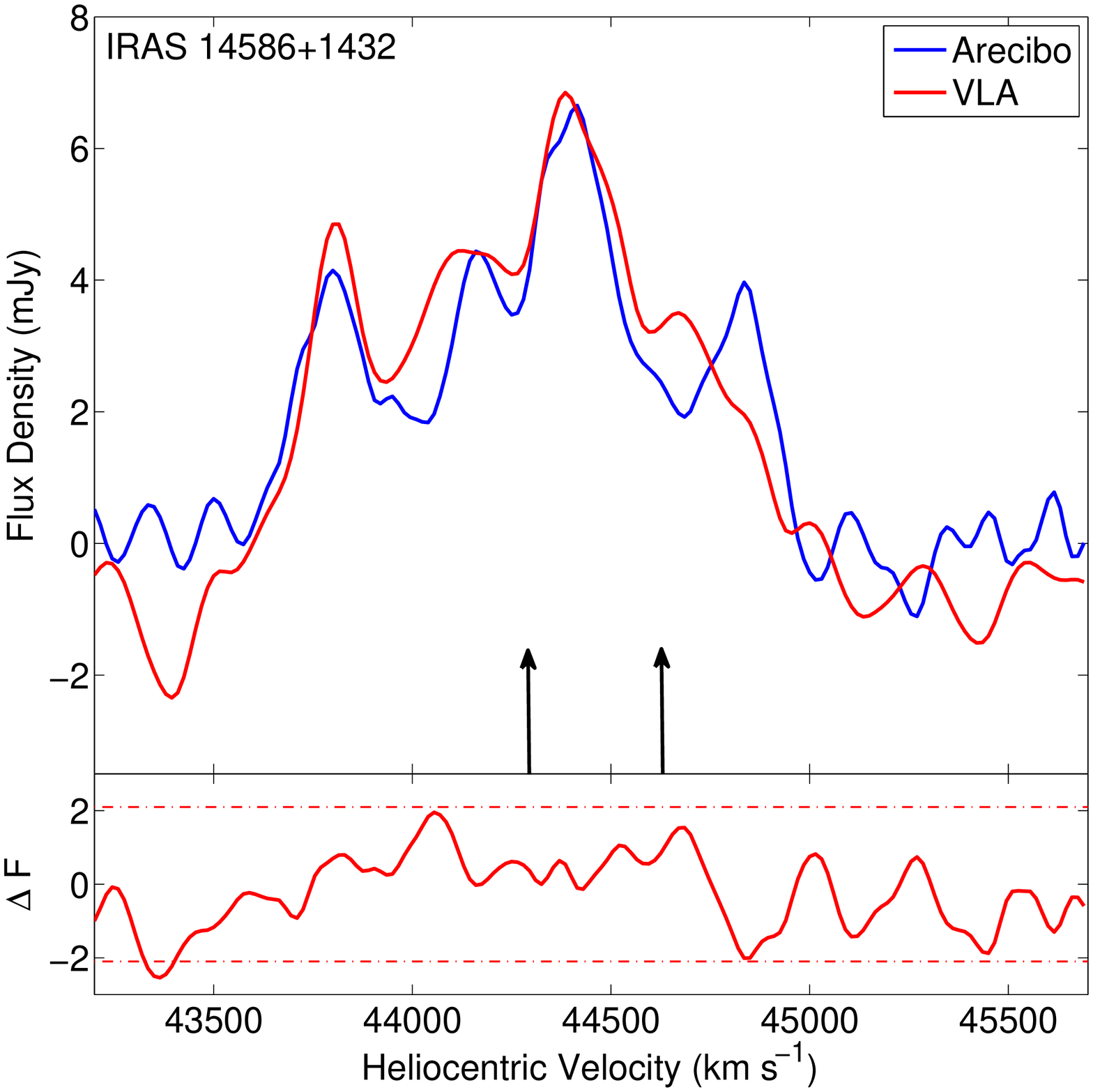}}
  \hfill
    \subfigure{ \includegraphics[width=0.485\textwidth,height=6.2cm]{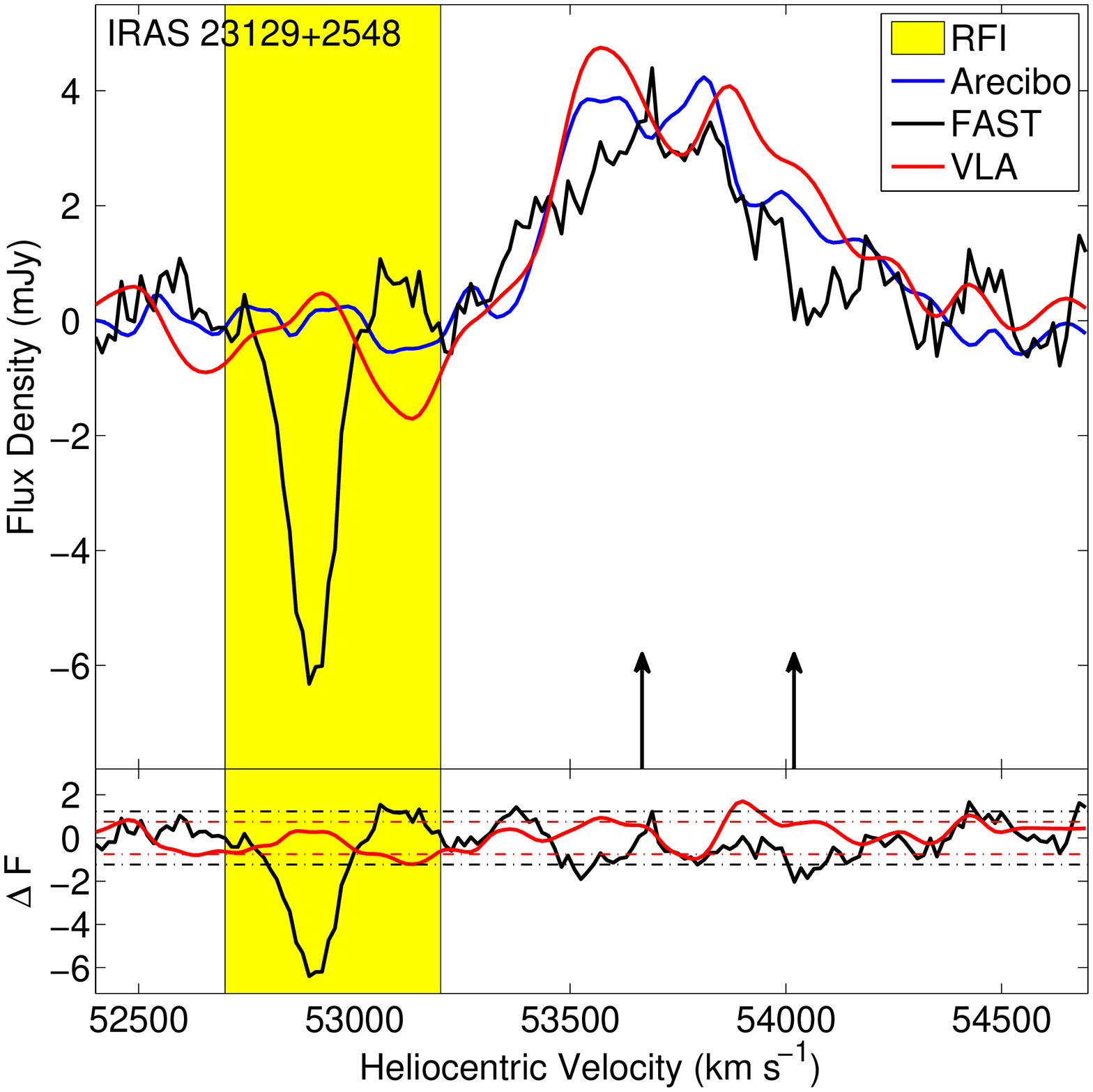}}
    \hfill
\subfigure{ \includegraphics[width=0.485\textwidth,height=6.2cm]{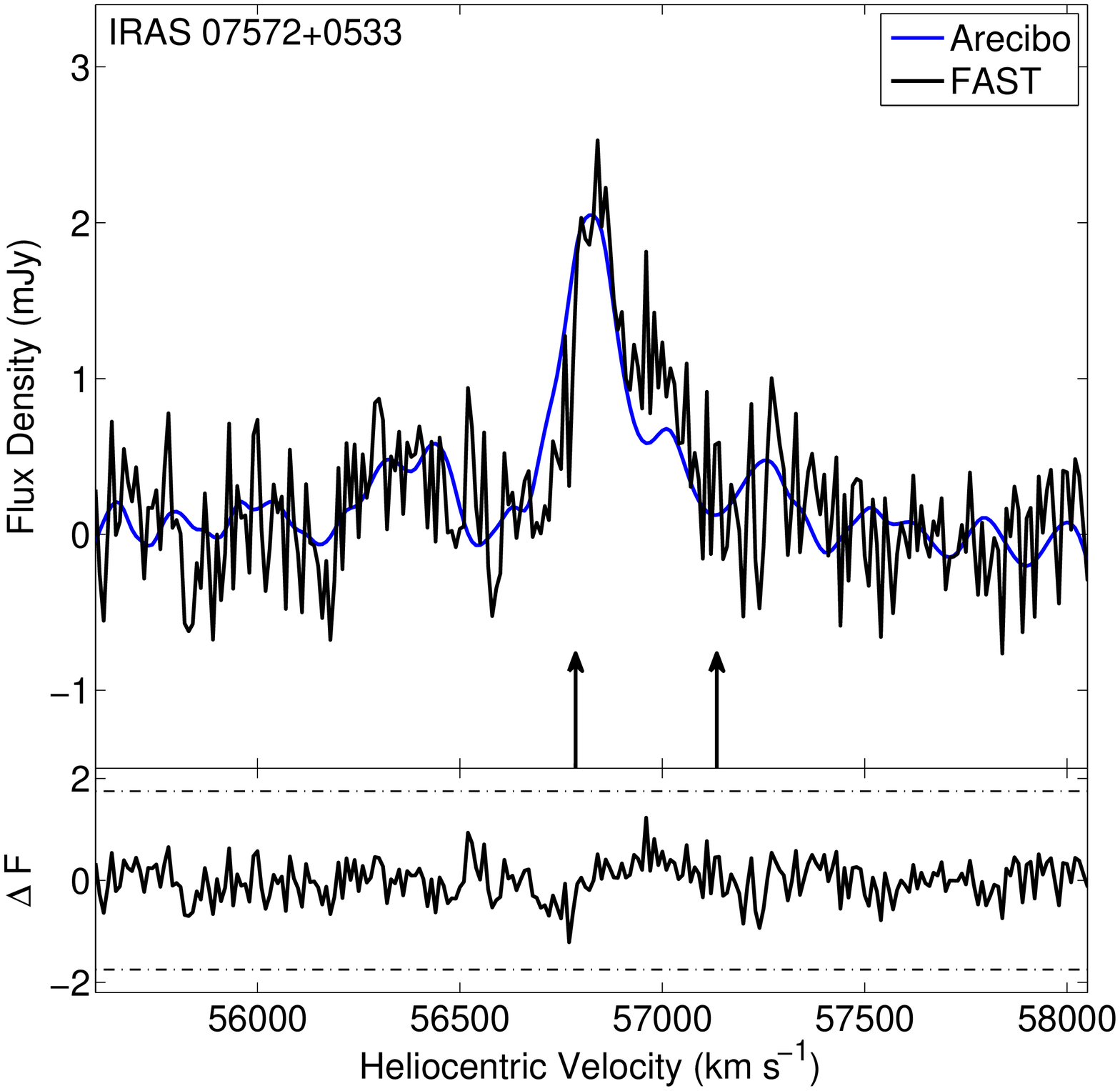}}
  
      \caption{The OH spectra use the 1667.359
MHz line as the rest frequency for the velocity scale. The lines and labels are same as showed in Fig. \ref{ohline1}
      }
         \label{ohline2}
   \end{figure*} 
\addtocounter{figure}{-1}
\begin{figure*}
   \centering

    \subfigure{ \includegraphics[width=0.485\textwidth,height=6.2cm]{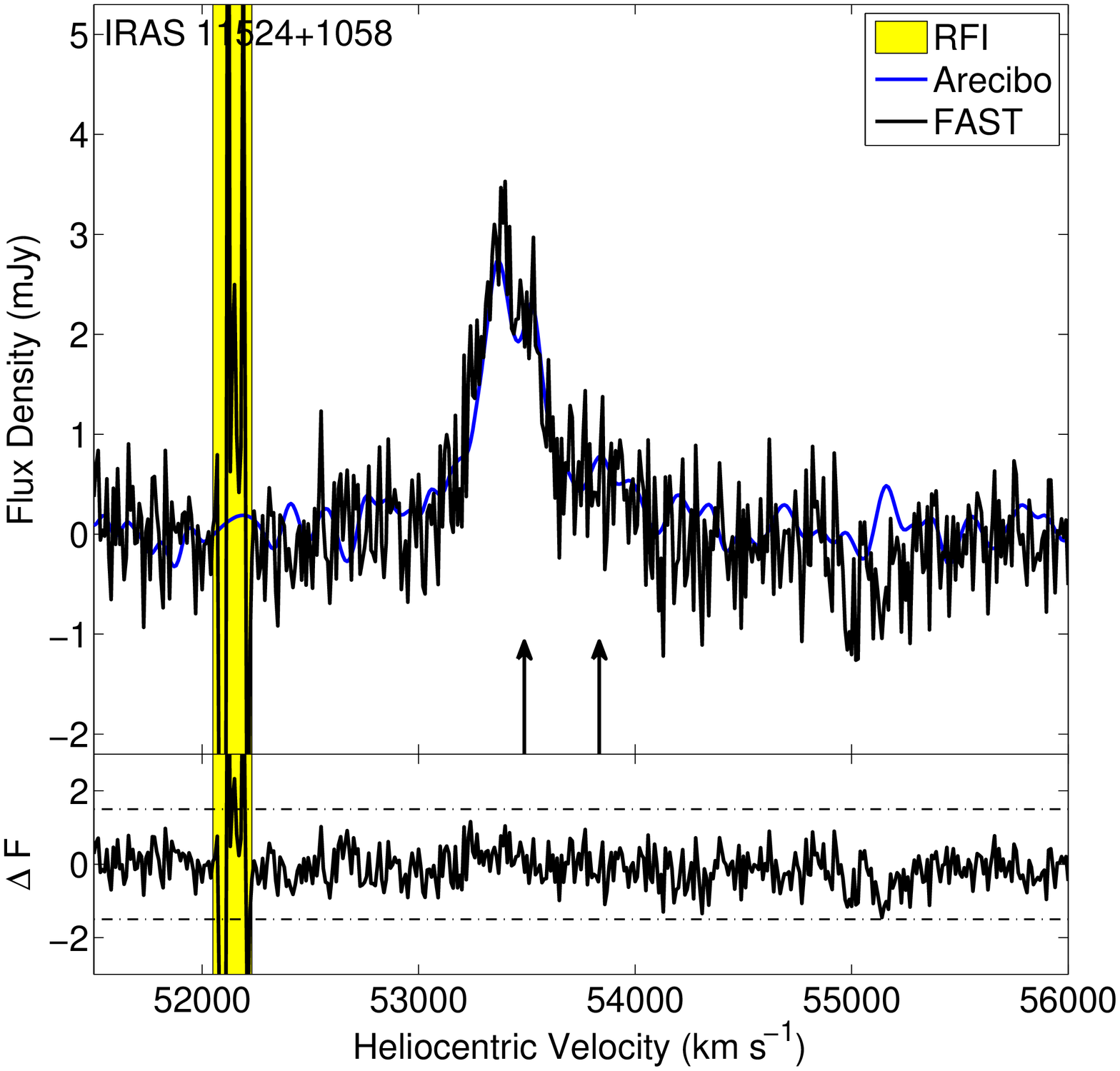}}
     \hfill
    \subfigure{  \includegraphics[width=0.485\textwidth,height=6.2cm]{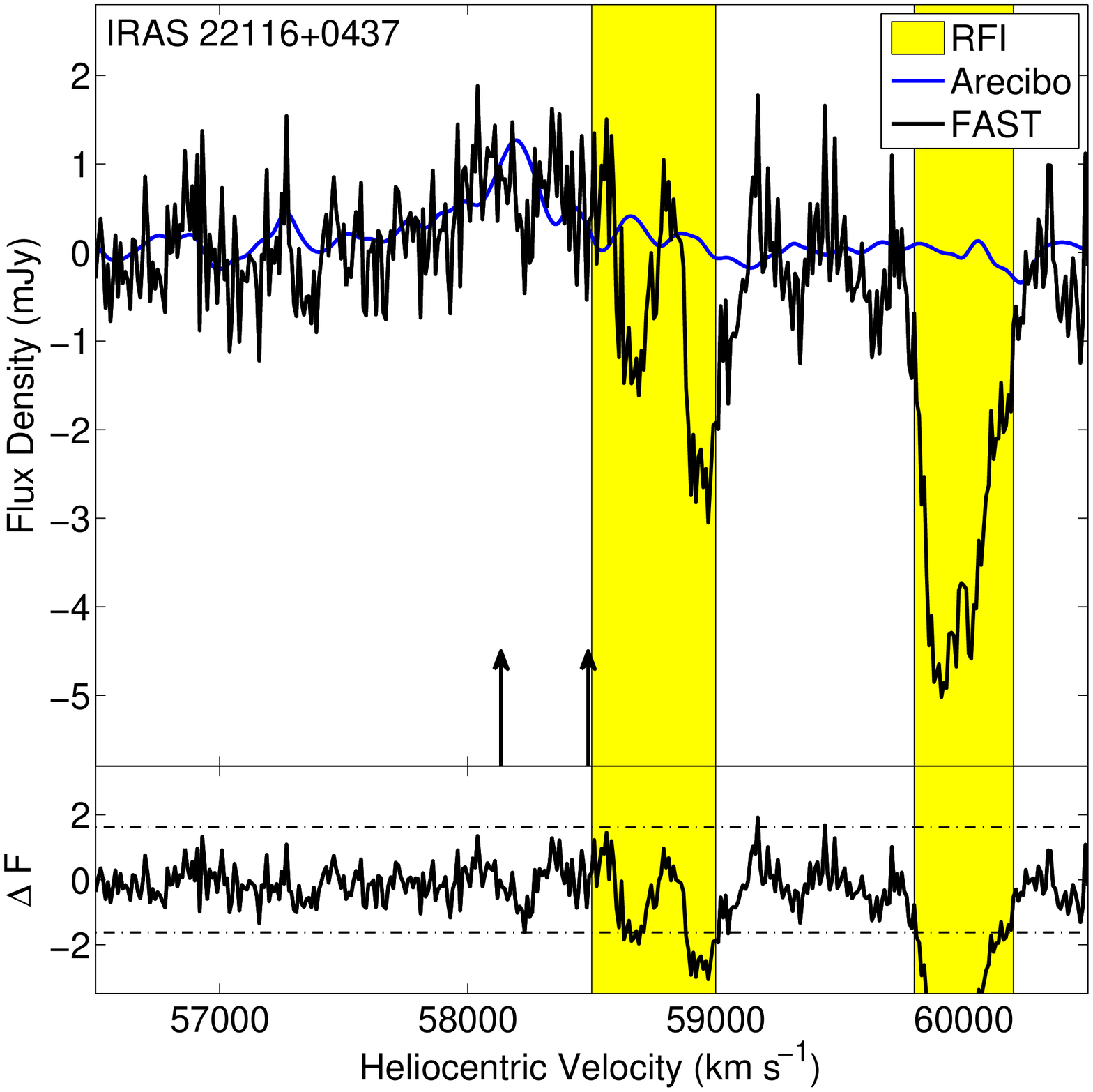}}
      \hfill
     \subfigure{ \includegraphics[width=0.485\textwidth,height=6.2cm]{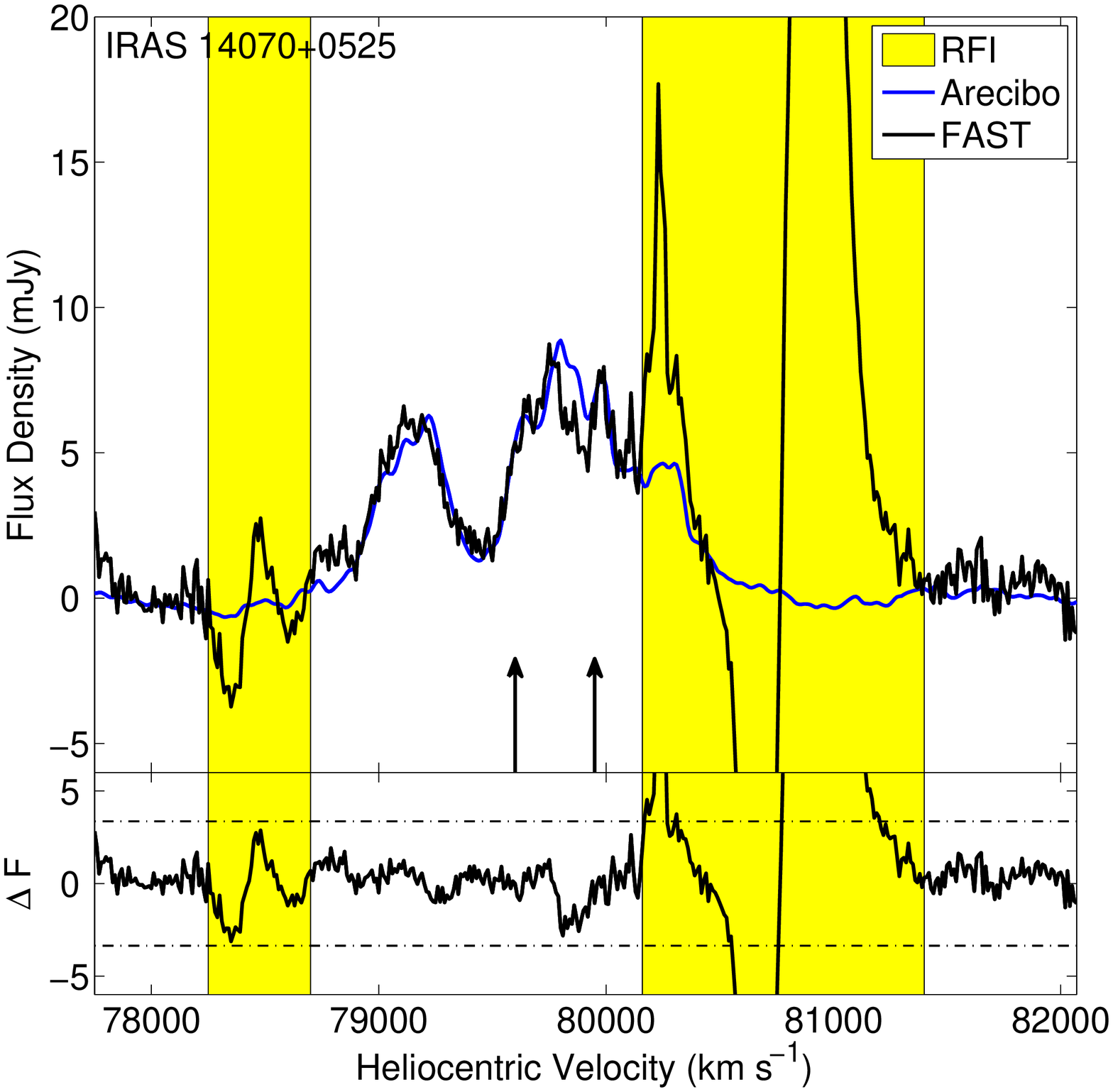}}
      \hfill
      \caption{The OH Spectra use the 1667.359
MHz line as the rest frequency for the velocity scale. The lines and labels are the same as shown in Fig. \ref{ohline1}. 
      }
         \label{ohline2}
   \end{figure*}

\begin{figure*}
   \centering
  
   \includegraphics[width=14.6cm,height=23.0cm]{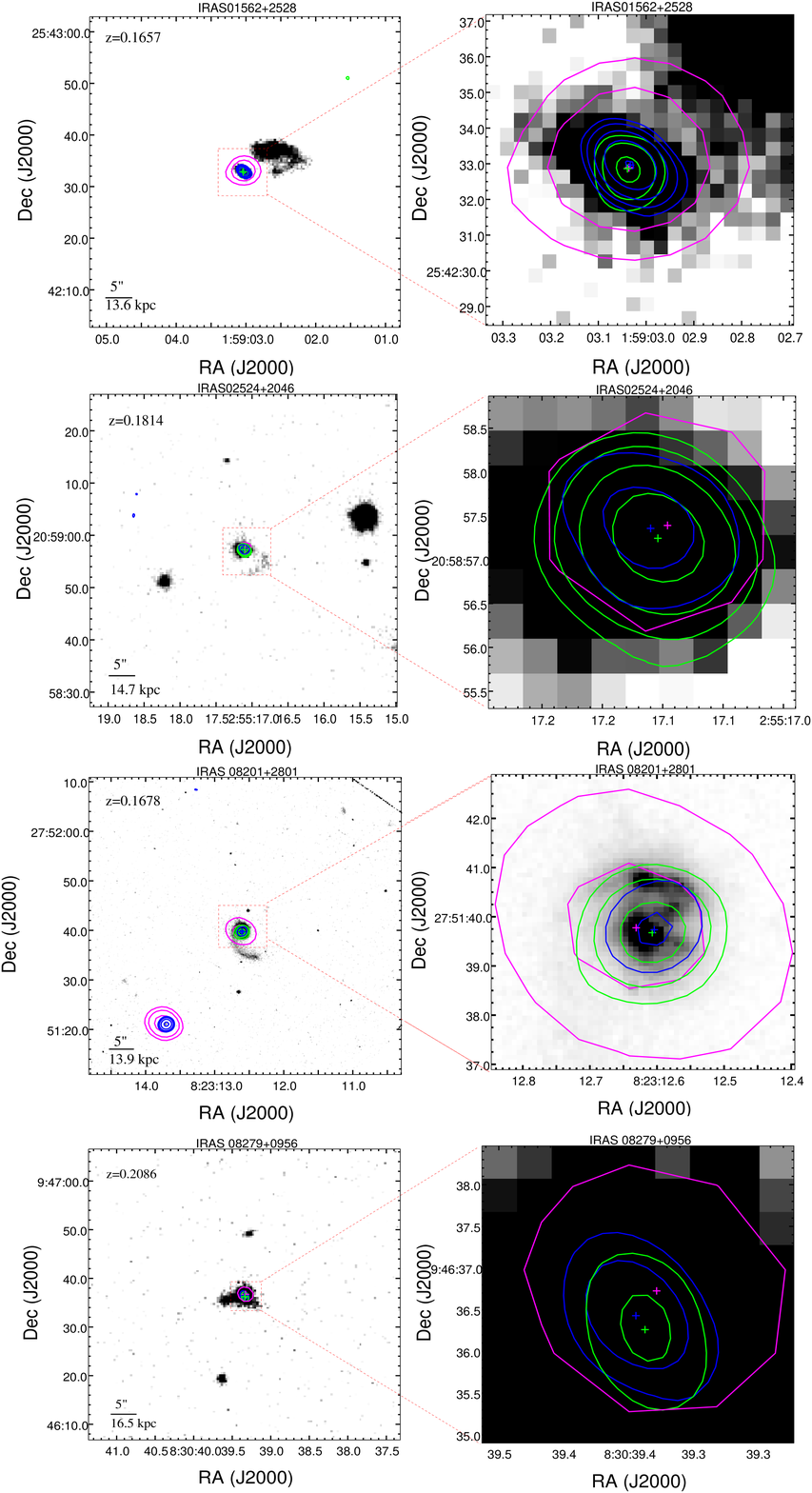}

      \caption{VLA contours overlaid on SDSS or HST R-band gray image of OHMs. IRAS 08201: HST image with F814W filter at 790.48 nm. IRAS 13218+0552: HST image from project 11604 with $\rm wfc\_total$ filter at 550 and 790.48 nm. Magenta: The contour map of the radio continuum emission from the VLASS survey. Blue: VLA-A L-band continuum emission. Green: The OH line emission with velocities ranges: IRAS 01562: 49474.6-50047.6 \kms  IRAS 02524: 54107.4-54352.1 \kms  IRAS 08201: 49802.6-50712.8 \kms  IRAS 08279: 62677.3-62164.7 \kms. The redshift and physical scale searched from NASA Extragalactic Database (NED\textsuperscript{2}) for each source are present in the left panels.
      }
         \label{lowfreq}
          \textsuperscript{2 http://ned.ipac.caltech.edu/forms/byname.html }
   \end{figure*}

\addtocounter{figure}{-1}
\begin{figure*}
   \centering
 
   \includegraphics[width=14.6cm,height=23.0cm]{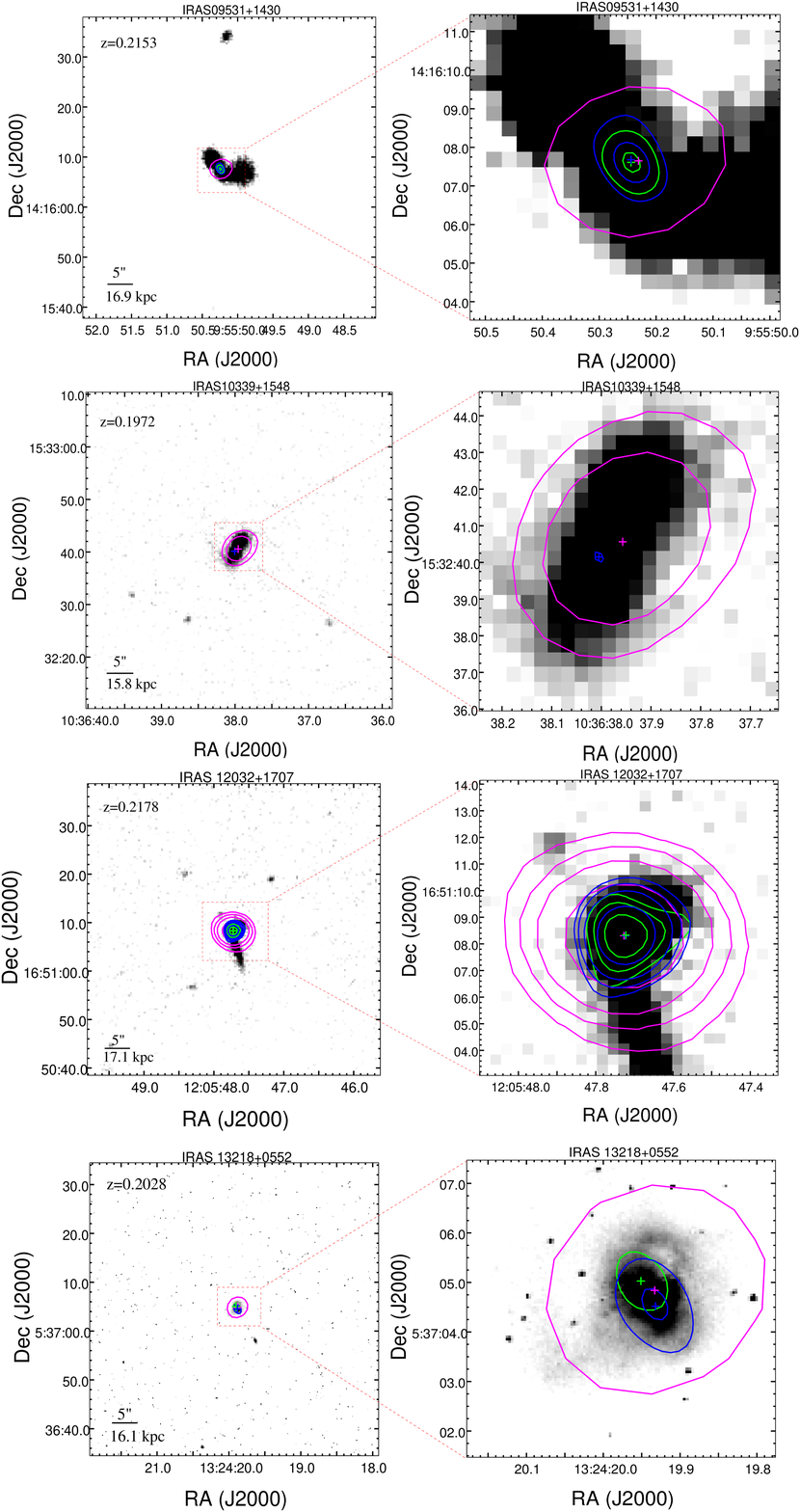}
      \caption{Continued. The OH line emission with velocities ranges: IRAS 09531: 63709.4-65057.0 \kms   IRAS 12032: 64351.9-65599.4 \kms   IRAS 13218: 60509.1-61477.0 \kms . 
      }
         \label{lowfreq}
   \end{figure*}
   
 \addtocounter{figure}{-1}
\begin{figure*}
   \centering  

   \includegraphics[width=14.6cm,height=23.0cm]{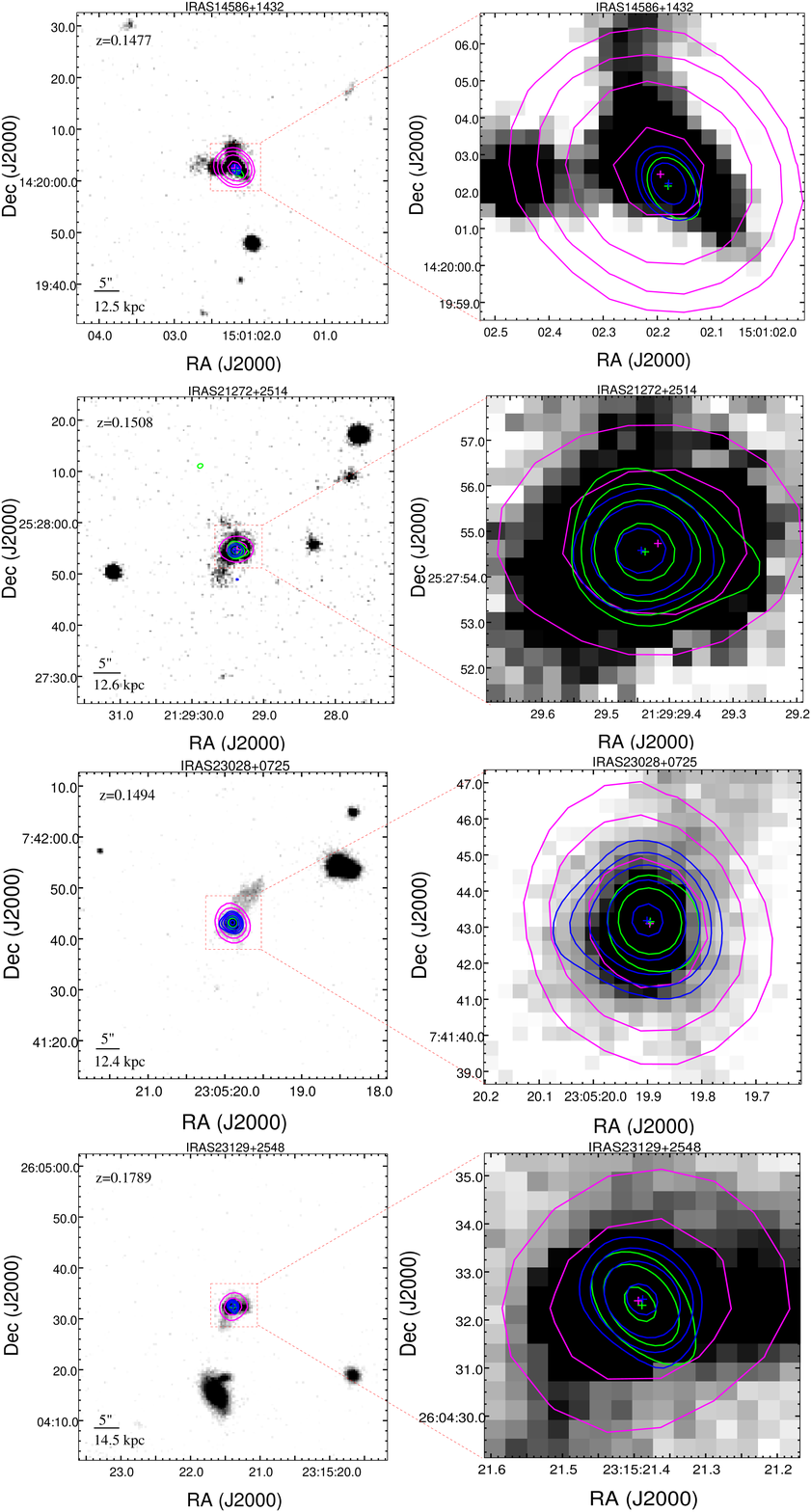} 
       \caption{Continued. The OH line emission with velocities ranges: IRAS 14586: 44892.7-43735.5 \kms  IRAS 21272: 44471.3-45493.4 \kms  IRAS 23028: 44386.7-45081.6 \kms. IRAS 23129: 53390.3-54023.9 \kms. 
      }
         \label{lowfreq}
   \end{figure*}

\begin{figure*}
   \centering   
   \subfigure[IRAS 02524+2046]{\includegraphics[width=0.485\textwidth,height=5.2cm]{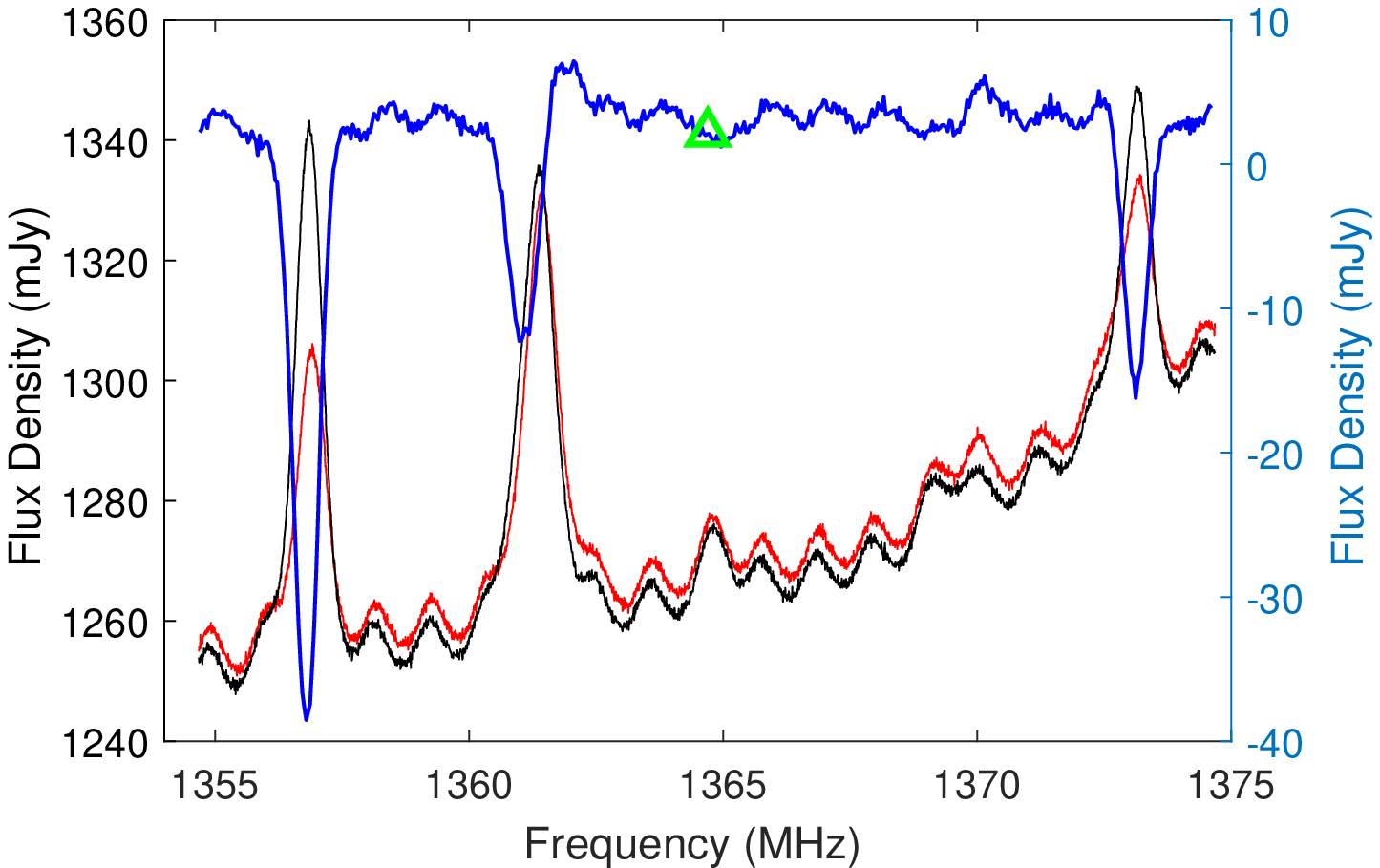}}
 \subfigure[IRAS 02524+2046]{\includegraphics[width=0.485\textwidth,height=5.2cm]{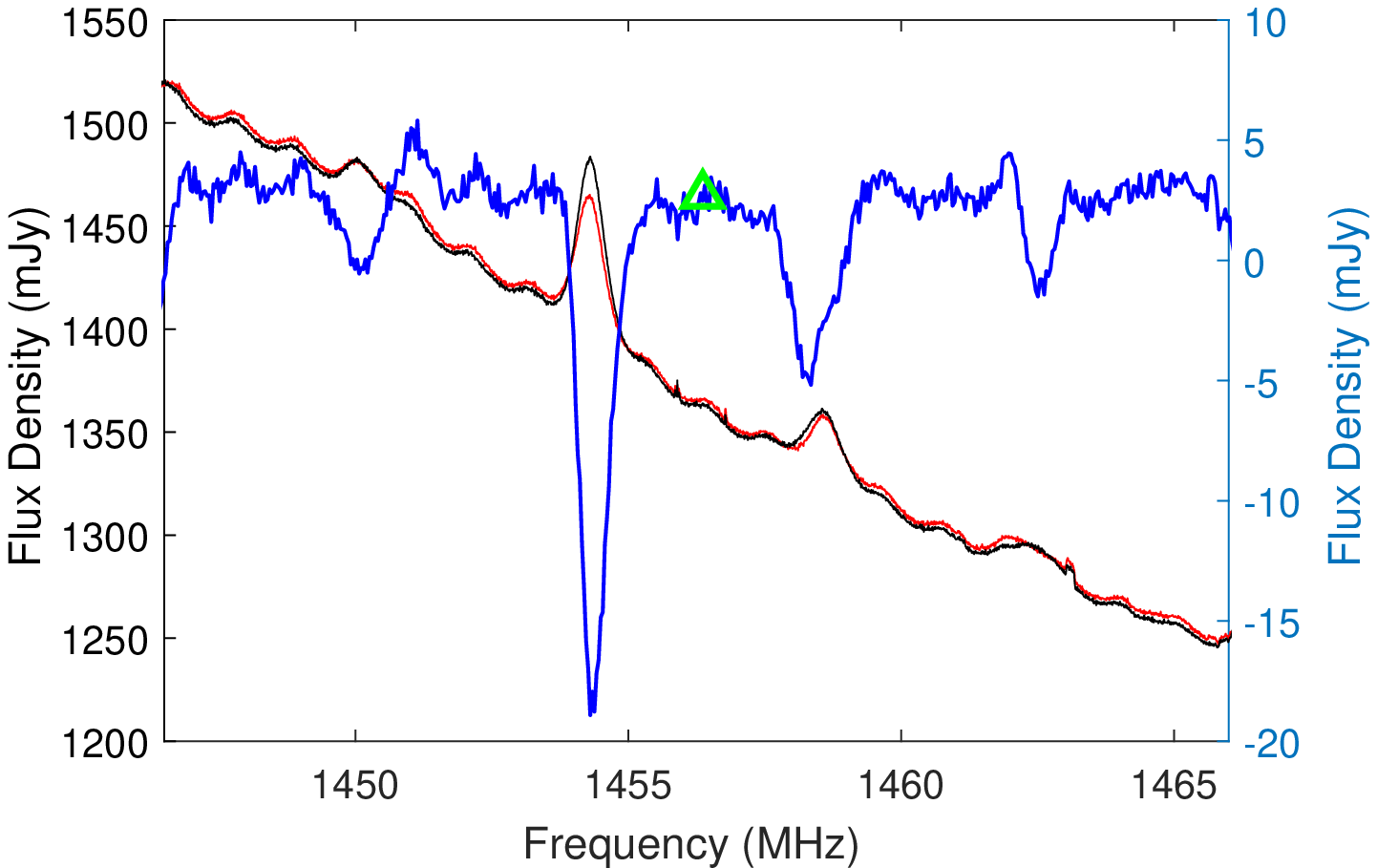}}
 \subfigure[IRAS 07572+0533]{\includegraphics[width=0.485\textwidth,height=5.2cm]{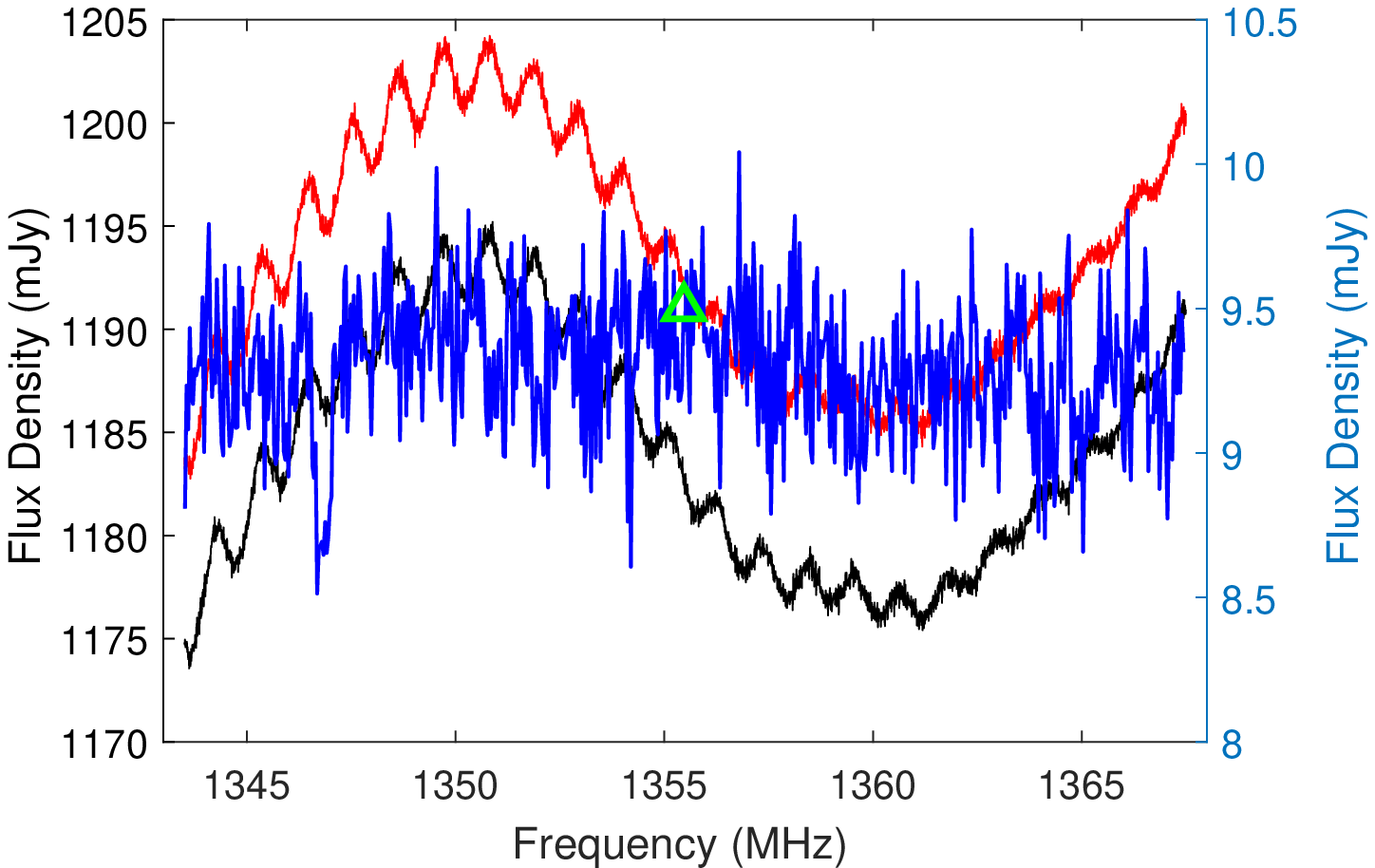}}
 \subfigure[IRAS 07572+0533]{\includegraphics[width=0.485\textwidth,height=5.2cm]{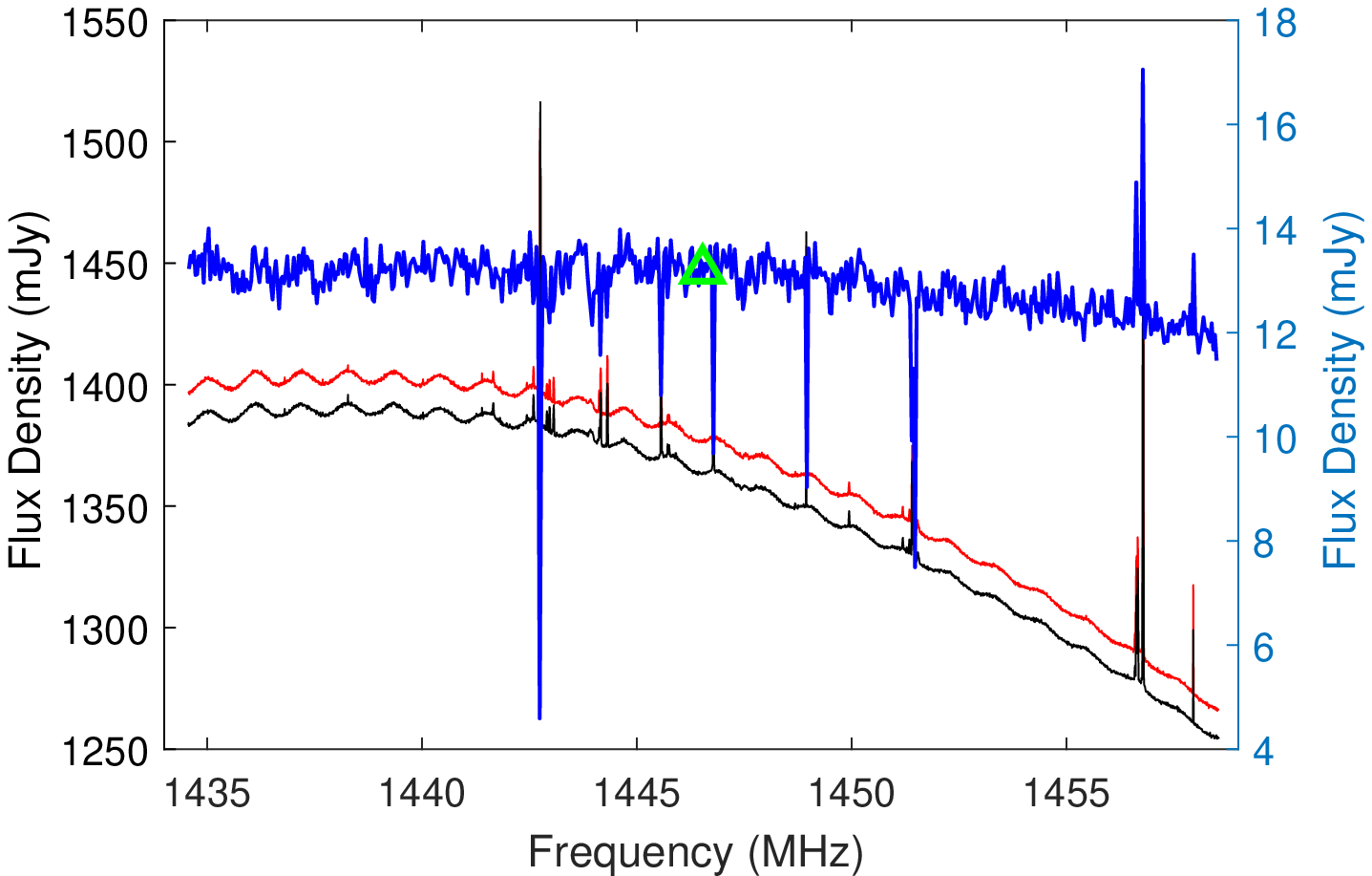}}
  \subfigure[IRAS 08279+0956]{\includegraphics[width=0.485\textwidth,height=5.2cm]{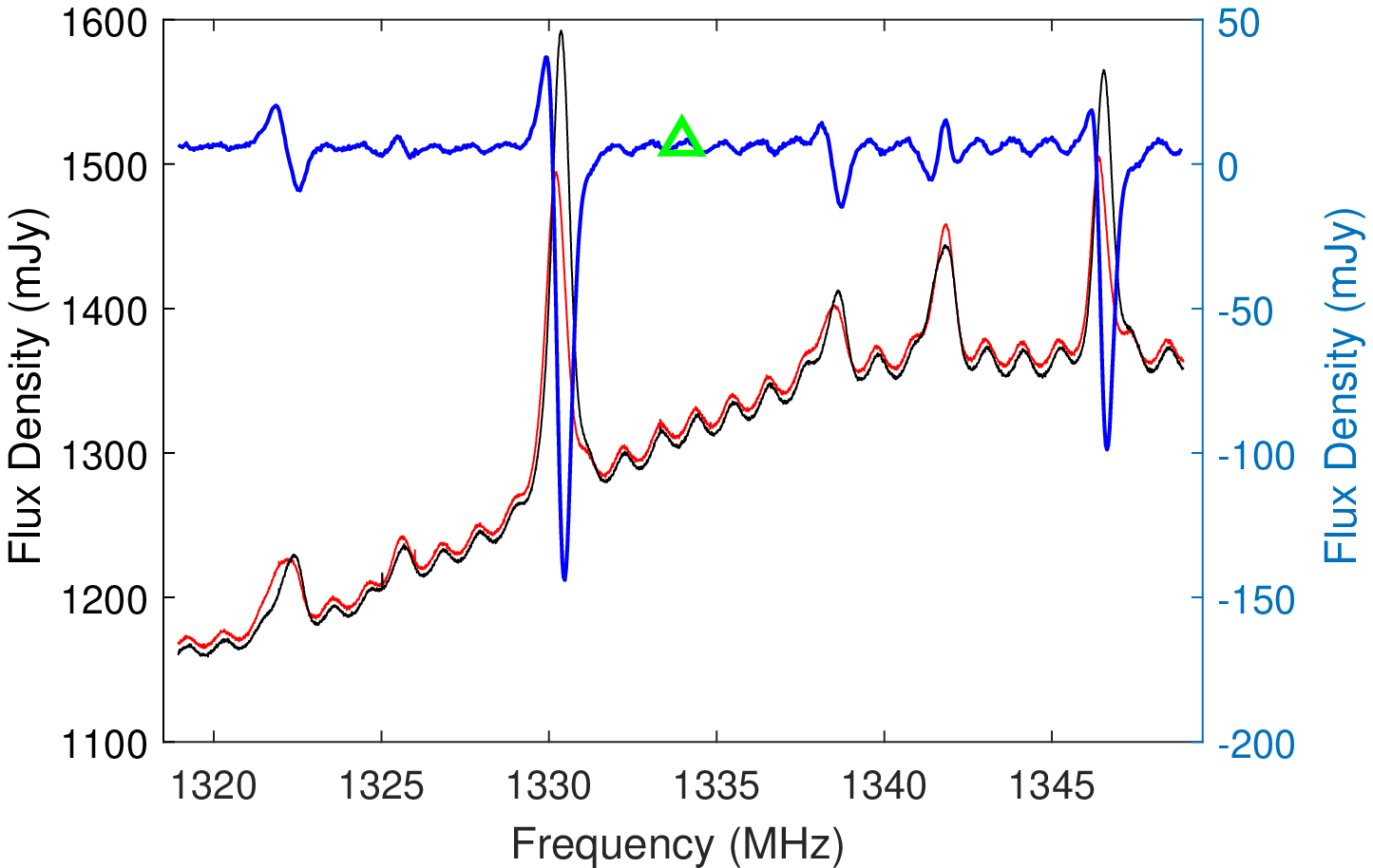}}
 \subfigure[IRAS 08279+0956]{\includegraphics[width=0.485\textwidth,height=5.2cm]{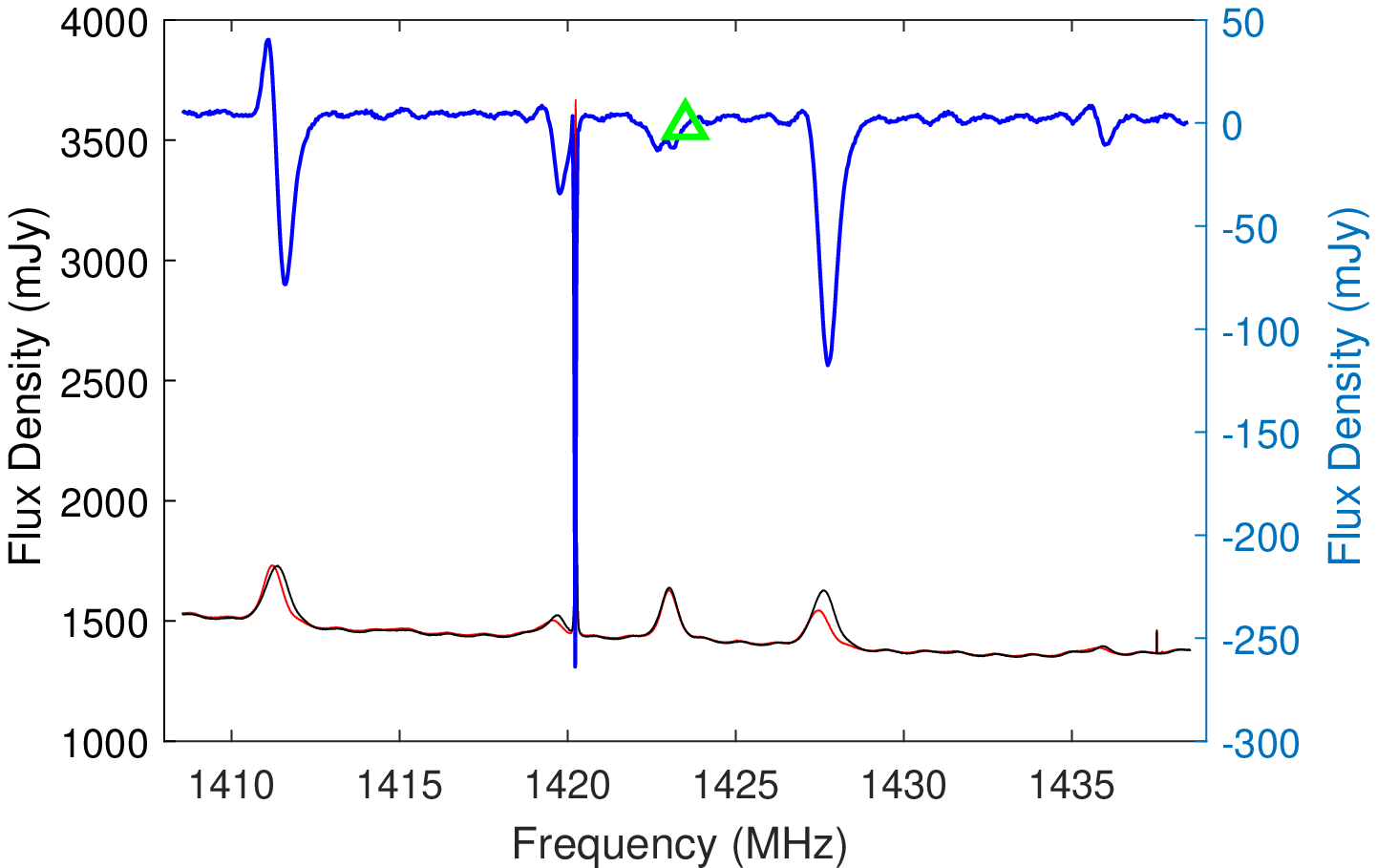}}
   \subfigure[IRAS 09531+1430]{\includegraphics[width=0.485\textwidth,height=5.2cm]{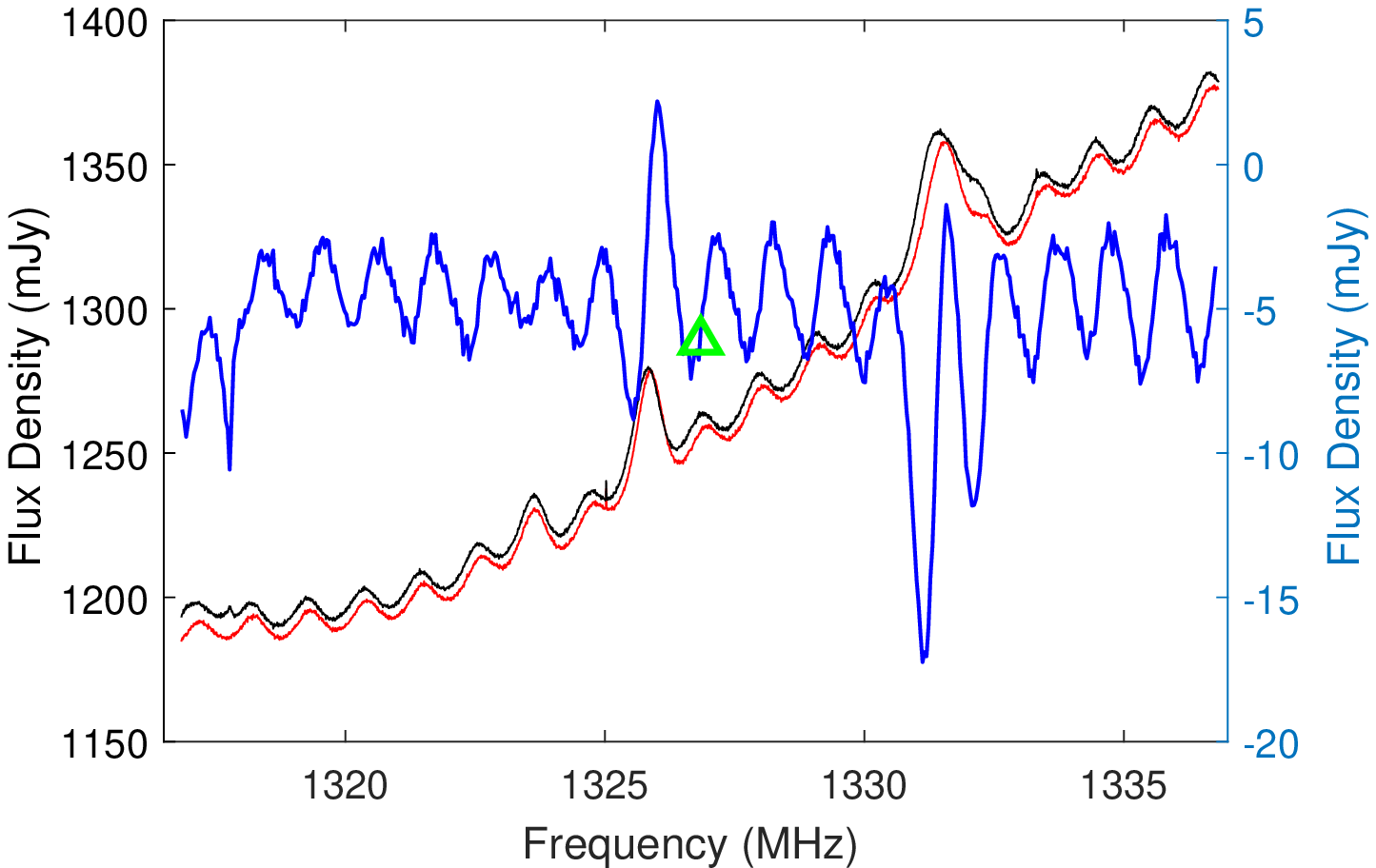}}
 \subfigure[IRAS 09531+1430]{\includegraphics[width=0.485\textwidth,height=5.2cm]{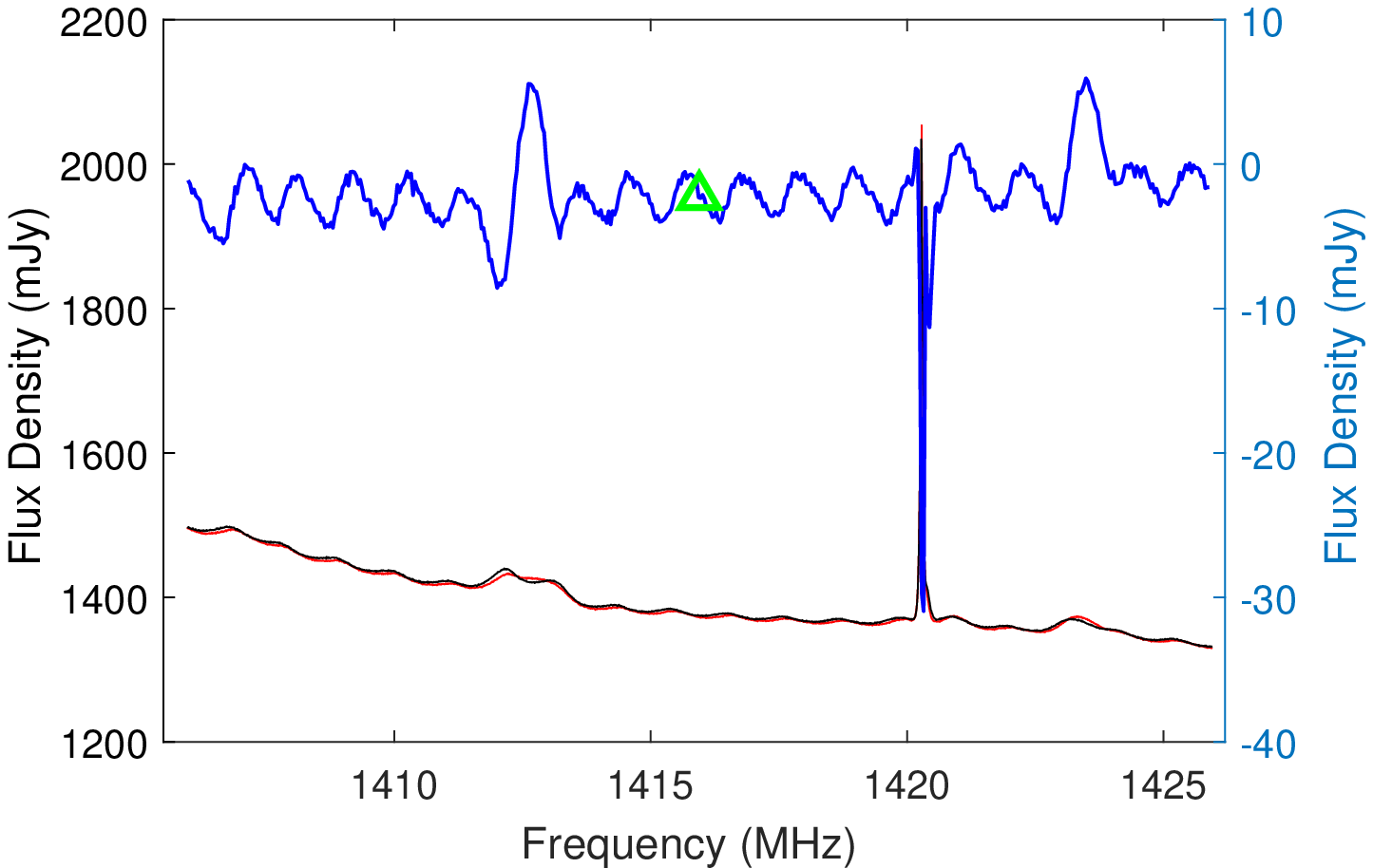}}
 \vspace{0.2in}
       \caption{The detection of OH satellite lines at 1612 MHz (Left) and 1720 MHz (Right) respectively. The line profiles of each panel are centered at the expected frequency (green triangle) based on the optical redshift. The red and black line profiles (axis left) are the ON and OFF spectra respectively. The profiles in blue color (axis right) are the binned ON-OFF spectra with velocity resolution about 10 \kms.}
         \label{OHsatellite}
  \end{figure*} 
\addtocounter{figure}{-1}  
\begin{figure*}
   \centering   
   \subfigure[IRAS 10339+1548]{\includegraphics[width=0.485\textwidth,height=5.3cm]{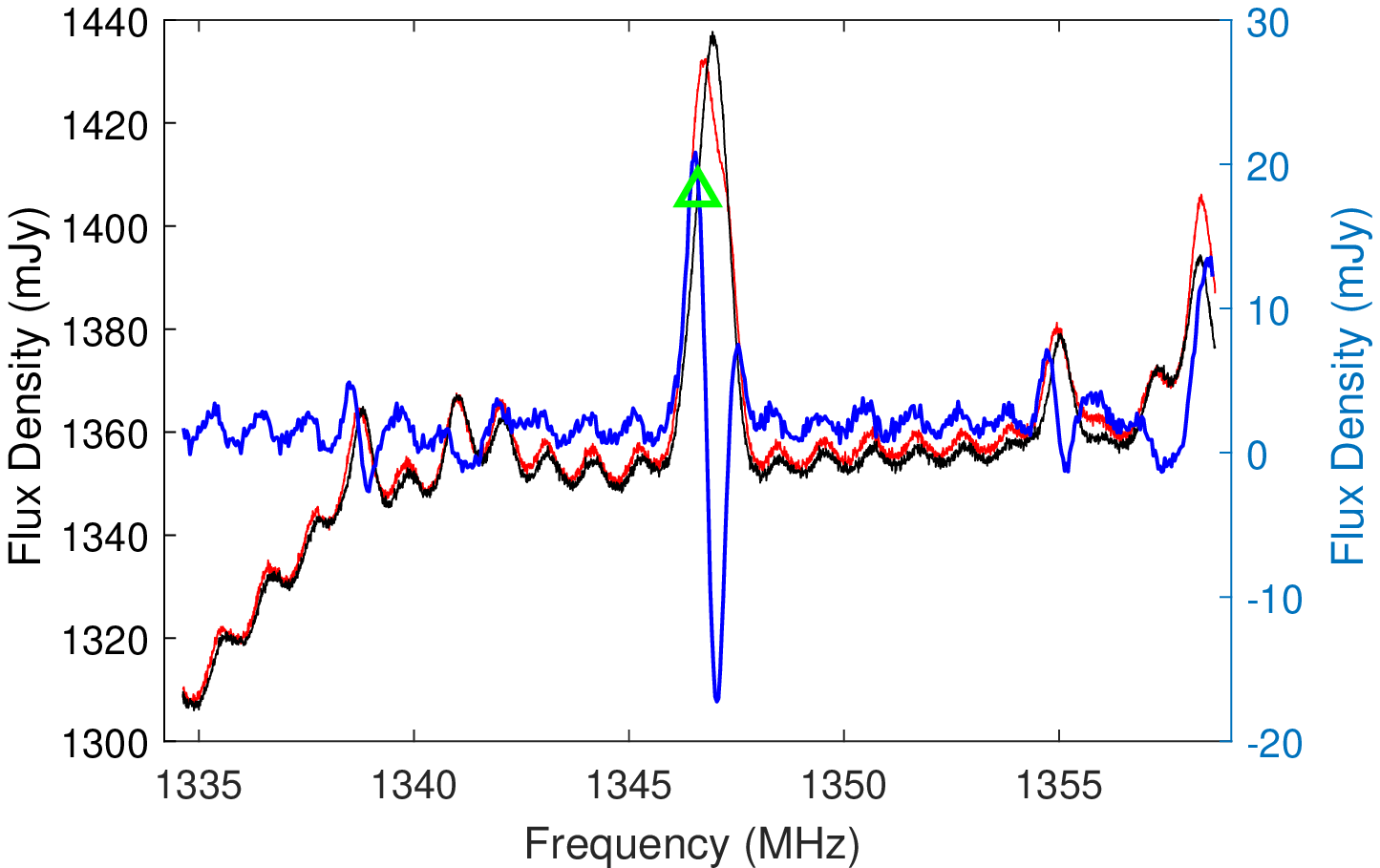}}
 \subfigure[IRAS 10339+1548]{\includegraphics[width=0.485\textwidth,height=5.3cm]{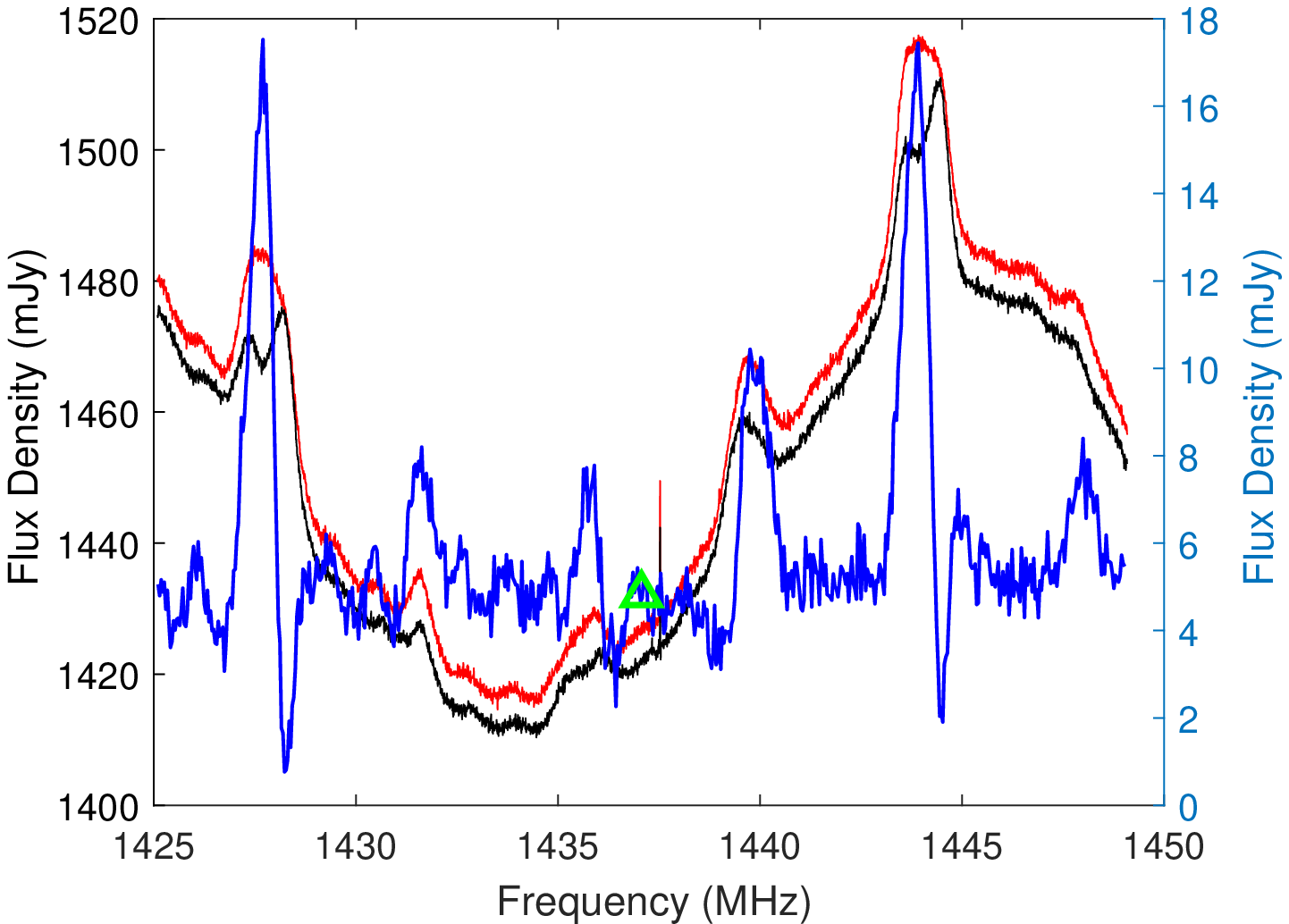}}
 \subfigure[IRAS 11028+3130]{\includegraphics[width=0.485\textwidth,height=5.3cm]{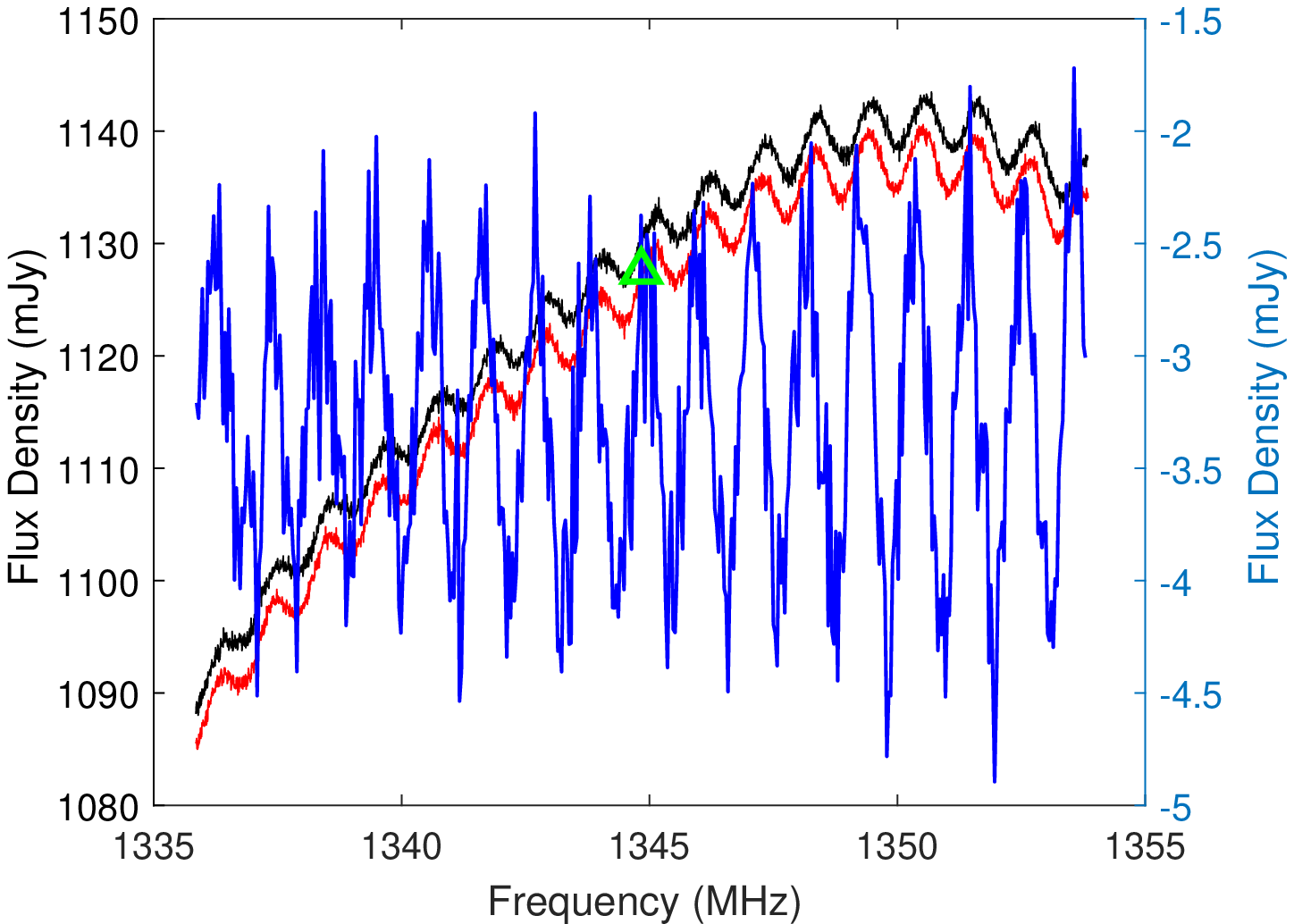}}
 \subfigure[IRAS 11028+3130]{\includegraphics[width=0.485\textwidth,height=5.3cm]{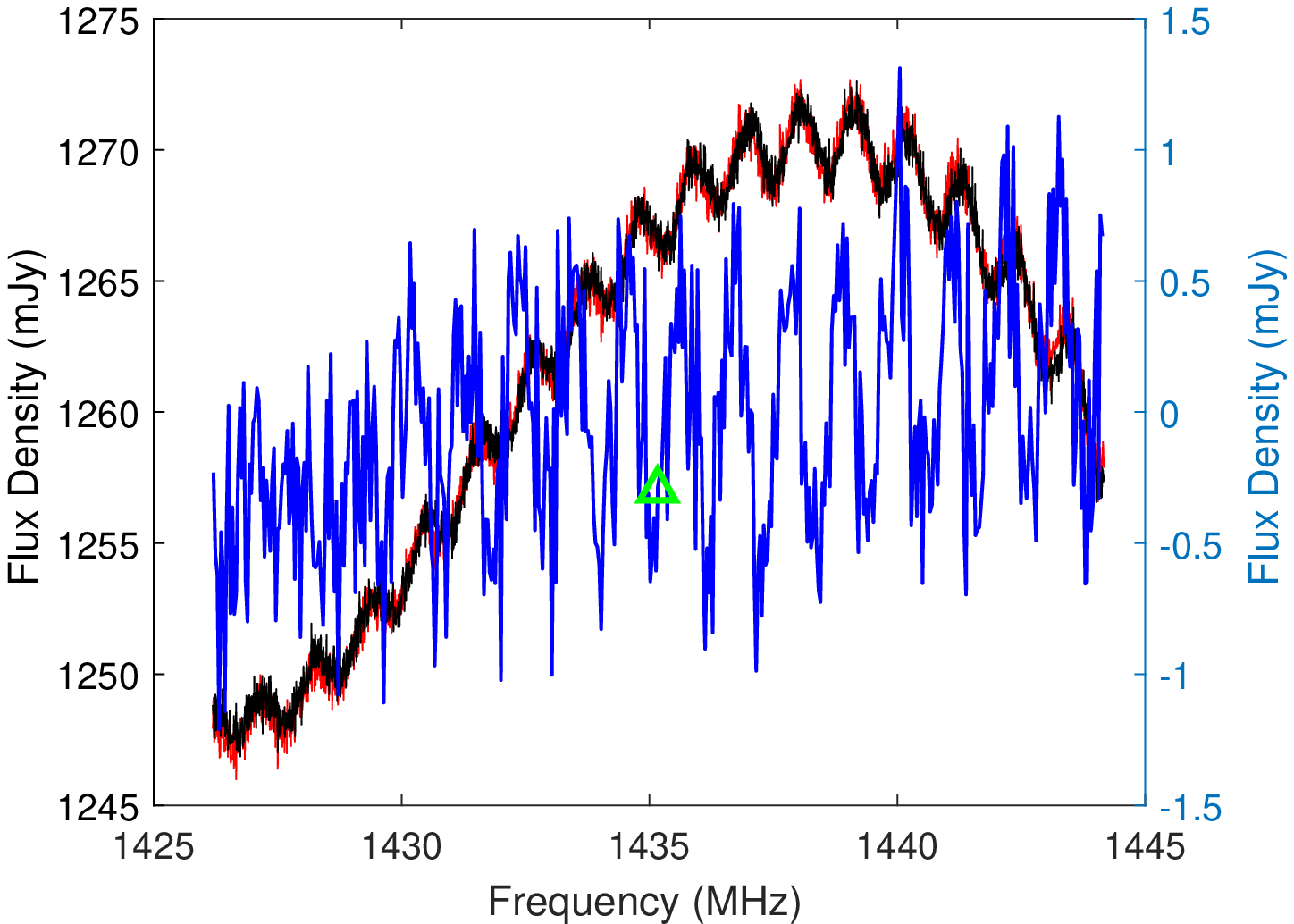}}
  \subfigure[IRAS 11524+1058]{\includegraphics[width=0.485\textwidth,height=5.3cm]{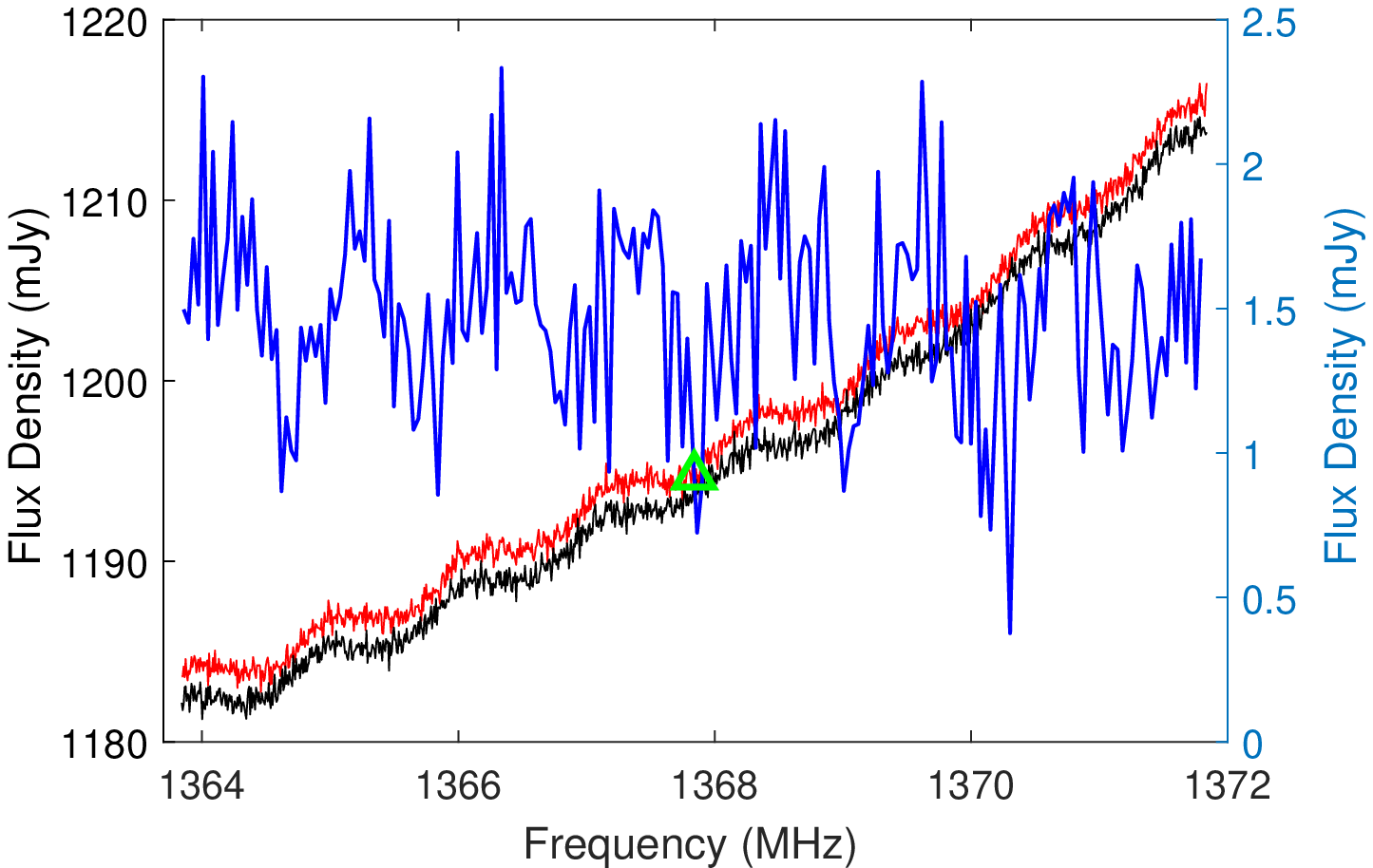}}
 \subfigure[IRAS 11524+1058]{\includegraphics[width=0.485\textwidth,height=5.3cm]{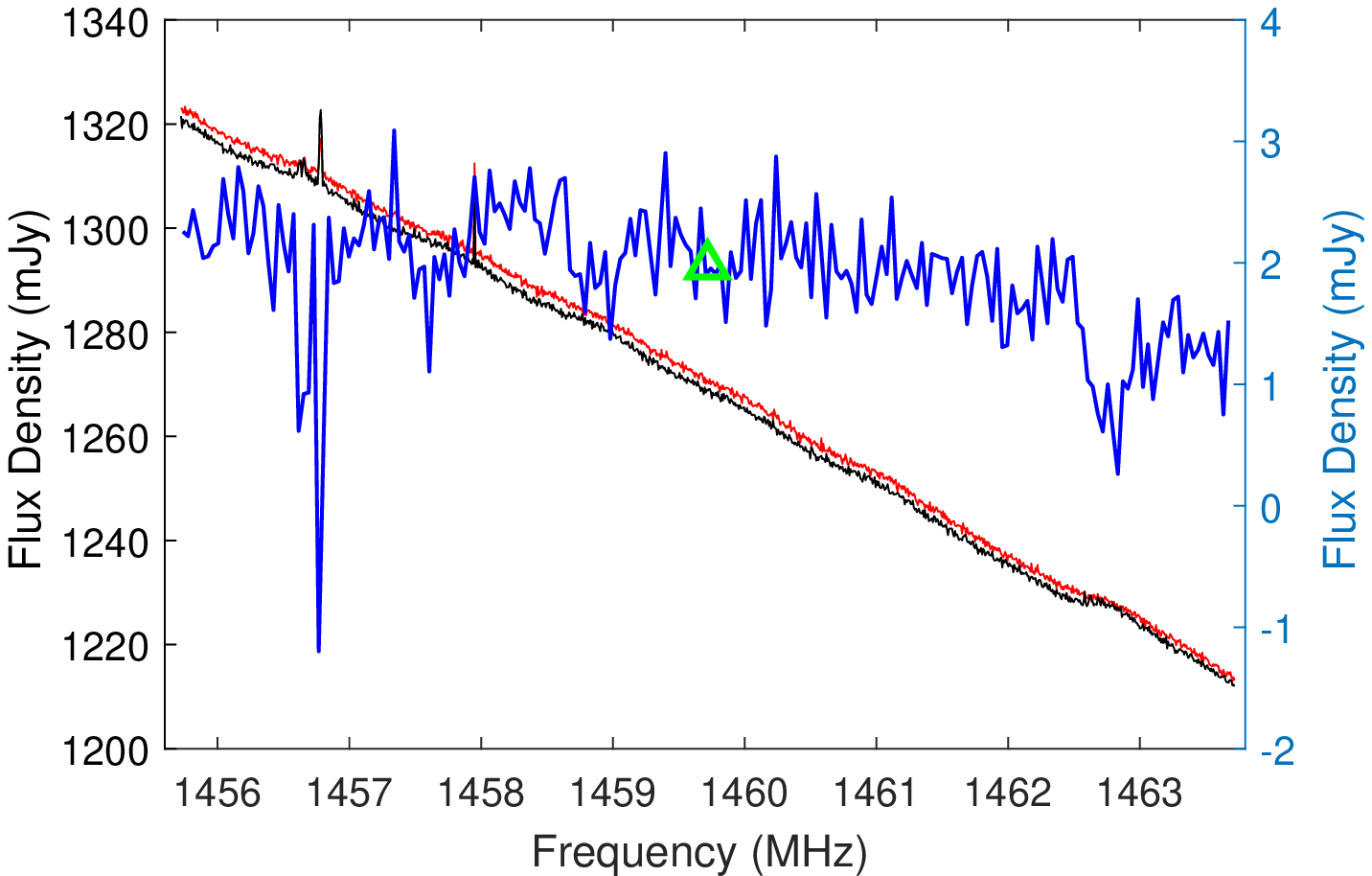}}
   \subfigure[IRAS 12032+1707]{\includegraphics[width=0.485\textwidth,height=5.3cm]{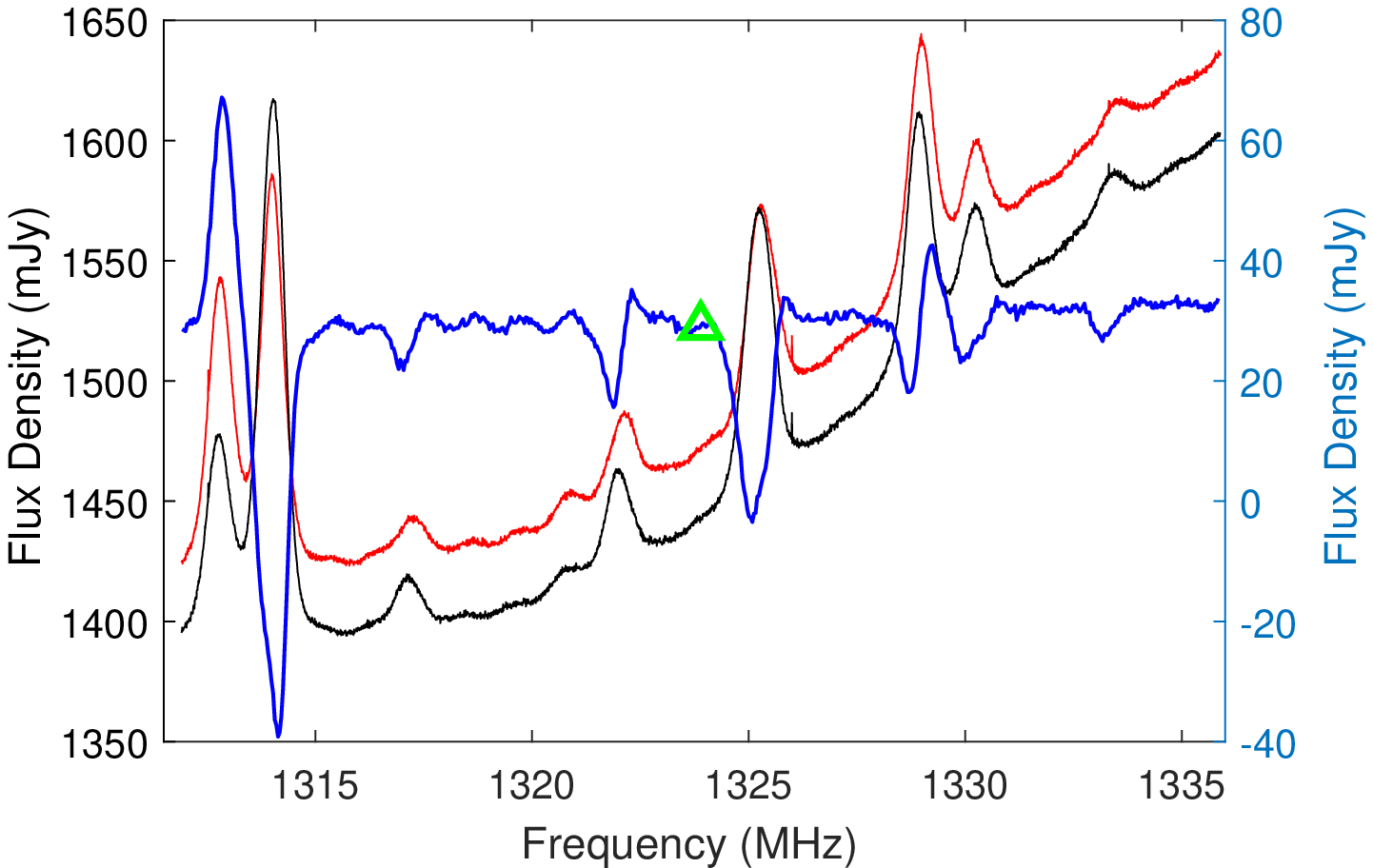}}
 \subfigure[IRAS 12032+1707]{\includegraphics[width=0.485\textwidth,height=5.3cm]{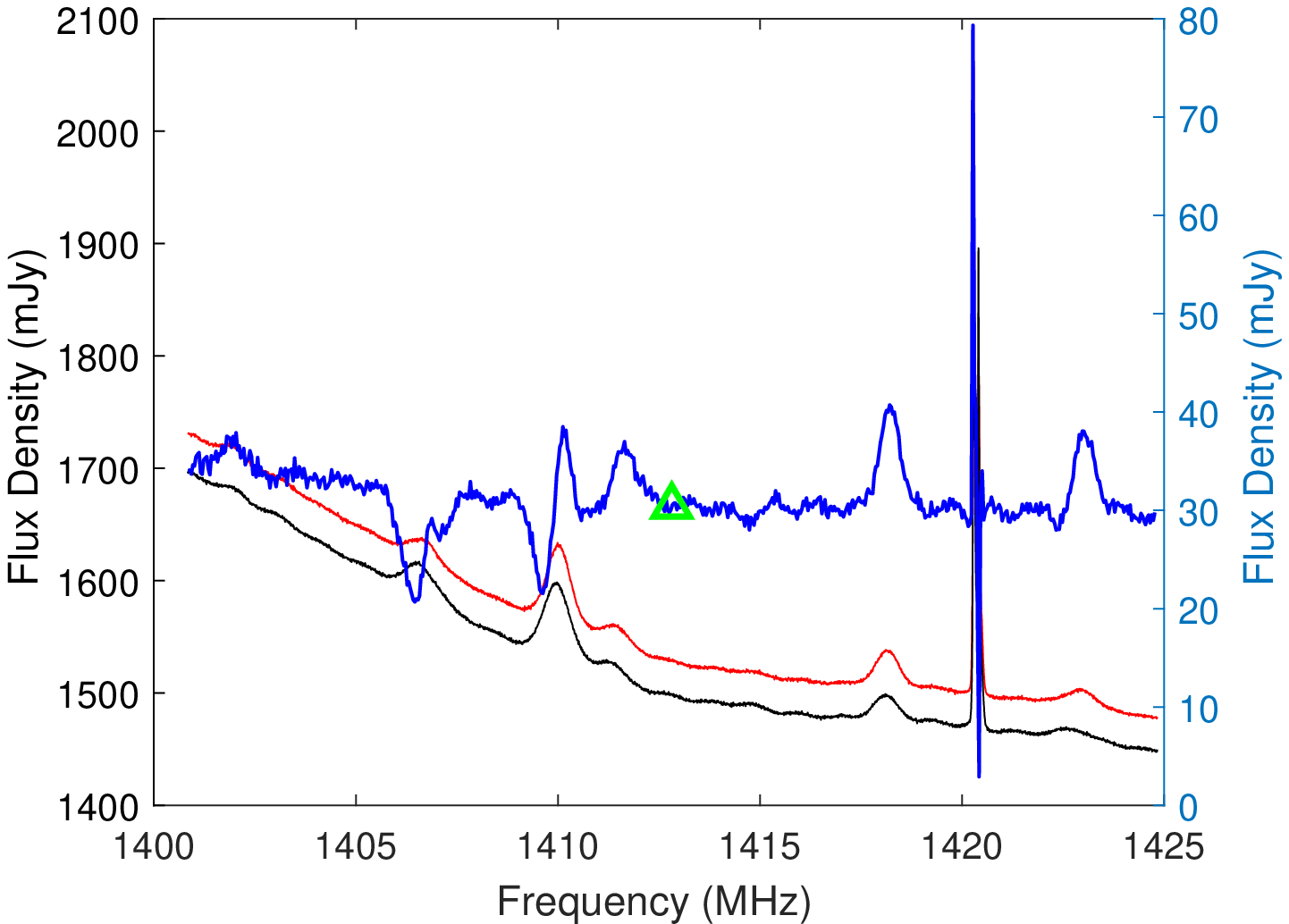}}
 \vspace{0.2in}
       \caption{Continued.
      }
         \label{OHsatellite}
  \end{figure*}
  \addtocounter{figure}{-1}
  \begin{figure*}
   \centering   
   \subfigure[IRAS 13218+0552]{\includegraphics[width=0.485\textwidth,height=5.0cm]{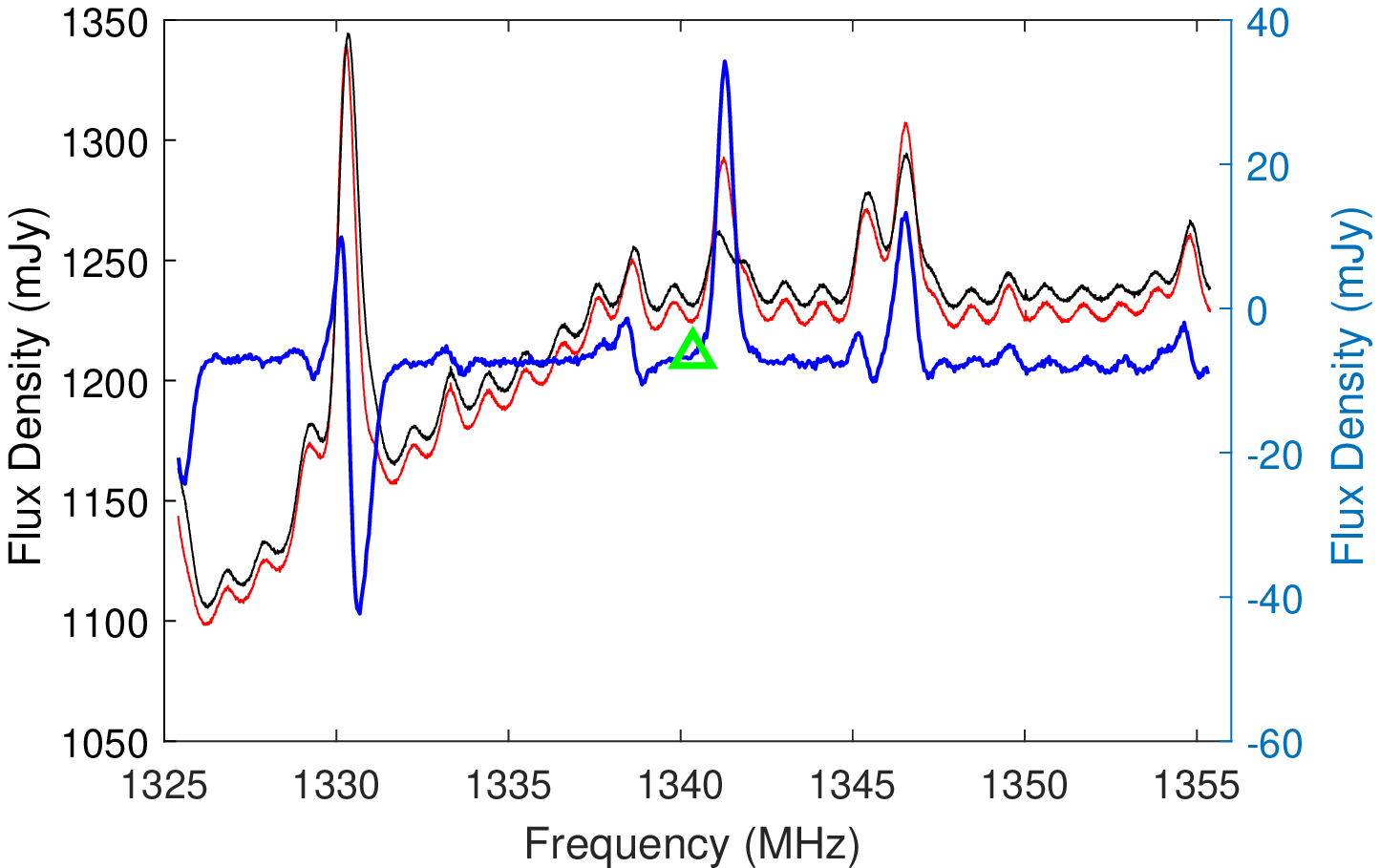}}
 \subfigure[IRAS 13218+0552]{\includegraphics[width=0.485\textwidth,height=5.0cm]{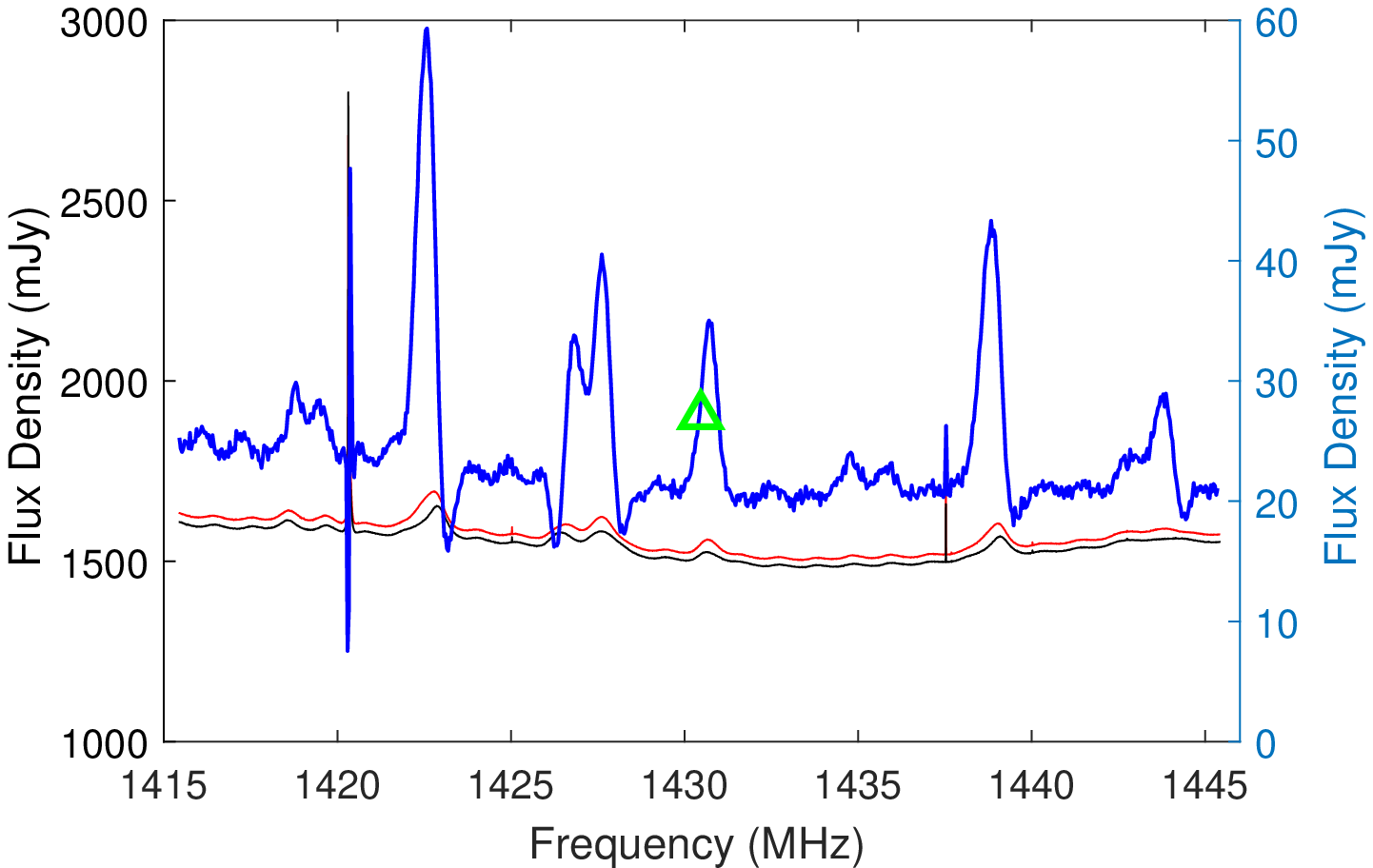}}
 \subfigure[IRAS 14070+0525]{\includegraphics[width=0.485\textwidth,height=5.0cm]{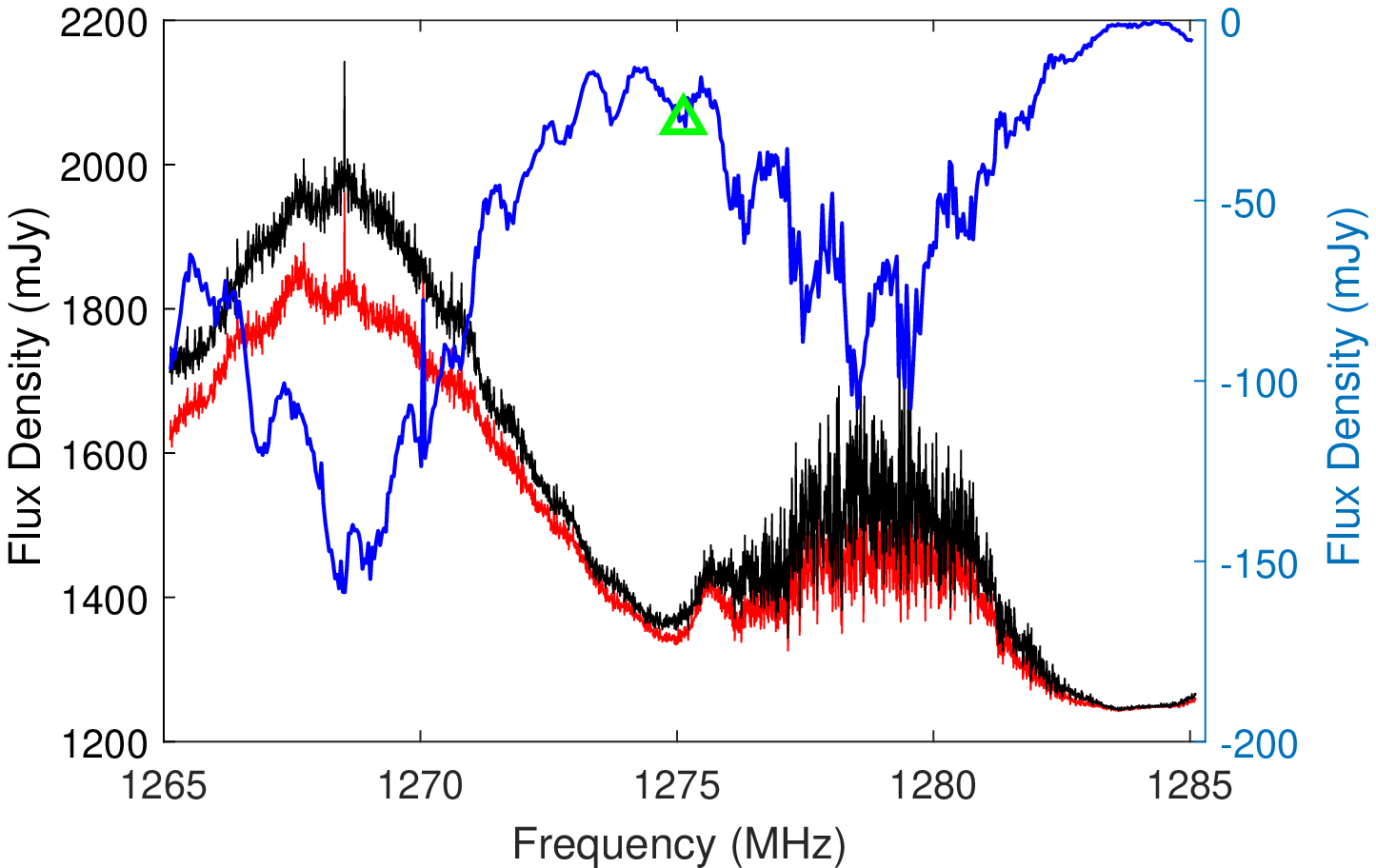}}
 \subfigure[IRAS 14070+0525]{\includegraphics[width=0.485\textwidth,height=5.0cm]{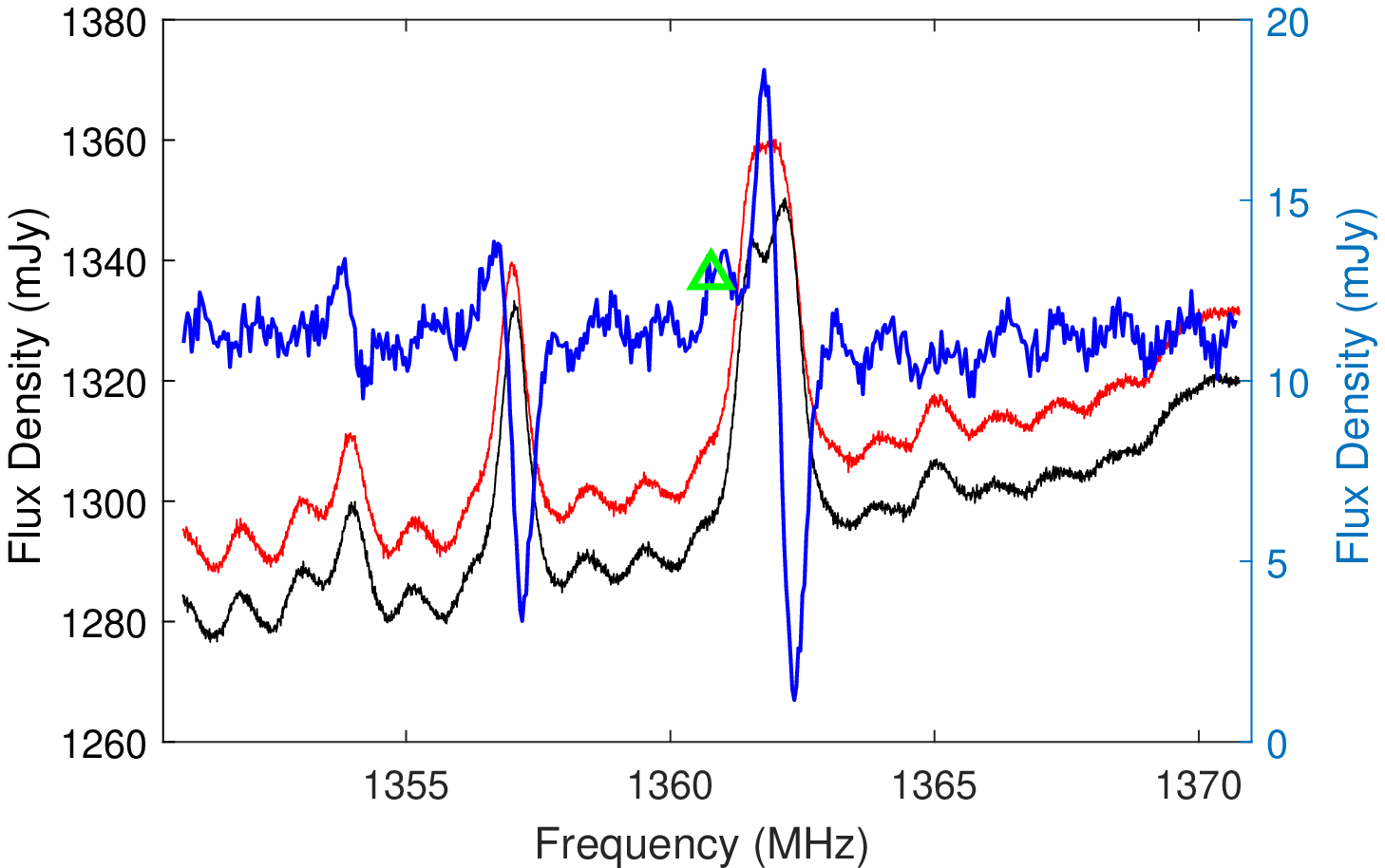}}
  \subfigure[IRAS 22116+0437]{\includegraphics[width=0.485\textwidth,height=5.0cm]{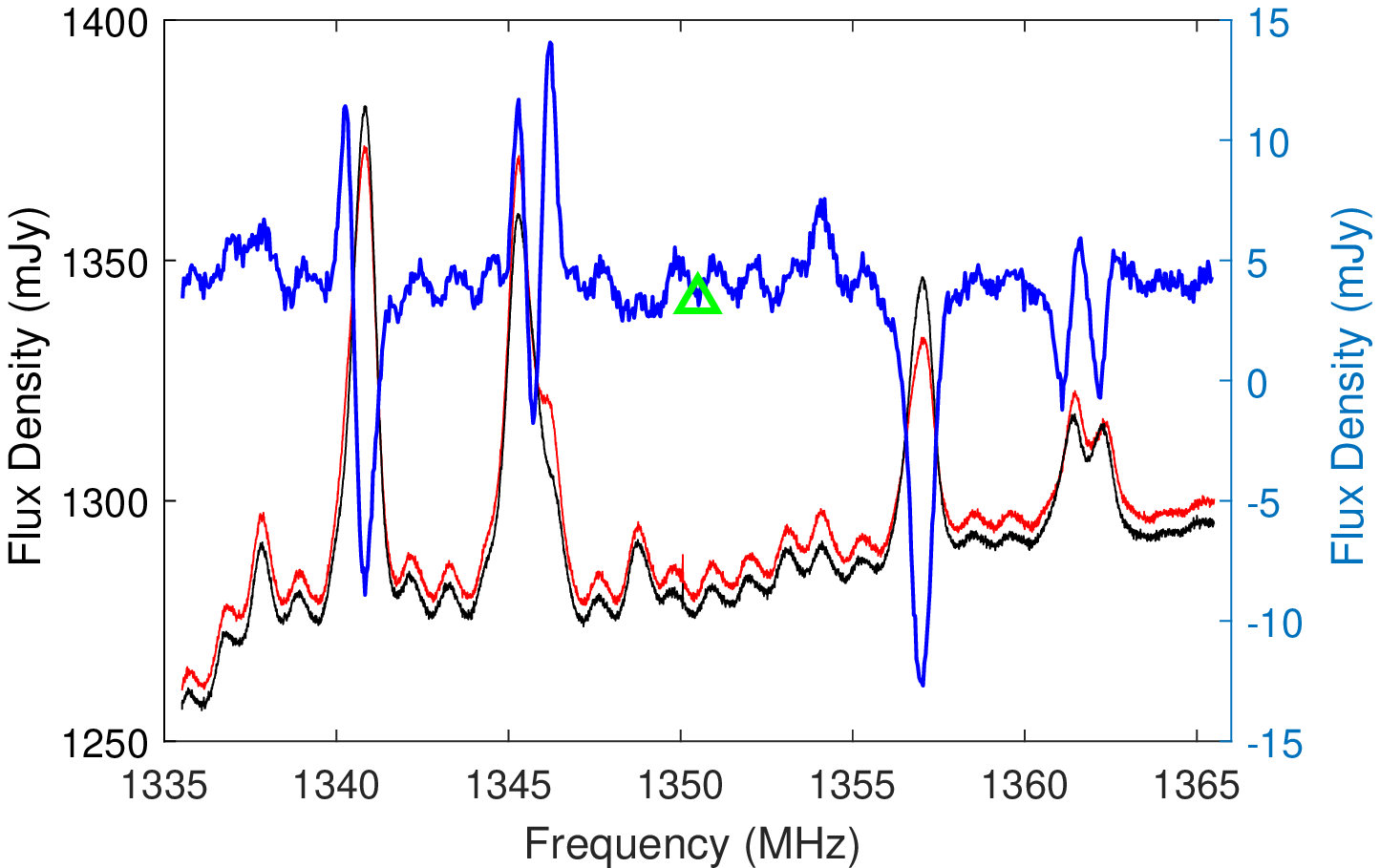}}
 \subfigure[IRAS 22116+0437]{\includegraphics[width=0.485\textwidth,height=5.0cm]{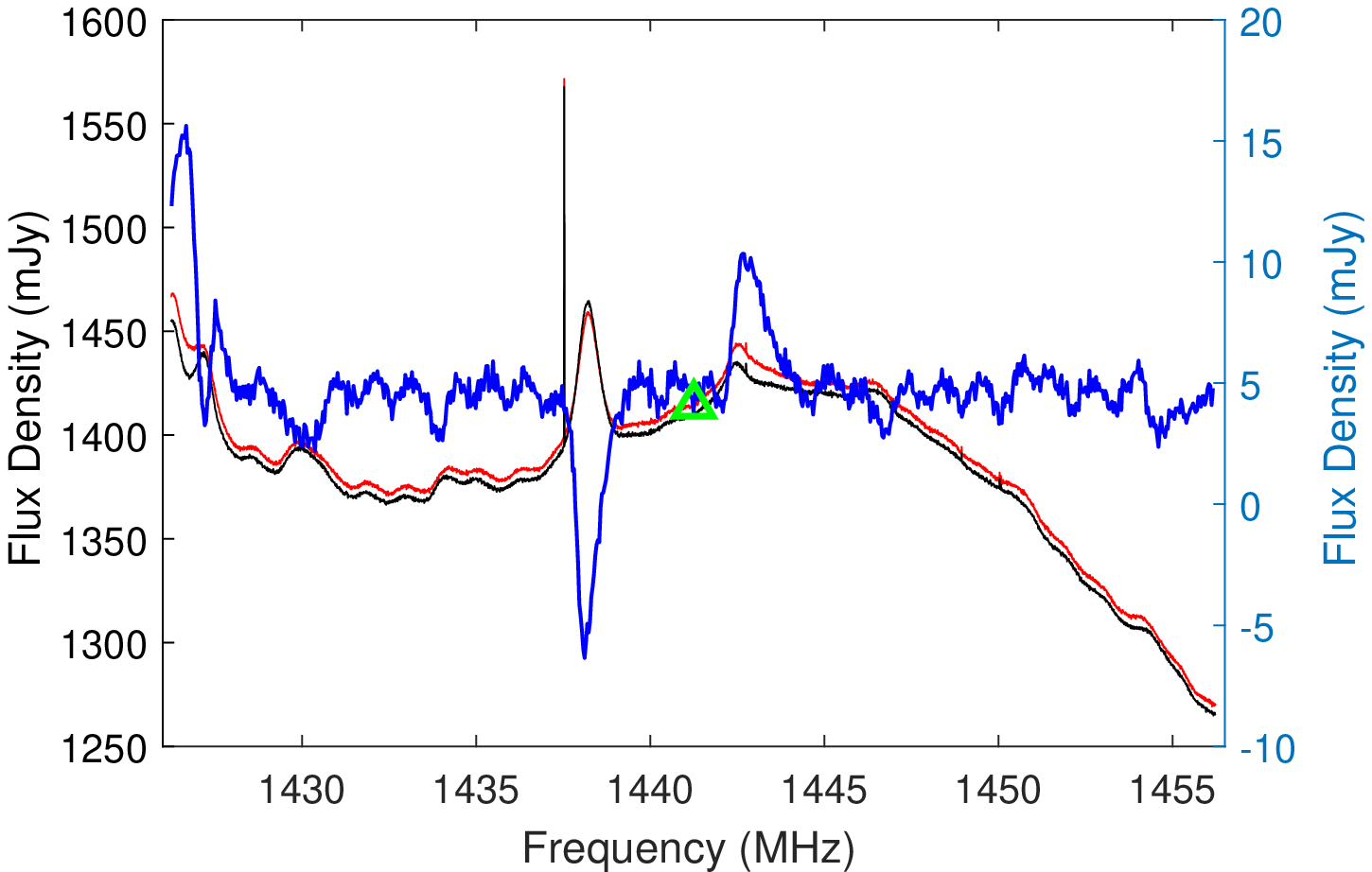}}
   \subfigure[IRAS 23129+2548]{\includegraphics[width=0.485\textwidth,height=5.0cm]{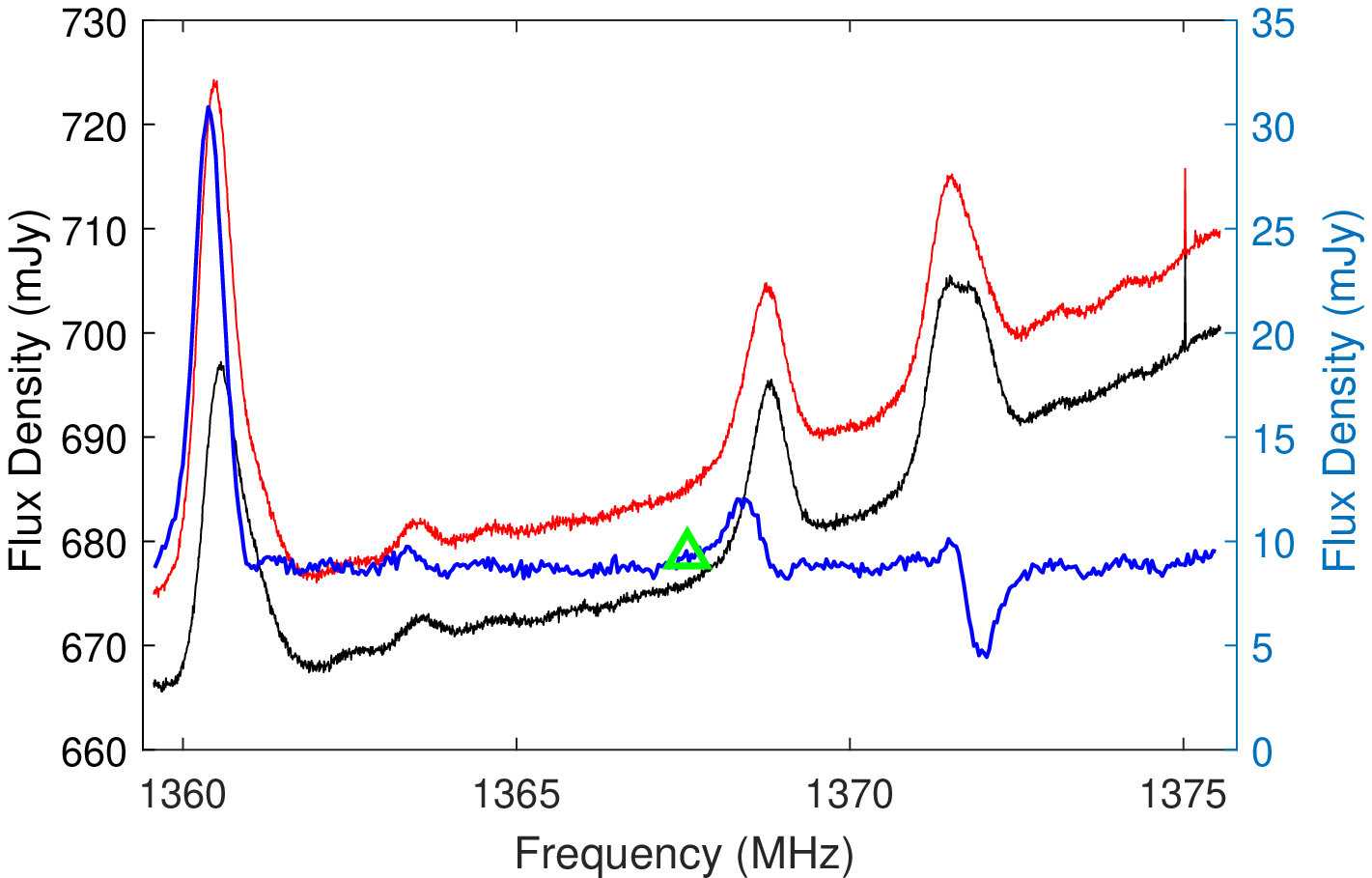}}
 \subfigure[IRAS 23129+2548]{\includegraphics[width=0.485\textwidth,height=5.0cm]{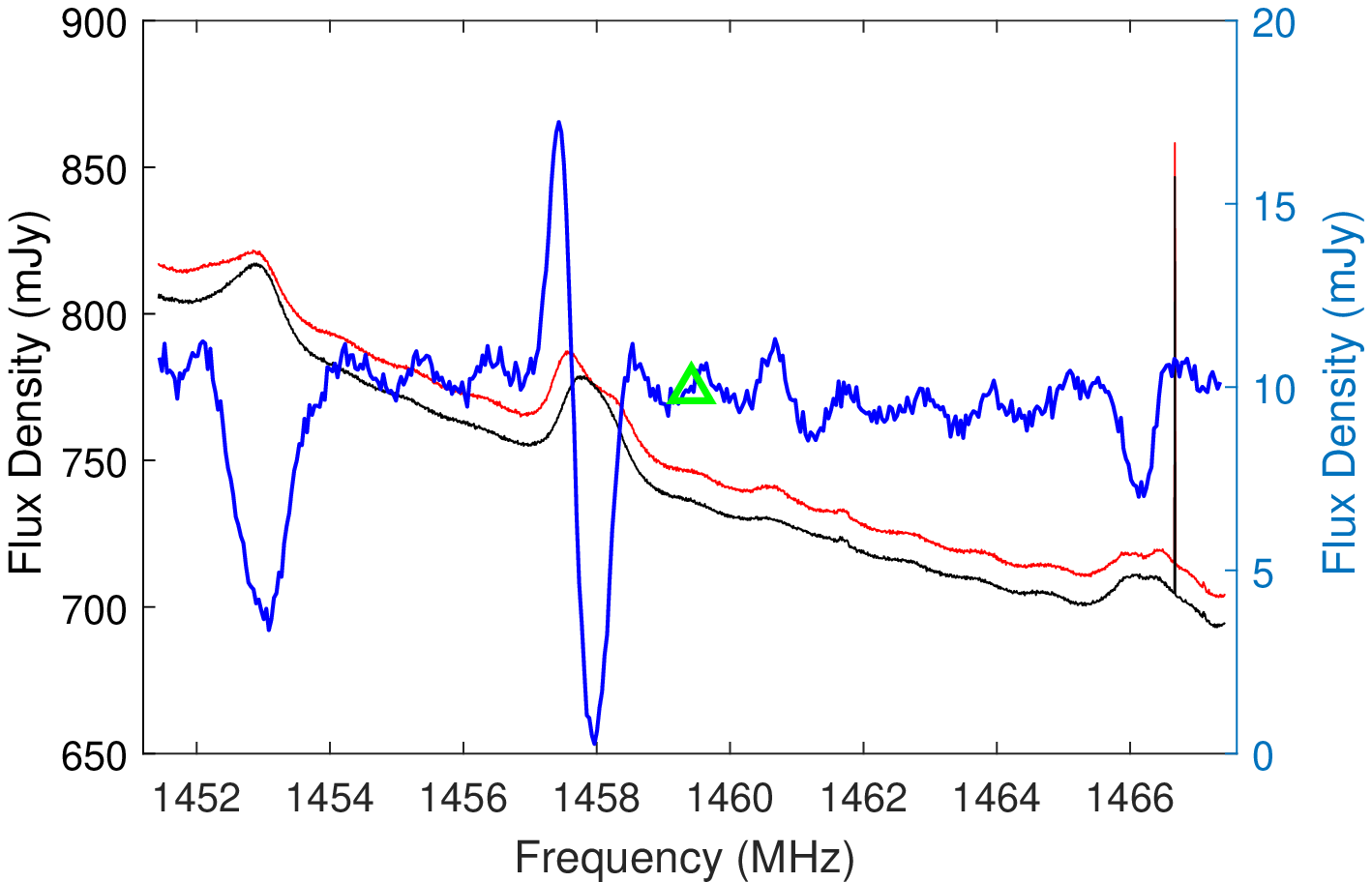}}
 \vspace{0.2in}
       \caption{Continued.
      }
         \label{OHsatellite}
  \end{figure*}
\begin{landscape}                                                                                                                                                                           \setlength{\tabcolsep}{0.01in}                                        
\begin{table}                                                                                                                                                                                                                        
       \caption{The image parameters and fitting results from VLA observations. }                                                                                                                                                     
     \label{imagepar}                                                                                                                                                                                                                 
  \centering                                                                                                                                                                                                                          
  \begin{tabular}{l c c c r r r r r c l r c l}     
  \hline\hline                                                                                                                                                                                                                        
   IRAS Name     & epoch           & projects    & Freq   & peak. Coords          & beam              &  PA    &  $\rm F_{peak}$  &  $\rm F_{tot}$   &  major         &  minor        &   PA     &  3 $\sigma$ &   $\rm T_{B}$     \\
              &                 &             & (GHz)  & (J2000)               & (arcsec)          & $ ^{\circ}$ & (mJy/b)    &   (mJy)          & (arcsec)       & (arcsec)  & $ ^{\circ}$ &  (mJy/b)    &   log(K)    \\      
    
\hline                                                                           01562+2528     & 2003Aug04       & $AD483^{1}$ & 1.43  & 015903.033+254232.964  &  1.49$\times$1.07 &  45.5  &  5.82$\pm$0.06   &  5.90$\pm$0.14   & -              & -             &   -      &  0.31       &  4.2       \\  
                  & 2003Aug04       & $AD483^{2}$ & 1.43  & 015903.039+254232.865  &  1.48$\times$1.07 &  45.6  &  5.81$\pm$0.08   &  5.92$\pm$0.10   & -              & -             &   -      &  1.05       &    -        \\  
                  &   2019Mar17              & VLASS       & 3.00  & 015903.036+254233.089  &     2.84$\times$2.57              &  75.9     &  3.90$\pm$0.16   &  4.04$\pm$0.28   & -              & -             &   -      &  0.42       & 2.7       \\  
                  & 2002Jun25       & AD461       & 4.86  & 015903.048+254233.130  &  0.46$\times$0.44 & 19.9    &  2.58$\pm$0.03   &  2.75$\pm$0.06   & 0.16$\pm$0.03  & 0.04$\pm$0.06 &   89     &  0.39       &  4.6 $\pm$ 0.8      \\  
  02524+2046      & 2003Aug04       & $AD483^{1}$ & 1.41  & 025517.110+205857.357  &  1.47$\times$1.03 &  51.0  &  2.19$\pm$0.09   &  3.19$\pm$0.21   & 1.03$\pm$0.18  & 0.70$\pm$0.19 &   81     &  0.76       &  3.7 $\pm$ 0.1      \\  
                  & 2003Aug04       & $AD483^{2}$ & 1.41  & 025517.104+205857.244  &  1.48$\times$1.10 &  51.2  & 14.58$\pm$0.65   & 14.78$\pm$0.38   & -              & -             &   -      &  1.11       &    -        \\  
                  &     2019Jun07            & VLASS       & 3.00  & 025517.096+205857.391  &    2.75  $\times$  2.23            &   -54.9    &  1.18$\pm$0.12   &  1.57$\pm$0.26   & 1.26$\pm$0.78  & 0.45$\pm$0.75 &   49     &  0.38       &  2.8  $\pm$ 0.8      \\  
                  & 2002Jun25       & AD0461      & 4.86  & 025517.110+205857.227  &  0.47$\times$0.43 &   2.3    &  0.57$\pm$0.03   &  0.91$\pm$0.06   & 0.50$\pm$0.05  & 0.15$\pm$0.11 &   59     &  0.14       &  3.0 $\pm$ 0.3       \\  
  08201+2801      & 2002Apr27       & $AD462^{1}$ & 1.43  & 082312.604+275139.736  &  1.61$\times$1.54 & -60.6  &  3.41$\pm$0.04   &  3.18$\pm$0.07   & -              & -             &   -      &  1.12       &  3.7       \\  
                  & 2002Apr27       & $AD462^{2}$ & 1.43  & 082312.607+275139.676  &  1.61$\times$1.51 & -65.8  &  9.57$\pm$0.11   & 10.51$\pm$0.21   & 0.600$\pm$0.11 & 0.32$\pm$0.29 &  122     &  1.59       &    -        \\  
                  &                 & VLASS       & 3.00  & 082312.631+275139.776  & 3.19 $\times$2.31                &  66.8    &  1.85$\pm$0.11   &  2.07$\pm$0.22   & 2.09$\pm$0.52  & 1.15$\pm$0.49 &   60     &  0.36       &  2.3 $\pm$ 0.2       \\  
   08279+0956     & 2003Aug04       & $AD483^{1}$ & 1.38  & 083039.343+094636.431  &  1.69$\times$1.08 &  27.9  &  2.51$\pm$0.06   &  2.26$\pm$0.11   & -              &  -            &   -      &  0.63       &  3.8       \\  
                  & 2003Aug04       & $AD483^{2}$ & 1.38  & 083039.337+094636.254  &  1.69$\times$1.05 &  28.1  &  2.43$\pm$0.12   &  4.70$\pm$1.30   &  -  & -   &  -     &  0.90       &     -       \\  
                  &    2018Apr28             & VLASS       & 3.00  & 083039.329+094636.616  &  2.62$\times$2.07                 &  -31.9     &  1.21$\pm$0.14   &  1.71$\pm$0.31   &     -   &   -          &      -   &  0.34       &  2.1       \\  
   09531+1430     & 2003Aug04       & $AD483^{1}$ & 1.37  & 095550.242+141607.645  &  1.48$\times$1.15 &  31.7  &  2.56$\pm$0.09   &  2.61$\pm$0.15   & -              &  -             &   -      &  0.65       &  3.9       \\  
                  & 2003Aug04       & $AD483^{2}$ & 1.37  & 095550.247+141607.512  &  1.48$\times$1.15 &  32.0  &  2.84$\pm$0.08   &  2.90$\pm$0.15   & -      &  -       &   -      &  0.80       &      -      \\  
                  &   2020Oct03              & VLASS       & 3.00  & 095550.223+141607.814  &   3.01 $\times$2.38                &  49.6     &  1.80$\pm$0.11   &  1.77$\pm$0.19   & 1.55$\pm$0.54  & 0.50$\pm$0.78 &   94     &  0.35       &  2.7$\pm$ 0.7       \\  
  10339+1548      & 2003Aug04       & $AD483^{1}$ & 1.39  & 103638.010+153240.256  &  1.23$\times$0.97 &  45.6  &  5.01$\pm$0.12   &  6.08$\pm$0.25   & 0.59$\pm$0.48  & 0.09$\pm$0.38 &   31     &  2.66       &  5$\pm$ 2       \\  
                  &     2017Oct06            & VLASS       & 3.00  & 103637.957+153240.562  &3.56 $\times$ 2.29                  &  -56.9  &  2.81$\pm$0.14   &  3.53$\pm$0.03   & 2.55$\pm$0.55  & 1.28$\pm$0.57 &  143     &  0.36       &  2.4$\pm$ 0.2       \\  
  12032+1707      & 2002Apr27       & $AD462^{1}$ & 1.37  & 120547.722+165108.327  &  1.74$\times$1.67 & -43.3  & 27.11$\pm$0.42   & 28.46$\pm$0.76   & 0.50$\pm$0.19  & 0.21$\pm$0.22 &   18     &  0.88       &  5.5$\pm$ 0.5       \\  
                  & 2002Apr27       &$ AD462^{2}$ & 1.37  & 120547.723+165108.270  &  1.72$\times$1.66 & -40.2  &  7.55$\pm$0.49   &  7.49$\pm$0.85   & 0.63$\pm$0.18 & 0.14$\pm$0.61 & 46 &  0.85 &   -         \\  
                  &       2019Mar21          & VLASS       & 3.00  & 120547.727+165108.305  &    3.37$\times$2.34               &  -87.4    & 11.66$\pm$0.15   & 12.48$\pm$0.28   & 1.02$\pm$0.21  & 0.78$\pm$0.42 &   41     &  0.42       &  3.6$\pm$ 0.3       \\  
                  & 2004Oct16       & AN0122      & 8.46  & 120547.723+165108.264  &  0.27$\times$0.24 &  -46.1     &  2.67$\pm$0.09   &  4.96$\pm$0.25   & 0.25$\pm$0.03  & 0.22$\pm$0.03 &   1.7    &  0.12       &  3.4 $\pm$ 0.1       \\  
  13218+0552      & 2003Aug04       & $AD483^{1}$ & 1.38  & 132419.882+053704.520  &  1.92$\times$1.04 &  27.1  &  3.30$\pm$0.22   &  3.11$\pm$0.38   & 0.99$\pm$0.33  & 0.74$\pm$0.71 &   175    &  1.20       &  3.7$\pm$ 0.4      \\  
                  & 2003Aug04       & $AD483^{2}$ & 1.38  & 132419.901+053705.029  &  1.93$\times$1.05 &  27.4  &  1.99$\pm$0.26   &  1.73$\pm$0.43   & 1.63$\pm$0.71  & 0.47$\pm$0.57 &   169    &  1.24       &    -        \\  
                  &     2019May13            & VLASS       & 3.00  & 132419.883+053704.840  &    2.43$\times$2.04               &  13.9     &  2.04$\pm$0.19   &  3.40$\pm$0.47   & 0.70$\pm$0.36  & 0.38$\pm$0.17 &    14    &  0.46       &  3.5 $\pm$ 0.3       \\  
                  & 1989Jan10       & AH0333      & 4.86  & 132419.890+053704.667  &  0.50$\times$0.41 &  -1.4     &  1.25$\pm$0.06   &  1.64$\pm$0.12   & 0.30$\pm$0.06 & 0.21$\pm$0.06 &   1.1    &  0.19       &  3.4$\pm$ 0.2       \\  
                  & 2014Mar10       & 14A-332     & 8.50  & 132419.887+053704.706  &  0.29$\times$0.24 &  9.6     &  0.79$\pm$0.02   &  0.92$\pm$0.04   & 0.13$\pm$0.023 & 0.09$\pm$0.03 &   175    &  0.04       &  3.4$\pm$ 0.2       \\  
                  & 2014Mar10       & 14A-332     & 9.50  & 132419.886+053704.715  &  0.26$\times$0.22 &  10.90     &  0.71$\pm$0.02   &  0.86$\pm$0.04   & 0.12$\pm$0.021 & 0.09$\pm$0.04 &   112    &  0.04       &  3.3 $\pm$ 0.2       \\  
  14586+1432      & 2003Aug03       & $AD483^{1}$ & 1.45  & 150102.179+142002.221  &  1.44$\times$0.87 &  27.5  & 10.01$\pm$0.36   & 11.04$\pm$0.69   & -              & -             &    -     &  1.09       &  4.5       \\  
                  & 2003Aug03       &$ AD483^{2}$ & 1.45  & 150102.165+142001.988  &  1.60$\times$0.96 &  27.5  &  4.74$\pm$0.40   &  7.70$\pm$1.00   & 1.58$\pm$0.39  & 0.53$\pm$0.29 &    39    &  2.35       &     -       \\  
                  &   2019Apr25              & VLASS       & 3.00  & 150102.194+142002.470  &  3.01$\times$2.19                 &  51.1     & 10.09$\pm$0.15   & 10.26$\pm$0.28   & 3.05$\pm$0.06  & 2.20$\pm$0.03 &    33    &  0.39       &  2.54 $\pm$ 0.02       \\  
                  & 2002Jan26       & AD0461      & 4.86  & 150102.177+142002.291  &  0.48$\times$0.42 &  -13.1     &  8.55$\pm$0.04   &  8.82$\pm$0.07   & 0.12$\pm$0.01  & 0.04$\pm$0.02 &    71    &  0.46       &  5.2 $\pm$ 0.2       \\  
  21272+2514      & 2002Apr27       & $AD462^{1}$ & 1.45  & 212929.394+252754.581  &  1.59$\times$1.47 &  50.0  &  3.04$\pm$0.06   &  3.15$\pm$0.11   & -              & -             &    -     &  0.45       &  3.7       \\  
                  & 2002Apr27       & $AD462^{2}$ & 1.45  & 212929.388+252754.550  &  1.59$\times$1.47 &  51.6  &  7.37$\pm$0.15   &  7.72$\pm$0.28   & 0.39$\pm$0.20  & 0.34$\pm$0.24 &   106    &  0.50       &    -        \\  
                  &    2019May07             & VLASS       & 3.00  & 212929.368+252754.733  &   3.26$\times$2.30                &  -86.0      &  2.54$\pm$0.06   &  2.48$\pm$0.15   & -              & -             &    -     &  0.35       &  2.5       \\  
  23028+0725      & 2002Apr27       & $AD462^{1}$ & 1.45  & 230519.901+074143.189  &  1.67$\times$1.53 &  4.1  & 13.18$\pm$0.25   & 16.50$\pm$0.51   & 0.92$\pm$0.11  & 0.67$\pm$0.15 &    51    &  0.56       &  4.4$\pm$ 0.1       \\  
                  & 2002Apr27       & $AD462^{2}$ & 1.45  & 230519.898+074143.155  &  1.65$\times$1.52 &  4.5  &  4.39$\pm$0.18   &  4.90$\pm$0.34   & 0.68$\pm$0.38  & 0.34$\pm$0.34 &   129    &  0.79       &    -        \\  
                  &        2020Aug12         & VLASS       & 3.00  & 230519.910+074143.501  &    3.12$\times$2.31               &   20.7     &  7.00$\pm$0.14   &  7.61$\pm$0.25   & 1.42$\pm$0.22  & 0.38$\pm$0.28 &    47    &  0.36       &  3.5 $\pm$ 0.3       \\  
                                                                                                               
    \hline                                                                                                                                                                                                                  
     \end{tabular}                                                                                                                                                                                                          
 \end{table}     
 \end{landscape}                                                                                                                                                                                          
 \addtocounter{table}{-1}                                                                                                                                                                                                   
 \begin{landscape}                                                                                                                                                  \setlength{\tabcolsep}{0.01in}                            
 \begin{table}                                                                                                                                                                                                             
       \caption{The image parameters and fitting results from VLA observations.  }                                                                                                                                          
     \label{imagepar}                                                                                                                                                                                                       
  \centering                                                                                                                                                                                                                
  \begin{tabular}{l c c c r r r r r c l r c l}     
  \hline\hline                                                                                                                                                                                                              
IRAS Name     & epoch           & projects    & Freq   & peak. Coords          & beam              &  PA    &  $\rm F_{peak}$  &  $\rm F_{tot}$   &  major         &  minor        &   PA     &  3 $\sigma$ &   $\rm T_{B}$     \\
              &                 &             & (GHz)  & (J2000)               & (arcsec)          & $^{\circ}$ & (mJy/b)    &   (mJy)          & (arcsec)       & (arcsec)  & $^{\circ}$ &  (mJy/b)    &   log(K)    \\      
           
\hline                                                                                                                                                                                                                      
23129+2548     & 2003Aug03       & $AD483^{1}$ & 1.41  & 231521.388+260432.436  &  1.51$\times$1.03 &  28.1  &  4.12$\pm$0.06   &  4.23$\pm$0.11 & -              & -             &    -      & 0.388       &   4.1  \\ 
               & 2003Aug03       & $AD483^{2}$ & 1.41  & 231521.389+260432.305  &  1.54$\times$1.03 &  28.2  &  3.49$\pm$0.14   &  3.53$\pm$0.25 & -              & -             &    -      & 0.64       &   -     \\ 
               &    2019May16             & VLASS       & 3.00  & 231521.358+260432.680  &   2.49$\times$2.19                &  -52.4     &  3.06$\pm$0.18   &  3.43$\pm$0.34 & 3.15$\pm$0.86  & 0.50$\pm$0.80 &   148     & 1.20       &   2.7$\pm$ 0.7  \\

 01355-1814          & 2018Feb09 &   VLASS     &3.00 &  013757.469-175921.285  &    2.84 $\times$1.69    &54.9   &  0.68 $\pm$0.15   &   1.29             $\pm$ 0.41  &  -$\pm$-    &   -$\pm$-   & - & 0.42 &2.4 \\ 
             & 2001May20    &  AL537   &  8.46 & 013757.469-175921.285 &  1.68    $\times$0.79   &   -24.6&  0.90 $\pm$0.03    &                  0.94$\pm$ 0.06   &   -   & -  & - &  0.099 &1.9 \\
03521+0028  & 2017Nov19 &   VLASS     &3.00 & 035442.262+003703.124   &    2.74 $\times$2.31    &11.4   &   3.69$\pm$0.13   &    3.59$\pm$0.22             &  -$\pm$-    &   -$\pm$ -  & - &0.39  &2.7 \\        
03521+0028A &2004Oct19  & AN144      &8.46 &035442.278+003703.220    &   0.36  $\times$0.26    & -46.3  & 4.30  $\pm$ 0.04  &  4.40               $\pm$0.08   & 0.06 $\pm$ 0.04  & 0.03$\pm$ 0.02  &  105& 0.10 &4.9$\pm$ 0.5 \\
            &2002Jan25  & AD461      &4.86 &035207.985+002817.436    &   0.55  $\times$0.43    & -10.6  & 3.90  $\pm$ 0.09  &  3.87 $\pm$0.15   & -   &  -   &  -& 0.15 &3.8 \\
03521+0028B &2004Oct19  & AN144      &8.46 &035442.174+003703.280    &   0.36  $\times$0.26    & -46.3  & 0.32  $\pm$ 0.03  &  0.43               $\pm$0.07   & -  & -   &  -& 0.10 & 2.7\\   
            &2002Jan25  & AD461      &4.86 &035207.885+002817.434    &   0.55  $\times$0.43    & -10.6  & 0.44  $\pm$ 0.05  &  0.44               $\pm$0.09   & -   &  -   &  -& 0.14 & 2.8\\             
 07572+0533  & 2017Nov25 &   VLASS     &3.00 & 075959.742+052451.499  & 2.79    $\times$2.12   &  -32.2 & 3.04 $\pm$ 0.18 &  2.86 $\pm$ 0.30   &- & - & - & 0.55&2.7 \\ 
             &2002Jan25 &   AD461     &4.84 & 075959.738+052451.583 &  0.72    $\times$0.45   &   -47.5 & 4.67 $\pm$ 0.08 &  5.40                 $\pm$ 0.15   & 0.35 $\pm$  0.04 & 0.04  $\pm$ 0.08 & -47.5 & 0.23& 4.5$\pm$ 0.9\\

08449+2332   & 2016Oct13 &   14A-332     &1.49 & 084750.257+232110.799   &   1.42  $\times$1.26    &20.1   &  2.36 $\pm$0.04   &                  3.32 $\pm$ 0.08  &0.96 $\pm$ 0.07   &   0.71$\pm$ 0.09  & 137 & 0.11 &3.7$\pm$0.1 \\                    
            & 2019Apr13 &   VLASS     &3.00 & 084750.258+232110.890   &   3.16  $\times$2.19    &-63.8   &  1.90 $\pm$0.12   &   2.28 $\pm$ 0.25  &1.77 $\pm$ 0.66   &   0.67$\pm$ 0.36  & 108 & 0.369 & 2.6$\pm$ 0.3\\
            & 2016Dec02 &   16B-063     &9.00 &  084750.260+232110.807   &   0.25  $\times$0.21    &11.7   &  0.50 $\pm$0.01   &   0.71 $\pm$ 0.03  &0.19 $\pm$ 0.02   &   0.11$\pm$ 0.03  & 131 & 0.03 & 2.9$\pm$ 0.1\\ 
  
11180+1623A             & 2021Nov21 &   VLASS     &3.00 & 112041.737+160656.914   &     2.57$\times$2.16    &  37.1 &   1.48$\pm$0.13   &                       2.11$\pm$0.29   &  2.27$\pm$0.58    &   0.81$\pm$0.45   & 29 & 0.32 &2.4$\pm$0.3 \\
                & 2004OCT16 &   AN122     &8.46 & 112041.751+160656.729   & 0.26   $\times$ 0.24   & -35.4  &  0.39 $\pm$0.03   &  1.33                $\pm$ 0.15  & 0.76 $\pm$ 0.09   &  0.08 $\pm$ 0.05  & 34 & 0.083 & 2.8$\pm$ 0.3\\
11180+1623B & 2021Nov21 &   VLASS     &3.00 & 112041.157+160656.050   &     2.57$\times$2.16    &   37.1&   1.21$\pm$0.11   &                            1.61$\pm$0.23   & - $\pm$-    &   -$\pm$-   & - & 0.32 & 2.4\\
              &2004OCT16  &   AN122     &8.46 & 112041.173+160656.430   &  0.26   $\times$ 0.24   & -35.4  &  0.93 $\pm$0.03   & 1.05  $\pm$ 0.05  & 0.12 $\pm$ 0.03   &  0.05 $\pm$ 0.03  & 133 & 0.08 &3.7$\pm$ 0.3 \\
21077+3358  & 2014May23 &   16B-063     &1.49 & 210950.656+341033.738   &   1.39  $\times$1.30    &80.6   &  3.10 $\pm$0.03   &   7.60             $\pm$ 0.11  &2.70 $\pm$ 0.04   &   0.66$\pm$ 0.03  & 77 & 0.096 & 3.61$\pm$0.02\\    
            & 2019Jun05 &   VLASS     &3.00 &210950.647+341033.863    &   2.44  $\times$2.23    &  -87.6 &   3.42$\pm$0.14   &   4.08 $\pm$0.27   & 1.39 $\pm$0.25    &  0.42 $\pm$0.51   & 26 &0.39  & 3.2$\pm$ 0.5\\
            & 2014May23 &   16B-063     &9.00 & 210950.654+341033.849   &   0.27  $\times$0.24    &86   &  0.49 $\pm$0.01   &   1.0 $\pm$ 0.04  &0.42 $\pm$ 0.02   &   0.12$\pm$ 0.01  & 80 & 0.038 &2.72$\pm$0.05 \\

    \hline                                                                                                                                                                                                                  
     \end{tabular}                                                                                                                                                                                                          
 \end{table}                                                                                                                                                                                                               
\vskip 0.1 true cm \noindent Notes. Column (1) IRAS name. Column (2) Observing date of VLA projects. Column (3) projects code. Column (4) Frequency of observations in GHz. Col (5)-(13) Fitted peak position, beam size and position angle, peak flux(mJy/beam), the integrated flux (mJy), the deconvolved major and minor diameter of the source in arcseconds and position angle, the 3 $\sigma$ noise level of the image. Column (14) The derived brightness temperature of the source (in K) using equation: $T_{\mathrm{B}}$ =1.8 $\times$ $10^{9}$(1+z)( $\frac{F_{\mathrm{tot}}}{1 mJy}$)( $\frac{\nu}{1  GHz})^{-2}$( $\frac{\theta_{max}\theta_{min}}{1 mas^{2}})^{-1}$ K \,,
    
   \noindent
   Where, $\theta$$_{max}$ and $\theta$$_{min}$ are the fitted major and minor of these sources (in mas) as listed in Column (10) and (11) of this table; The major and minor can not be well fitted for several sources because they are likely much smaller than the beam size, we adopted half of the beam FWHM as their value. $z$  is the optical redshift.

 \end{landscape}

\begin{figure*}
   \centering
      \includegraphics[width=0.4\textwidth,height=6cm]{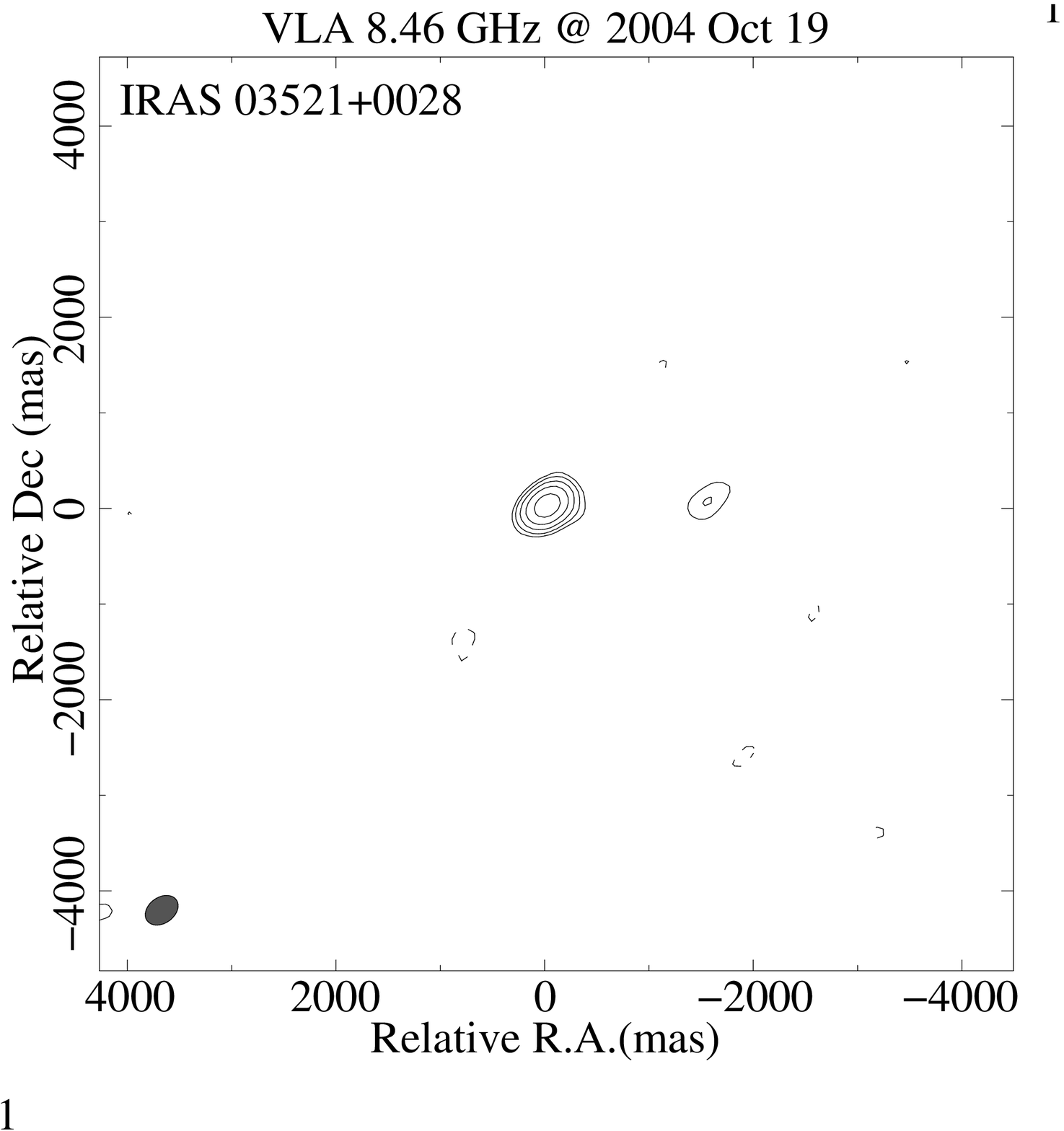}
   \includegraphics[width=0.4\textwidth,height=6cm]{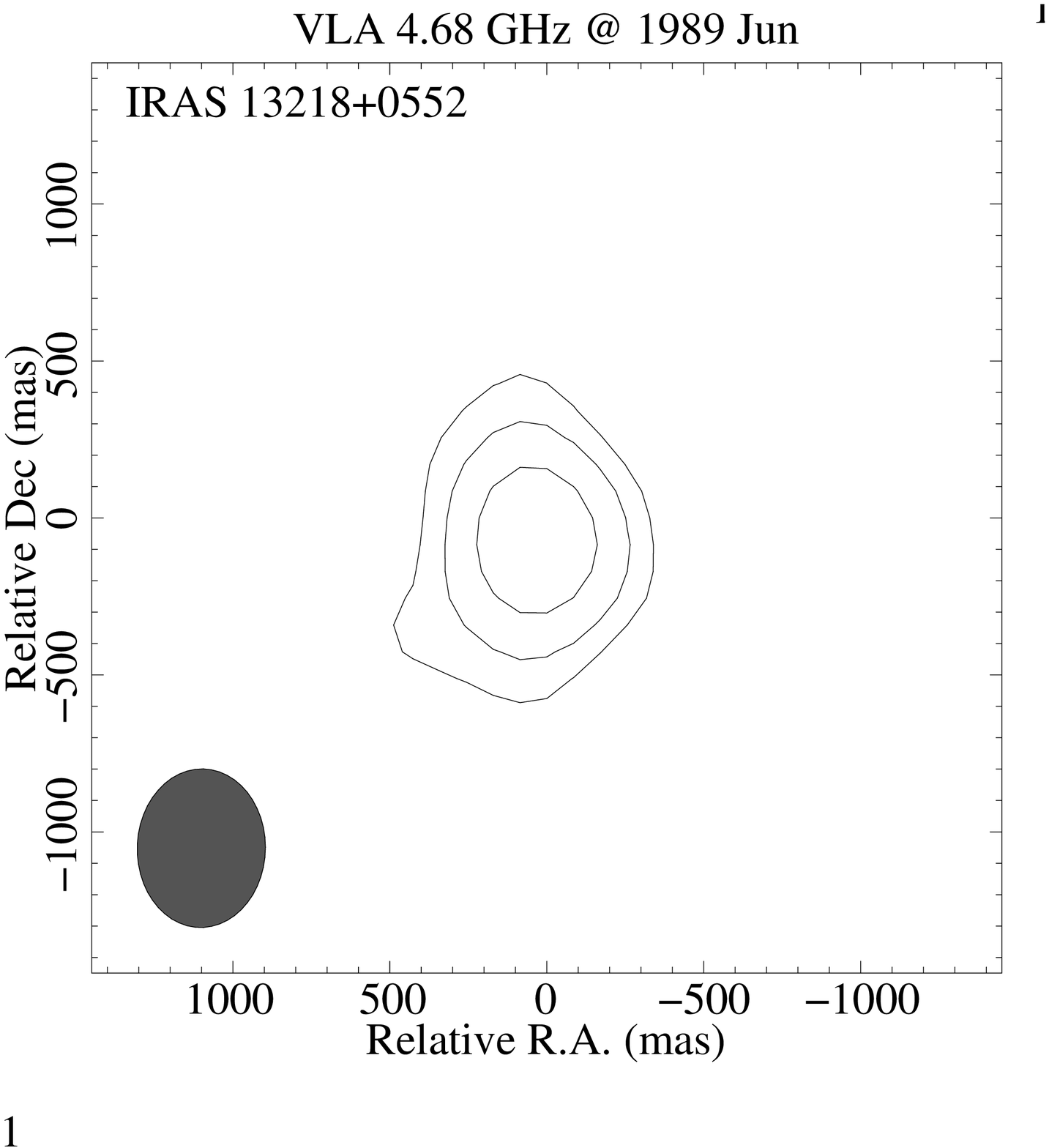}
   \includegraphics[width=0.4\textwidth,height=6cm]{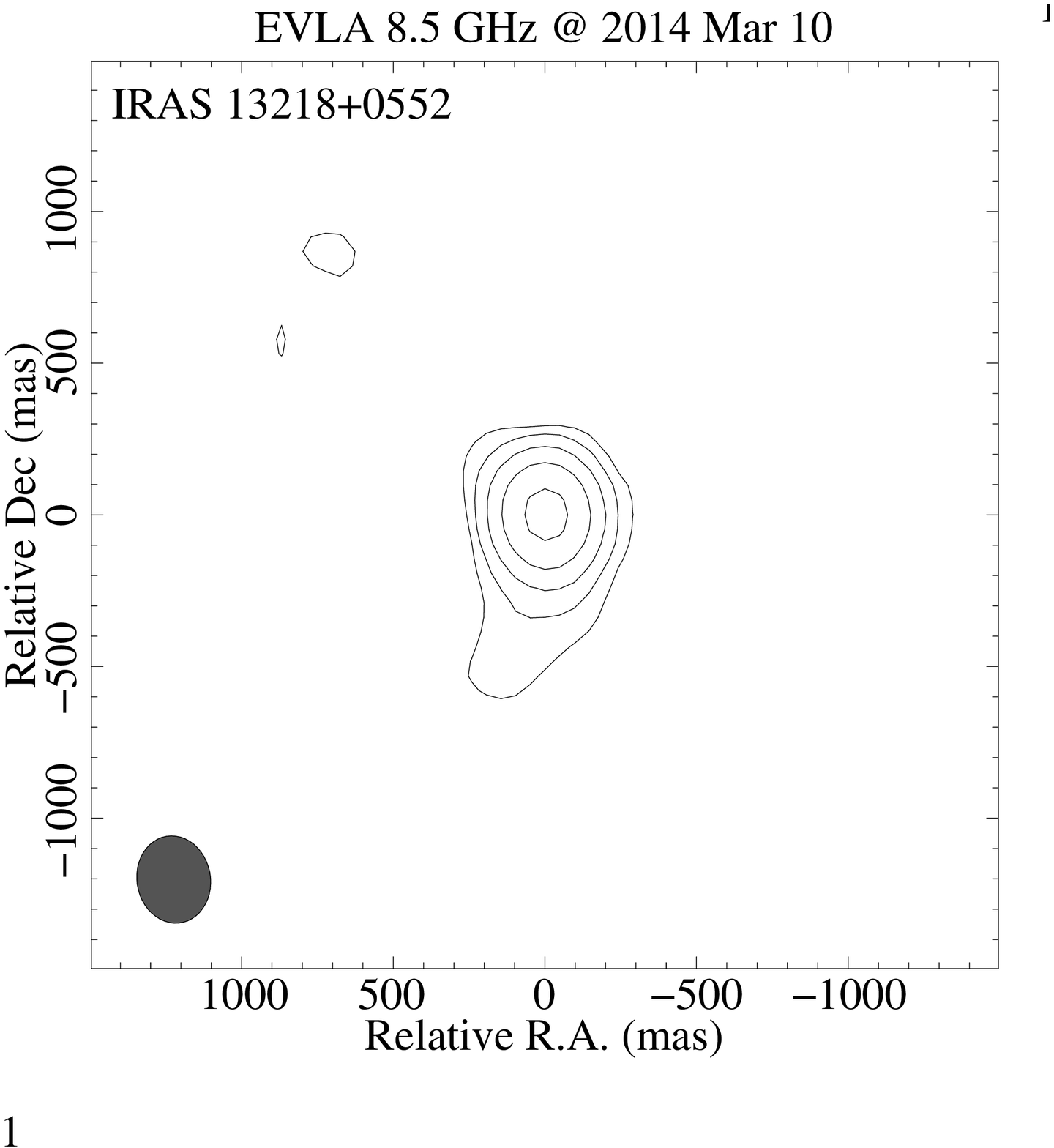}
    \includegraphics[width=0.4\textwidth,height=6cm]{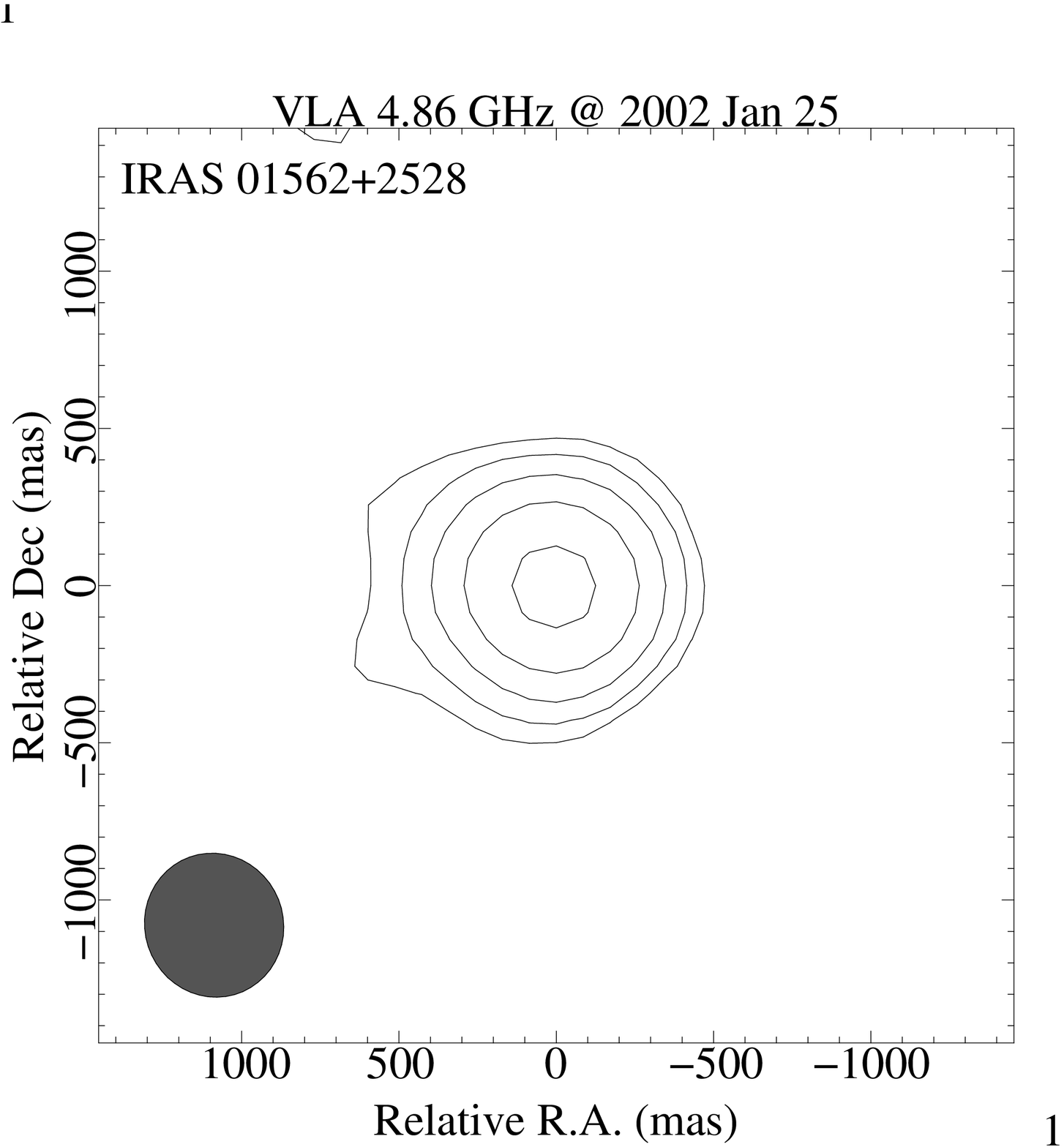}
   \includegraphics[width=0.4\textwidth,height=6cm]{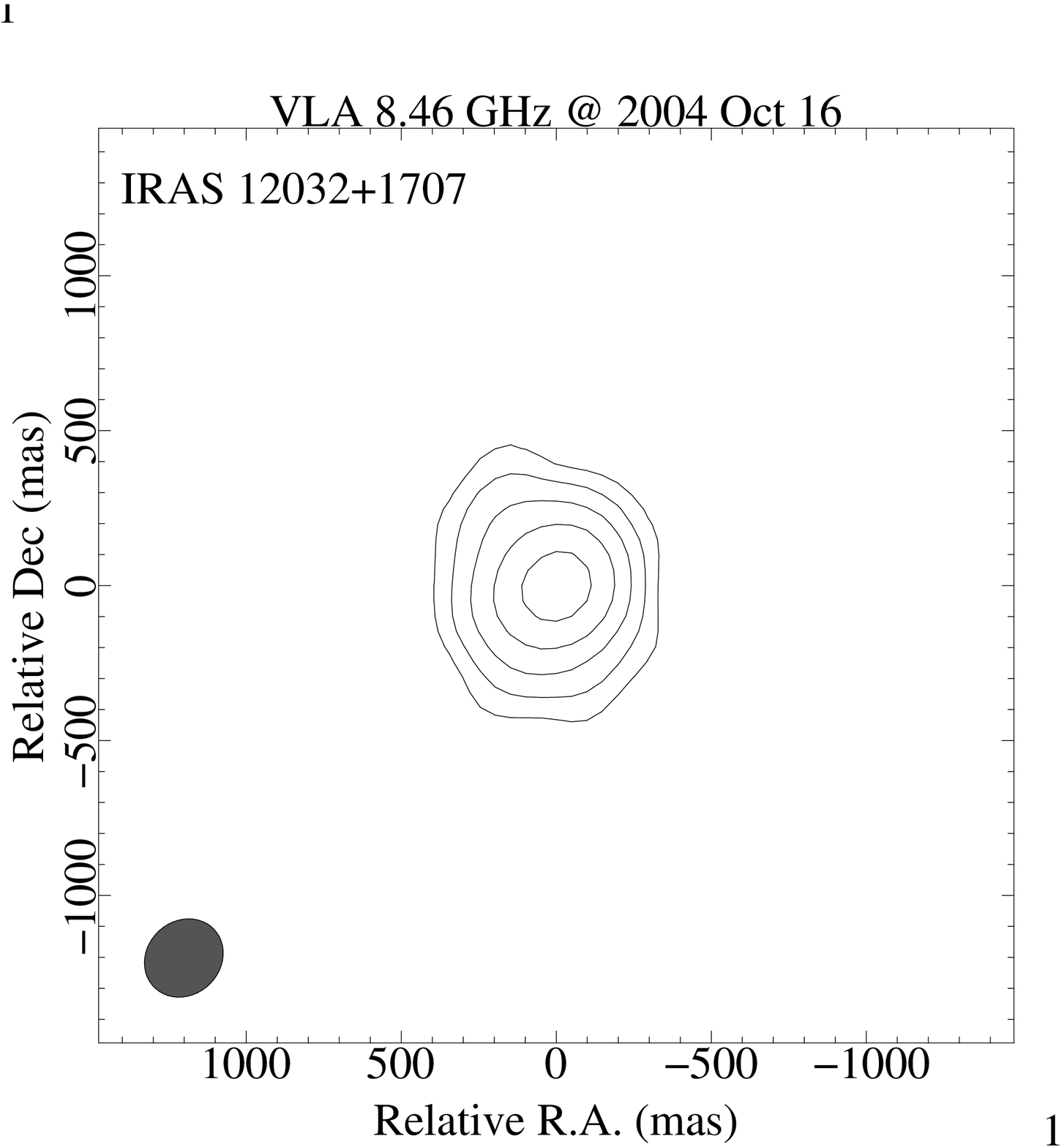}
   \includegraphics[width=0.4\textwidth,height=6cm]{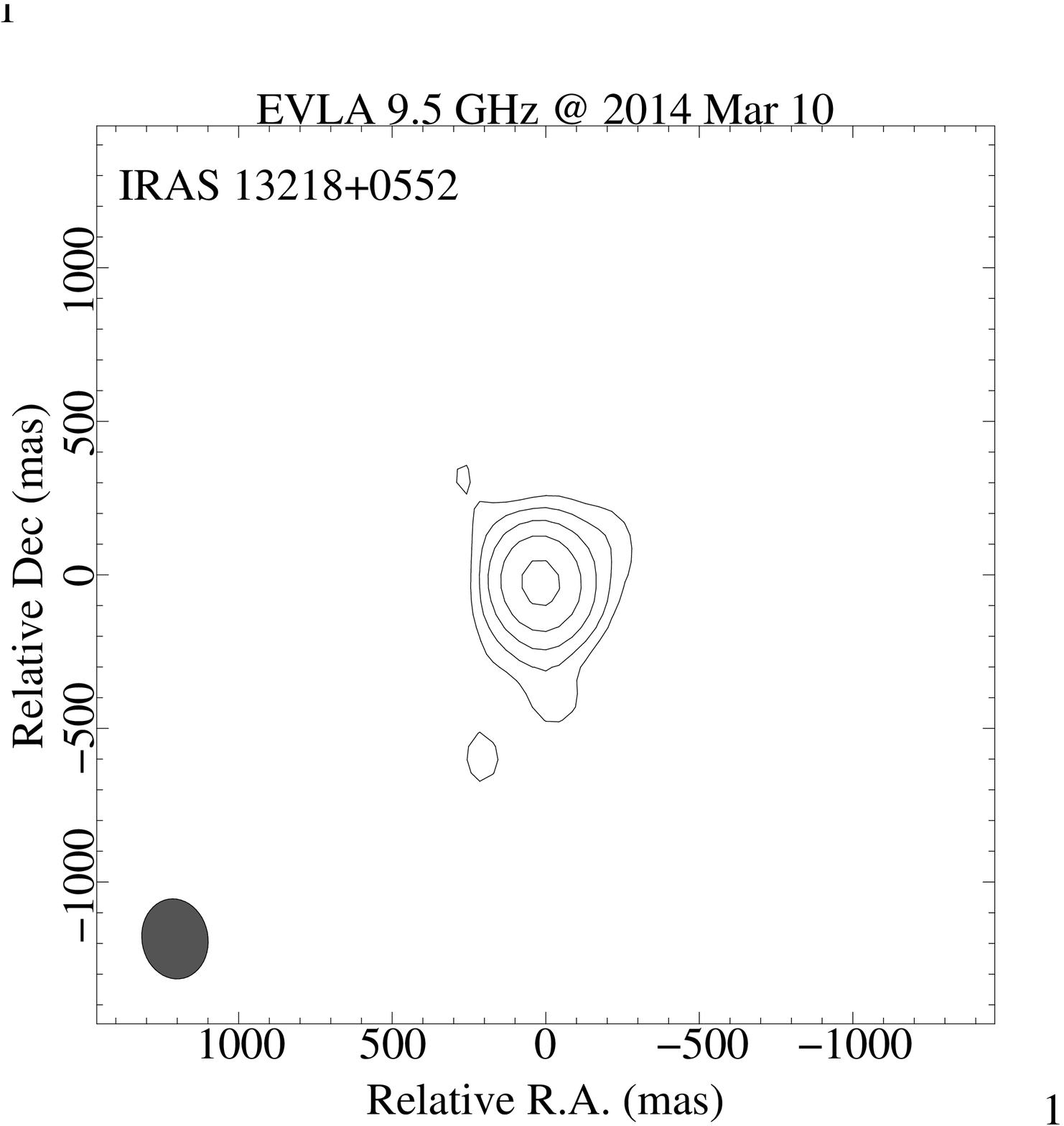}

      \caption{VLA continuum images of the OHMs at C-band and X-band. The images parameter including map peak, beam FWHM and  3 $\sigma$ noise level were present in Table A.1.
      }
         \label{highfreq2}
   \end{figure*}
\end{document}